\documentclass[algorithms,article,accept,moreauthors,pdftex,10pt,a4paper]{mdpi} 
\usepackage[]{algorithm2e}
\usepackage{booktabs} 
\usepackage{multirow}
\usepackage{soul} 
\usepackage{microtype}
\usepackage{subfigure} 
\makeatletter
\renewcommand{\@thesubfigure}{\normalsize(\textbf{\alph{subfigure}})}
\makeatother

\firstpage{1} 
\articlenumber{93}
\doinum{10.3390/a10030093}
\pubvolume{10}
\pubyear{2017}
\copyrightyear{2017}
\externaleditor{}
\history{Received: 5 June 2017; Accepted: 15 August 2017; Published: 19 August 2017}

\pdfoutput=1

\usepackage{bm}
\usepackage{color}

\newif\ifcomments
\commentstrue
\ifcomments 
        \newcommand{\todo}[1]{{\color{red}\noindent $\bullet$ #1}}
        \newcommand{\pjm}[1]{{\color{magenta}{#1}}}
        \newcommand{\drt}[1]{{\color{cyan}{#1}}}
        \newcommand{\whw}[1]{{\color{green}{#1}}}
\else 
        \newcommand{\todo}[1]{}
        \newcommand{\pjm}[1]{}
        \newcommand{\drt}[1]{}
        \newcommand{\whw}[1]{}
        \renewcommand{\st}[1]{}
\fi

\Title{Post-Processing Partitions to Identify Domains of Modularity Optimization}

\Author{William H. Weir $^{1,2,*}$, Scott Emmons $^{1}$, Ryan Gibson $^{1}$, Dane Taylor $^{1}$ and Peter J. Mucha $^{1,2}$}

\AuthorNames{William H. Weir, Scott Emmons, Ryan Gibson, Dane Taylor and Peter J. Mucha}

\address[]{%
$^{1}$ \quad Carolina Center for Interdisciplinary Applied Mathematics, Department of Mathematics, 
University of North Carolina, Chapel Hill, NC 27599, USA; scott3@live.unc.edu (S.E.); ragibson@live.unc.edu~ (R.G.); dane.r.taylor@gmail.com (D.T.); mucha@unc.edu (P.J.M.)\\
$^{2}$ \quad Curriculum in Bioinformatics and Computational Biology,  University of North Carolina, Chapel Hill, NC~27599, USA}

\corres{Correspondence: wweir@med.unc.edu}

\abstract{We introduce the Convex Hull of Admissible Modularity Partitions (CHAMP) algorithm to prune and prioritize different network community structures identified across multiple runs of possibly various computational heuristics. 
Given a set of partitions, CHAMP identifies the domain of modularity optimization for each partition {---i.e., the parameter-space domain where it has the largest modularity relative to the input set---discarding partitions with empty domains to obtain the subset of partitions that are ``admissible'' candidate community structures that remain potentially optimal over indicated parameter domains.}
Importantly, CHAMP can be used for multi-dimensional parameter spaces, such as those for multilayer networks where one includes a resolution parameter and interlayer coupling. Using the results from CHAMP, a user can more appropriately select robust community structures by observing the sizes of domains of optimization and the pairwise comparisons between partitions in the admissible subset. We demonstrate the utility of CHAMP with several example networks.  {In these examples, CHAMP focuses attention onto pruned subsets of admissible partitions that are 20-to-1785 times smaller than the sets of unique partitions obtained by community detection heuristics that were input into CHAMP}.}


\begin{document}


\section{Introduction}
Networks are a natural and powerful representation for relational data, providing access to a large repertoire of analytic tools that may be leveraged to better understand the underlying data in numerous applications. Among the many popular methods developed throughout social network analysis and network science, community detection provides a valuable vehicle for exploring, visualizing, and modeling network data. The identification and characterization of community structures also highlights subgraphs that may be of special interest, depending on the application. Many methods for community detection are available and have been employed meaningfully in applications (see, e.g., reviews \cite{Porter:2009,Fortunato:2010,Fortunato:2016,Abbe:2017,Schaub:2017,Shai:2017}). 

Some of the most heavily used computational heuristics for finding communities involve optimizing a quantity known as modularity, which was introduced by Newman and Girvan \cite{NewmanGirvan:2004} and measures the total weight of within-community edges relative to the expected weight in a corresponding ``null-model'' random graph. 
For (possibly weighted) undirected networks that are compared to the configuration null model, modularity is given by \cite{NewmanGirvan:2004}

\begin{equation}
\label{eq:mod}
Q(\gamma)=\frac{1}{2m}\sum_{i,j}{\left( A_{ij}-\gamma \frac{k_ik_j}{2m}\right)\delta(c_i,c_j)}\,,
\end{equation}

\noindent
where $A_{ij}$ are the elements of the adjacency matrix describing the presence (and possibly weights) of edges between nodes $i$ and $j$, $k_i=\sum_j A_{ij}$ is the strength (weighted degree) of node $i$, $m=\frac{1}{2}\sum_i k_i$ is the total edge weight, $c_i$ indexes the community to which node $i$ has been assigned, $\delta(c_i,c_j)=1$ if $c_i=c_j$ and $0$ otherwise, and $\gamma$ is a resolution parameter introduced by Reichardt and Bornholdt \cite{Reichardt:2006} to influence the number and sizes of communities obtained, whereas $\gamma=1$ in the original formulation~\cite{NewmanGirvan:2004}. Tuning the resolution parameter $\gamma$ can reveal community structures at multiple scales, which offers one strategy to overcome the ``resolution limit'' of modularity \cite{Fortunato:2007} wherein small communities in sufficiently large networks cannot be detected via Equation~\eqref{eq:mod} given fixed $\gamma$. See also \cite{Arenas:2008,Granell:2011} for an alternative approach for resolving multiple scales.

Formulae analogous to Equation~\eqref{eq:mod} exist to define modularity for a variety of other network types, including directed \cite{Leicht:2008}, bipartite \cite{Barber:2007}, signed \cite{Gomez:2009,Traag:2009} and multilayer networks \cite{Mucha:2010vk}, with corresponding replacements for the $\frac{k_ik_j}{2m}$ term to account for the expected weights under different null models. Motivating our present contribution, some of these models introduce additional parameters beyond the resolution parameter $\gamma$.
We emphasize that throughout this work we will use the term ``modularity'' in its broadest sense to include any of these generalizations as applied appropriately to a given data set. Such generalizations include the use of resolution parameter $\gamma$, multiple resolution parameters  for signed networks, and 
including one or more interlayer-coupling parameters for multilayer networks.
Regardless of the network type, the primary goal in modularity maximization is to determine the community labels $\{c_i\}$ that maximize $Q$.
Finding the partition with a guarantee of globally optimizing modularity is not computationally feasible except in the smallest networks \cite{Brandes:2008}, and there may be many nearly-optimal partitions \cite{Good:2010}. 

It is worth noting that in initial exploration of a new data set, there is no \emph{a priori} notion of what constitutes a ``good'' value of $Q$. Furthermore, real community structures may be more complex than those describable by a hard partition of nodes into communities, which fails to account for overlapping communities and insists on assigning every node to a community. 	

Even with its problems, maximizing modularity remains a highly-used method for community detection, with many software packages available, some of which are computationally very efficient in practice. Moreover, maximizing modularity is one of the few approaches for community detection in networks that has been extended in a principled way \cite{Mucha:2010vk} to multilayer networks \cite{Kivela:2014},  {though we also call attention to} the multilayer extension \cite{Rosvall:2015} of Infomap \cite{Rosvall:2008} and recent developments extending stochastic block models [SBMs] to multilayer networks (see \cite{Airoldi:2015,Stanley:2016,Taylor:2016} and, for a general update of developments in SBMs, \cite{Abbe:2017}). While multilayer modularity provides a means for community detection in multilayer networks using many of the same heuristics and applying some of the same conventional wisdom developed for single-layer networks, the generalization admits at least one more parameter to control the contribution of interlayer connections to modularity relative to that from intralayer connections, e.g., the interlayer coupling $\omega$ in \cite{Mucha:2010vk}. The same multilayer modularity framework can be applied generally to include multiple interlayer coupling parameters controlling the relative contributions of different parts of the multilayer structure, e.g., for data that is both temporal and multiplex. As such, multilayer modularity  {requires exploring} a two-dimensional parameter space in its simplest setting, and higher dimensions in more general cases. For present purposes, we will here only explicitly consider the case of a single interlayer coupling parameter $\omega$ in Sections \ref{subsec:multilayer} and \ref{subsec:rollcall};
but this does not put constraints on the coupling topology or relative values, only that there is some selected interlayer coupling tensor that is multiplied by $\omega$. Meanwhile, the approach we develop here can be naturally generalized to higher dimensions. 

 {Identifying appropriate values for parameter $\gamma$ (and in the multilayer setting, $\omega$)} involves running one or more heuristics at various parameter values and comparing the results.  {Because identifying globally optimal community structure is computationally intractable (both for modularity and most other approaches), these algorithms are usually run stochastically or with random initial conditions to account for entrapment in local extrema.}  The possibly different community structures found by computational heuristics at a particular $\gamma$ parameter point [$(\gamma,\omega)$ for multilayer networks] are then typically assessed only at that point before moving on to generate results at other parameter values. For instance, one might select the partition with greatest modularity found at that specific value of $\gamma$ or measure some statistic over the partitions that were generated at that $\gamma$ (see, e.g.,~\cite{Pajek}). \linebreak In order to determine whether the obtained community structures are ``robust'' to the $\gamma$ selection in any sense, one might look for stable plateaus in the number of communities (see, e.g., \cite{Fenn:2009,Fenn:2012,Granell:2011,Macon:2012}), consider another metric such as significance \cite{Traag:2013}, directly visualize the different community assignments across parameters (as in \cite{Lewis:2010,Macon:2012}), or compare obtained communities with other generally-acceptable labels by some measure such as pairwise counting scores (see, e.g., the discussion in \cite{Traud:2011}) or information-theoretic measures like Variation of Information \cite{Meila:2007} and Normalized Mutual Information~\cite{FredJain:2003}. A more computationally-demanding approach that directly attacks the problem that there is no \emph{a priori} notion of what constitutes a ``good'' value of modularity is to compare the obtained best modularity at each $\gamma$ with the distribution of modularities obtained by running community detection across some selected random-graph model, either on realizations from a model or from permutations of the data, repeating this process at different $\gamma$ to identify parameter values where the obtained communities are strongest relative to the random cases \cite{Bassett:2013}.  {Additionally, one may use a given set of partitions to generate a new partition by ensemble learning \cite{Ovelgonne:2012} or consensus clustering~\cite{Lancichinetti:2012,Bassett:2013}.}

Importantly, in each of these approaches for exploring the parameter space, the optimal partitions associated with each $\gamma$ value are typically computed independently of those at other $\gamma$ values [and, again, in the multilayer case, $(\gamma,\omega)$].  {Variation in the structure of these partitions and their corresponding modularity can arise from both adjusting the input parameters and importantly, from~the stochasticity of the algorithm itself.  Often, for close enough values of $\gamma$, the variation in modularity of identified partitions is driven more by the stochasticity of the algorithm rather than the difference in the value of $\gamma$.} Because of this independent treatment of the results from different $\gamma$ values, a large amount of information that might be useful for further assessing the quality of the obtained partitions is typically thrown away.   We propose a different approach, which we call CHAMP, that uses the union of all computed partitions to identify the Convex Hull of Admissible Modularity Partitions in the parameter space.
CHAMP identifies the domains of optimality across a set of partitions by ignoring the $\gamma$ that was used to compute each partition, finding instead the full domain in $\gamma$ for which each partition is optimal relative to the rest of the input partitions (hereafter, we always use the word ``optimal'' in this restricted sense relative to the set of partitions at hand). Visualizing the geometry of this identification process, each partition is represented as a line in $(\gamma,Q)$ for single-layer networks, and as a plane in the $(\gamma,\omega,Q)$ space in the multilayer case with a single interlayer coupling parameter $\omega$.  {We find the intersection of the half-spaces above the linear subspaces by computing the convex hull of the dual problem. By identifying the convex hull of the dual problem,} we prune that set of partitions to the subset wherein each partition has at least some non-empty domain in the parameter space over which it is has the highest modularity. This pruned subset contains all of the partitions admitted  {through the dual convex hull calculation. Visually, plotting $Q$ as a function of the parameters, the~pruned subset is that which remains in the upper envelope of $Q$, so that each partition appears along the boundary of the convex space above the envelope in the domain where it provides the optimal $Q$ relative to the input set. The partitions removed by CHAMP do not provide optimal $Q$ for any range of parameters. Meanwhile, the partitions that remain in the pruned set are `admissible' candidates for the true but unknown upper envelope of $Q$; that is, each of these `admissible' partitions corresponds to some non-empty parameter domain of optimality relative to the input set of partitions. We propose an algorithm to find this convex intersection of half-spaces} for single-layer networks and demonstrate its ability to greatly reduce the number of  {partitions under consideration}. We also propose an algorithm for mapping out the two-dimensional $(\gamma,\omega)$ domains of optimal modularity for multilayer networks  { in terms of the dual convex hull problem.}

The rest of this paper is organized as follows. We first define the CHAMP algorithm in Section \ref{sec:CHAMP}. We then apply CHAMP to example networks in Section \ref{sec:results}, including several, single-layer examples with resolution parameter $\gamma$ and a multilayer network with a $(\gamma,\omega)$ parameter space (Section \ref{subsec:rollcall}, with~additional figures in the Appendix). We conclude with a brief Discussion (Section \ref{sec:discussion}).

\section{The CHAMP Algorithm (Convex Hull of Admissible Modularity Partitions)}
\label{sec:CHAMP}
Consider a set of $\Sigma>0$ unique network partitions encoded by the 
node community assignments $\{c_{i\sigma}\}$ with $\sigma\in\{1,\dots,\Sigma\}$. 
By construction, $\delta(c_{i\sigma},c_{j\sigma})=1$ if nodes $i$ and $j$ are in the same community in partition $\sigma$ (i.e., $c_{i\sigma}=c_{j\sigma}$), and $0$ otherwise.
Let $Q_\sigma(\gamma)$ denote the value of Equation~\eqref{eq:mod} for given $\gamma$ under partition $\sigma$.
Ignoring the constant multiplicative factor in front of the summation
(alternatively, absorbing that factor into the normalization of $A_{ij}$ and $P_{ij}$), Equation~\eqref{eq:mod} can be written as
\begin{align}
\label{eq:Qsigma}
Q_\sigma(\gamma) 
&= \sum_{i,j} \left( A_{ij} - \gamma P_{ij} \right) \delta(c_{i\sigma},c_{j\sigma})\nonumber\\
&= \sum_{i,j} A_{ij} \delta(c_{i\sigma},c_{j\sigma}) -
\gamma \sum_{i,j} P_{ij} \delta(c_{i\sigma},c_{j\sigma}) \nonumber\\
&= \hat{A}_\sigma - \gamma\hat{P}_\sigma
\end{align}
where the quantities $\hat{A}_\sigma$ and $\hat{P}_\sigma$ are the respective within-community sums over $A_{ij}$ and $P_{ij}$ for partition~$\sigma$.
Importantly, $\hat{A}_\sigma$ and $\hat{P}_\sigma$ are scalars that depend only on 
the network data (i.e., $A$), null model (i.e.,~$P)$, and partition $\sigma$.
Thus, for a given partition $\sigma$, Equation~\eqref{eq:Qsigma} is 
a linear function of $\gamma$, which can be visualized as a line in the $(\gamma,Q)$ plane. 
(See Figure~\ref{modAF}B in Section~\ref{subsec:football} for an illustration of lines $\{Q_\sigma(\gamma) \}$ for several partitions of the 2000 NCAA Division I-A college football network  \cite{Evans:2010ga,Girvan:2002ez}.)

We now compare the partitions' modularity lines $\{Q_\sigma(\gamma) \}$, seeking to identify the optimal partitions that yield the largest modularity values across the $\gamma$ values---that is,  {the upper-envelope boundary $\{Q_\sigma(\gamma) \}$ for the set}. We will additionally obtain $\gamma$-domains over which a given partition is optimal (discarding partitions that are never optimal).
Given a finite set of partitions $\{\sigma\}$, the~coefficients $\hat{A}_{\sigma}$ and $\hat{P}_{\sigma}$ can be computed individually, independent of how those partitions were obtained. 
Therefore, a given value of $\gamma$ admits an optimal partition $\sigma^*$ corresponding to the maximum $Q_{\sigma^* }(\gamma) \geq Q_{\sigma}(\gamma)$ from the given set of partitions $\{\sigma\}$. At most values of $\gamma$, only a single partition provides the maximum (i.e., ``dominant'') modularity.
When two partitions $\sigma$ and $\sigma'$ correspond to identical modularity values [i.e., $Q_\sigma(\gamma)=Q_{\sigma'}(\gamma)$], 
it is typically because this is the unique intersection of the two corresponding lines. (It is possible to have the case where two different partitions have identical $\hat{A}_\sigma$ and $\hat{P}_\sigma$ coefficients, and thus have equal $Q_\sigma(\gamma)$ for all $\gamma$; but in practice we have not observed this situation in our examples. We hereafter ignore this possibility; but if it were to occur in practice, it merely indicates two partitions of equal merit, in the sense of modularity, across all scales.)
For a pair of partitions $\sigma$ and $\sigma'$, the intersection point $(\gamma_{\sigma\sigma'},Q_{\sigma\sigma'})$ indicates the resolution $\gamma_{\sigma\sigma'}$ at which one partition becomes more (less) optimal
over the other with increasing (decreasing) $\gamma$. That~is, one partition dominates 
when $\gamma<\gamma_{\sigma\sigma'}$, 
while the other dominates when $\gamma>\gamma_{\sigma\sigma'}$.
It immediately follows that the $\gamma$-domain of optimality for a partition must be simply connected. (We note that in higher dimensions, such as for signed or multilayer networks, the same linearity requires that domains of optimality must be convex \cite{Mucha:2010vk}.)

We leverage these intersections to efficiently identify the  {upper envelope} of modularity for a given set of partitions, and the corresponding dominant partitions (relative to the set) for all $\gamma\geq 0$ as follows. Starting at $\gamma_0=0$, the partition with maximum $\hat{A}_\sigma$ is optimal. For networks with a single connected component, this partition is a single community containing all nodes; for multiple disconnected components, any union of connected components gives the same $\hat{A}_\sigma$, but we select the partition wherein each separate component defines a community.  
Denoting the optimal partition at $\gamma_0$ by $\sigma^*_0$, we calculate the intersection points $\{\gamma_{\sigma^*_0\sigma} \}$ with the other partitions $\{\sigma\}$ where
$Q_{\sigma^*_0 }(\gamma)=Q_{\sigma}(\gamma)$. Substituting Equation~\eqref{eq:Qsigma} into this constraint yields 

\begin{equation}\label{eq:intersect}
\gamma_{\sigma^*_p\sigma} = \frac{\hat{A}_{\sigma^*_p}-\hat{A}_{\sigma}}{\hat{P}_{\sigma^*_p}-\hat{P}_{\sigma}} ,
\end{equation}

\noindent
where $p\ge0$ for generality.
Starting with partition $\sigma^*_p$ for $p=0$, we identify the smallest intersection point $\gamma_{\sigma^*_p\sigma}>\gamma_p$, which we define as $\gamma_{p+1}$. We denote the associated partition by $\sigma^*_{p+1}$. That is, partition $\sigma^*_p$ is optimal for the $\gamma$-domain $\gamma\in[\gamma_p,\gamma_{p+1})$, above which partition $\sigma^*_{p+1}$ becomes optimal. In the unlikely event that multiple partitions are associated with the $\gamma_{p+1}$ intersection point, the one with smallest $\hat{P}_\sigma$ becomes $\sigma^*_{p+1}$.
Setting $p$ to $p+1$, we iteratively repeat this process until there are no intersections points satisfying $\gamma_{\sigma^*_p\sigma}>\gamma_p$. 
We thus obtain an ordered sequences of optimal partitions, $\{\sigma^*_p\}$, and intersection points $\{\gamma_p\}$ for $p=0,1,\dots$.
The optimal modularity curve for $\gamma>0$, given by the  {upper envelope} of the set $\{Q_\sigma(\gamma)\}$, is then given by the piecewise linear function
\begin{align}
\tilde{Q}(\gamma) 
&= Q_{\sigma_p^*}(\gamma)\,,\quad \gamma\in[\gamma_p,\gamma_{p+1}).
\end{align}

Of course, this procedure can be started at any selected $\gamma$ of interest, and the analogous procedure for identifying intersections for decreasing $\gamma$ could be used to obtain the  {upper envelope} for $\gamma<0$; but in practice here we restrict our attention to $\gamma\geq 0$.

\subsection{MultiLayer Networks and Qhull}
\label{subsec:multilayer}

As noted previously, modularity has also been extended to multilayer networks {\cite{Mucha:2010vk}}, for detecting communities across layers in a way that respects the disparate nature of intralayer v.\ interlayer edges. In order to keep {our notation as simple as possible, here we} let each node in a layer be indexed by a single subscript, $i$ or $j$. (See \cite{DeDomenico:2013,Kivela:2014} for broader discussion about different notations and their advantages.) The formulation developed in \cite{Mucha:2010vk} is then written as follows for the case of a single intralayer coupling parameter with general intralayer null models and interlayer connectivity (again, ignoring multiplicative prefactors in the definition of modularity):

\begin{equation}
Q(\gamma,\omega)=\sum_{i,j} \left( A_{ij} - \gamma P_{ij} + \omega C_{ij} \right)\delta(c_{i},c_{j})
\end{equation}

\noindent
where $A_{ij}$, $P_{ij}$, and $C_{ij}$ represent the (possibly weighted) edges, null model, and interlayer connections, respectively, between the node-in-a-layer indexed by $i$ and that indexed by $j$; and $c_{i}$ indicates the community assignment.  {For example, in the `supra-adjacency' representation of a simple multilayer network of multislice type where the same $N$ nodes appear in each of $L$ layers, one might order the indices so that $i\in\{1,\ldots,N\}$ corresponds to the first layer, $i\in\{N+1,\ldots,2N\}$ corresponds to the second layer, and so on. To emphasize that the formulation of CHAMP is independent of the details of the multilayer network under study, we note here that the only distinction used presently is that $A_{ij}$ encodes all of the edges, $P_{ij}$ specifies the within-layer null model contributions, and $C_{ij}$ describes the known interlayer connections. The key fact here is that $P_{ij}$ and $C_{ij}$ make distinct contributions to multilayer modularity, as controlled by two different parameters, $\gamma$ and $\omega$. As such, we need to extend CHAMP to simultaneously address both parameters.} We will not assume anything here about the values or the topology of the elements of $C_{ij}$, only that the role of these interlayer connections in determining multilayer modularity is controlled by a single interlayer coupling parameter, $\omega$. Larger values of $\omega$ promote partitions with larger total within-community interlayer weight, encouraging the identification of partitions with greater spanning across layers (for a detailed analysis of behavior across $\omega$, see \cite{Bazzi:2016}). We use the GenLouvain \cite{GenLouvain} generalized implementation of the Louvain \cite{Blondel:2008vn} heuristic to identify partitions at selected $(\gamma,\omega)$ parameter values in the multilayer network example in Section~\ref{subsec:rollcall}.

Coupling the communities across layers is conceptually intuitive. Unfortunately, introduction of the additional parameter, $\omega$ makes the previous methods for parameter selection via visual inspection difficult to employ in practice and would seem to greatly complicate the challenge of selecting good values of the parameters. (See \cite{Bassett:2013} for one approach to addressing this challenge.)

However, because the multilayer modularity function is linear in the parameters $\gamma$ and $\omega$, we~can again apply the general approach of CHAMP, albeit now in a larger dimensional parameter space. For  {each} partition $\sigma$, we again define  {the scalar quantities} $\hat{A}_\sigma$ and $\hat{P}_\sigma$ to be the within-community sums over the adjacency matrix and null model, respectively, and now include a similar sum over the interlayer connections, $\hat{C}_\sigma$:
\begin{equation}
\hat{A}_\sigma = \sum_{i,j} A_{ij}\delta(c_{i\sigma},c_{j\sigma})\,,\qquad
\hat{P}_\sigma = \sum_{i,j} P_{ij}\delta(c_{i\sigma},c_{j\sigma})\,,\qquad
\hat{C}_\sigma = \sum_{i,j} C_{ij}\delta(c_{i\sigma},c_{j\sigma})\,.
\end{equation}

In this notation, the multilayer modularity of partition 
$\sigma$ becomes simply
\begin{equation}
\label{eq:hats}
Q_\sigma(\gamma,\omega) = \hat{A}_\sigma - \gamma \hat{P}_\sigma + \omega \hat{C}_\sigma\,.
\end{equation}

Thus, the  partition $\sigma$ is represented by the plane $Q_\sigma$ in $(\gamma,\omega,Q)$. Analogous to the single-layer case, each point in the two-dimensional $(\gamma,\omega)$ parameter space admits an optimal $Q_{\sigma^*}$. 

Given a set of partitions $\{\sigma\}$, CHAMP calculates the coefficients of the $Q_\sigma(\gamma,\omega)$ planes in Equation~\eqref{eq:hats} and  {solves a convex hull problem to find the convex intersection of the half-spaces above} these partition-representing planes. That is, each partition is represented by a plane dividing $(\gamma,\omega,Q)$ in two, thereby defining a half-space. The intersection of the half-spaces above all of these planes is the  {convex space of $Q(\gamma,\omega)$ values greater or equal to all observed quality values, with~the boundary specifying} the maximum modularity surface of the set. In single-layer networks, we considered ordered $\gamma\geq 0$ and iteratively identified the next intersection and associated partition for increasing $\gamma$. In the presence of multiple parameter dimensions here, we instead apply the Qhull implementation~\cite{qhull,pyhull} of Quickhull \cite{Barber:1996iv} to  {solve the dual convex hull problem.}
In practice, multiple partitions of the network can be identified in parallel, calculating and saving each set of $\hat{A}$, $\hat{P}$, and $\hat{C}$ coefficients.  These coefficients defining the planes are then input into Qhull. CHAMP thereby prunes $\{\sigma\}$ to the subset admitted to the convex hull and identifies the convex polygonal domain in $(\gamma,\omega)$ where each partition is optimal (relative to $\{\sigma\}$). 

 {We note that in practice the runtime for finding the pruned subset of admissible partitions and associated domains of optimality is typically insignificant compared to that of identifying the input set of partitions in the first place. In particular, computing the scalar coefficients of the linear subspace of each partition is a direct $O(M)$ calculation for $M$ edges in the network. Meanwhile, the subsequent convex hull problem has no explicit dependence on the network size, depending instead on the number of partitions in the input set.}

While we assume here that there is a single interlayer coupling parameter $\omega$, we emphasize again that we do not restrict ourselves here to a particular form of the interlayer coupling, which might connect nearest-neighbor layers, all-to-all layers, connect only some nodes in one layer to those in another, and might have multiple different weights along different interlayer edges. Rather, we only require here that there is some selected interlayer coupling tensor $C$ that is multiplied by $\omega$. 

Even more complicated interlayer couplings with multiple parameters (e.g., data that is both multiplex and temporal with the freedom to vary the relative weights between these couplings) can in principle be treated analogous to the above in the appropriate higher-dimensional space. With the notation $\vec{\gamma}=(\gamma,\omega)$ and $\mathbf{\hat{P}}_\sigma=(\hat{P}_\sigma,-\hat{C}_\sigma)$, we can write Equation~\eqref{eq:hats} as $Q_\sigma(\vec{\gamma}) = \hat{A}_\sigma - \vec{\gamma}\cdot\mathbf{\hat{P}}_\sigma$, specifying linear subspaces of codimension one in higher-dimensional parameter spaces, given appropriate definitions of $\vec{\gamma}$ and $\mathbf{\hat{P}}_\sigma$. However, we do not go beyond two parameters $(\gamma,\omega)$ in our example results~here.

For convenience we have implemented and distributed a python package for running and visualizing both the single layer and multilayer CHAMP found at \cite{champ_software}.

\section{Results}
\label{sec:results}
We explore the results of running CHAMP on community structures found in various network data sets. In Section \ref{subsec:football}, we consider a network of NCAA Division I-A college football teams from the 2000 season \cite{Evans:2010ga,Girvan:2002ez}. We then look at results of applying CHAMP to a Human Protein Reactome (Section \ref{subsec:HPR}) and a Caltech Facebook network \cite{Traud:2011} (Section \ref{subsec:caltech}). All three of these undirected networks are studied using the Newman-Girvan null model with a resolution parameter as in Equation~\eqref{eq:mod}. Finally, in Section \ref{subsec:rollcall} we apply CHAMP to communities found using the multilayer generalization of modularity in the multilayer network of roll call similarities across time, where each layer is a different two-year Congress \cite{Mucha:2010vk}.

 {For each example, we input into CHAMP a set of partitions identified by the Louvain heuristic \cite{Blondel:2008vn}, as implemented by \cite{traag:louvain} for our three single-layer examples and by GenLouvain \cite{GenLouvain} for our multilayer example. Because of the modest sizes of these example networks, we perform large numbers of runs of the heuristic (between 20,000 and 240,000, as indicated for each example). Each run of the heuristic is performed at a resolution parameter $\gamma$ (including also a parameter $\omega$ in the multilayer example) selected uniformly from a preselected range of the parameter, as indicated for each example. Node indices were randomly permuted for each run to ensure different order of considering nodes in the heuristic, to allow for possibly different partitions to be found at identical parameters. CHAMP makes no requirement that so many partitions be generated, nor about the way in which those partitions were generated, assuming only that multiple partitions have been obtained by one means or another. CHAMP then prunes the input partitions down to the admissible subset; as such, the overall quality of the final subset of course depends on the input set. In practice, one's tolerance for the computational burden will be dictated by the cost of running the community detection heuristics employed on the network of interest. Once the input set of partitions is identified, CHAMP reduces each partition to its scalar coefficients---$\hat{A}_\sigma$, $\hat{P}_\sigma$, and for multilayer networks, $\hat{C}_\sigma$---and then prunes down to the admissible subset in a trivial additional computational cost relative to that already expended to obtain that input~set.}

\subsection{NCAA Division I-A College Football Network}
\label{subsec:football}
\begin{figure}[H]
\centering
\includegraphics[width=0.9\linewidth]{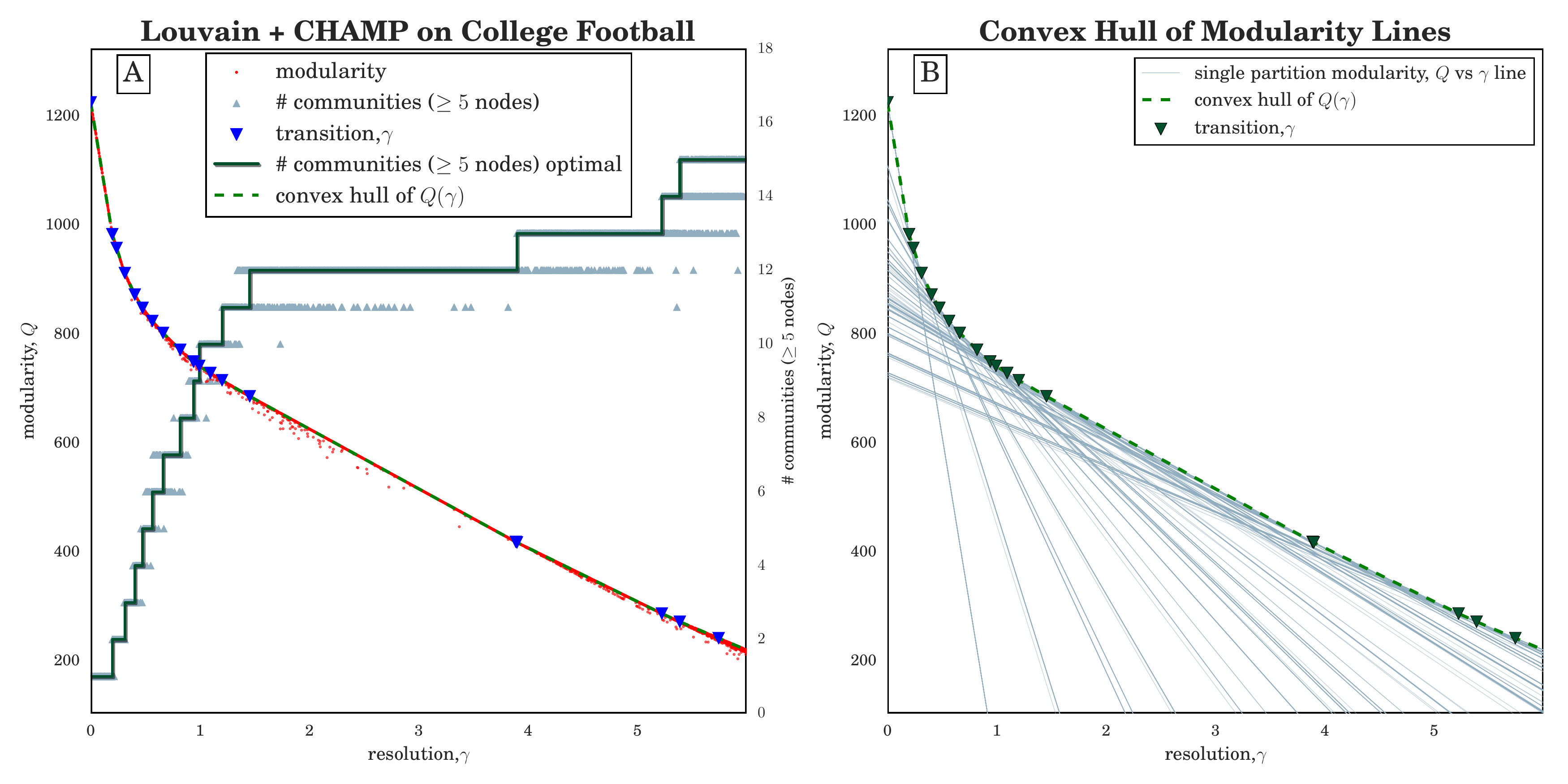}
\caption{(\textbf{A}) Modularity $Q(\gamma)$ given by Equation~\eqref{eq:mod} versus resolution parameter $\gamma$ for $50,000$ runs ($10\%$ of results displayed here) of the Louvain algorithm \cite{Blondel:2008vn,traag:louvain} at different $\gamma$ on the unweighted NCAA Division I-A (2000) college football network \cite{Evans:2010ga,Girvan:2002ez}. Grey triangles indicate the number of communities that include $\geq 5$ nodes in each run, while the  {green} step function shows the number in the optimal partition in each domain; (\textbf{B}) Graphical depiction of CHAMP algorithm (see Section~\ref{sec:CHAMP}). Each~line indicates $Q_\sigma(\gamma) $ given by Equation~\eqref{eq:Qsigma} for a particular partition $\sigma$.
Both panels show the convex hull of these lines as the dashed green piecewise-linear curve, with the transition values represented by downward triangles.}
\label{modAF}
\end{figure}

Figure \ref{modAF}A visualizes a computational scan of the $\gamma$ resolution domain for the Division I-A college football network of 115 nodes representing teams and 613 (unweighted) edges representing that at least one game was played between two teams. Additionally, each team has a label identifying its athletic conference, a subgroup of teams that generally share a geographic region and compete for a conference championship. One would expect that a good partition of the network reflects the conference structure.

For input to CHAMP, we ran the Louvain heuristic \cite{Blondel:2008vn,traag:louvain} 50,000 times on the network. The modularity and number of communities found for each run is plotted at the $\gamma$ resolution parameter used, which were uniformly spaced on $\gamma\in[0,6]$. We observe in particular the wide range of $\gamma$ over which one finds 12-community partitions, but note that the range also includes results with other numbers of communities, with ambiguity about which partition is the better choice.  

By considering each partition as a line over the full domain of $\gamma$ as shown in Figure~\ref{modAF}B, we find the set of lines that form the convex hull of all the modularity functions and the intervals in which each partition is optimal, indicated by the red step function in Figure~\ref{modAF}A, with the steps at the transition values of $\gamma$ indicated by blue triangles in Figure~\ref{modAF}B. These 50,000 runs of the heuristic generated 384 unique partitions,  {with the average run time for each cycle of the Louvain heuristic was 0.02 s.} After~application of CHAMP, there were only 19 partitions in the pruned admissible subset associated with the original parameter search space ($\gamma\in[0,6]$). Moreover, CHAMP identifies a wide $\gamma$-domain of optimality of the 12-community partition, running from $\gamma\doteq 1.45$ to just below $4$. 	

	 {Throughout the paper, we use an information theoretic metric  to assess how much the partitions are varying across the dominant domains.  Mutual information (see also normalized mutual information \cite{FredJain:2003}) quantifies the decrease in entropy for one random variable that comes from knowing the value of a second random variable.  Here the two random variables are discrete, community labels on the nodes.  If two partitions are highly similar, knowledge of the community label of node i in the first partition drastically reduces the uncertainty of the label of node i in the second partition.  In this paper, we use a more stringent, normalized version of the metric introduced by Vinh et al. \cite{Vinh:2009} called Adjusted Mutual Information (AMI),
\begin{equation}
AMI(X,Y)=\frac{MI(X,Y)-E(MI(X,Y))}{\max{\left( H(X),H(Y) \right)}-E(MI(X,Y))}\,,
\end{equation}
where $MI(X,Y)$ is the mutual information between random variables $X$ and $Y$ and $H(X)$ is the entropy of the random variable $X$. The expected value, $E(MI(X,Y))$, is calculated over random partitions sampled from a hypergeometric null distribution (see \cite{Vinh:2009} for details).  The AMI between two partitions equals 1 to indicate perfect concordance, with the value 0 representing alignment no better than random. There are various other clustering and label prediction metrics that could also be used, including pairwise counting scores such as Adjusted Rand Index (ARI) \cite{Hubert1985} (see also the discussion in \cite{Traud:2011}), Variation of Information (VI) \cite{Meila2003,Meila:2007}, and~$\text{F}_1$-score \cite{Rijsbergen:1979}.}

\begin{figure}[H]
\centering
\includegraphics[width=0.9\linewidth]{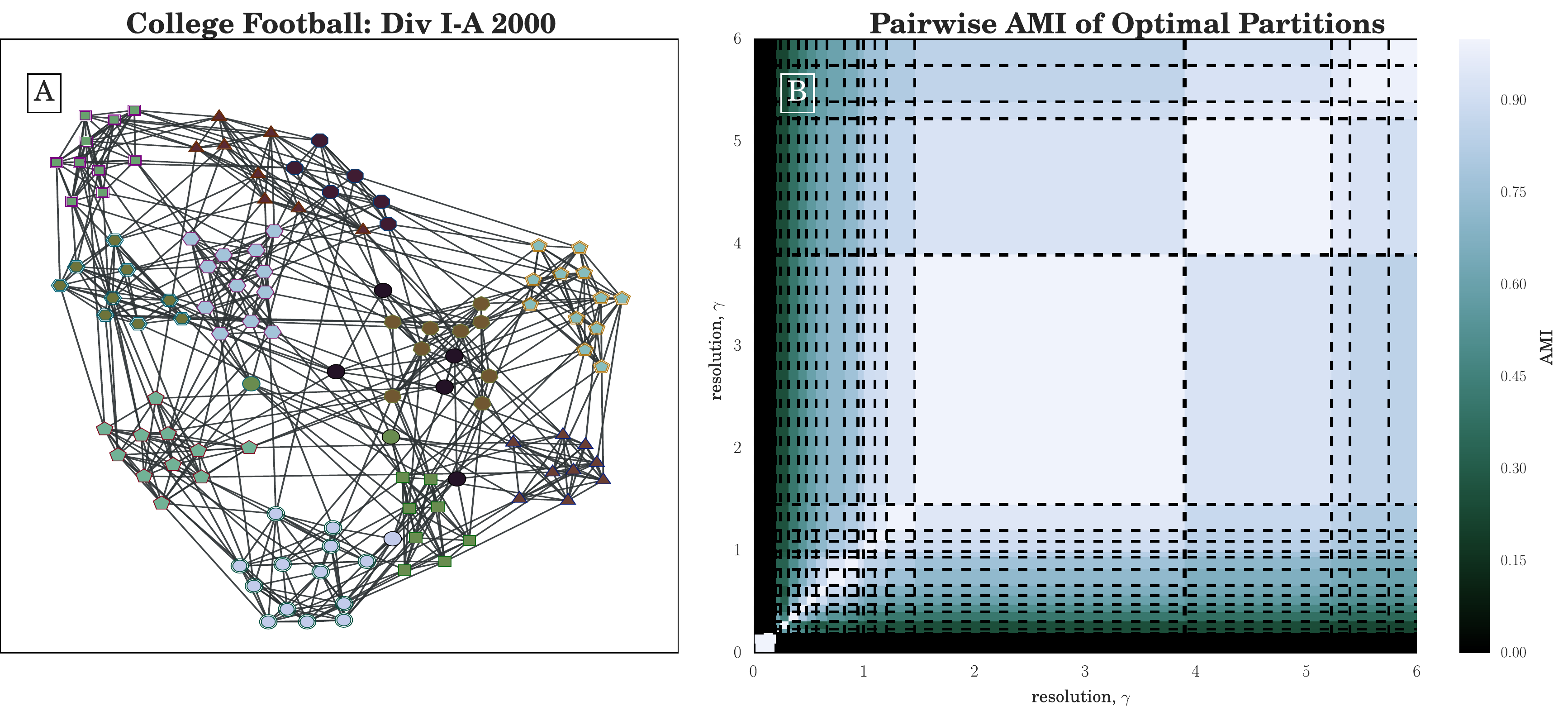}
\caption{(\textbf{A}) ForceAtlas2 \cite{jacomy_et_al} layout, created with \cite{peixoto_graph-tool_2014}, of the unweighted NCAA Division I-A (2000) college football network. Nodes are colored according to the dominant 12-community partition with the widest $\gamma$-domain $\gamma\in [1.45,3.89]$, with node shapes and border indicating their conference labels;
(\textbf{B}) Pairwise adjusted mutual information (N=AMI) between all partitions in the admissible subset identified by CHAMP, arranged by their corresponding $\gamma$-domains of optimality. Dashed lines indicate the transition values of $\gamma$ identified by CHAMP.}
\label{AFnmi} 
\end{figure}

This 12-community partition, visualized in Figure~\ref{AFnmi}B, aligns very closely with the conference labels of the teams as measured by Adjusted Mutual Information (AMI$\doteq 0.92$). Further increasing $\gamma$, we see this 12-community partition domain is followed immediately by a smaller (but still sizeable) domain of optimality for a 13-community partition. Note that while partitions with 11 communities are repeatedly returned by the heuristic, CHAMP indicates the corresponding domain of optimal $\gamma$ to be quite small.

Figure \ref{AFnmi}A shows the pairwise adjusted mutual information (AMI) of the admissible partitions, as organized by their domains of optimality. That is, the large white blocks on the diagonal of the figure are $\mathrm{AMI}=1$ agreement between each partition and itself. In particular, we observe that the 12-community partition (visualized in Figure~\ref{AFnmi}B) is fairly similar to the next few partitions in increasing~$\gamma$, suggesting stability of some main features as communities break up into smaller communities with increasing $\gamma$. At lower values of $\gamma < 1$, we see another possible grouping of domains with reasonable pairwise AMI to one another but who have much lower AMI with the partitions found at higher $\gamma$.  These partitions could represent additional large-scale network structure.

\subsection{Human Protein Reactome Network}
\label{subsec:HPR}
\begin{figure}[H]
\centering
\includegraphics[width=0.9\linewidth]{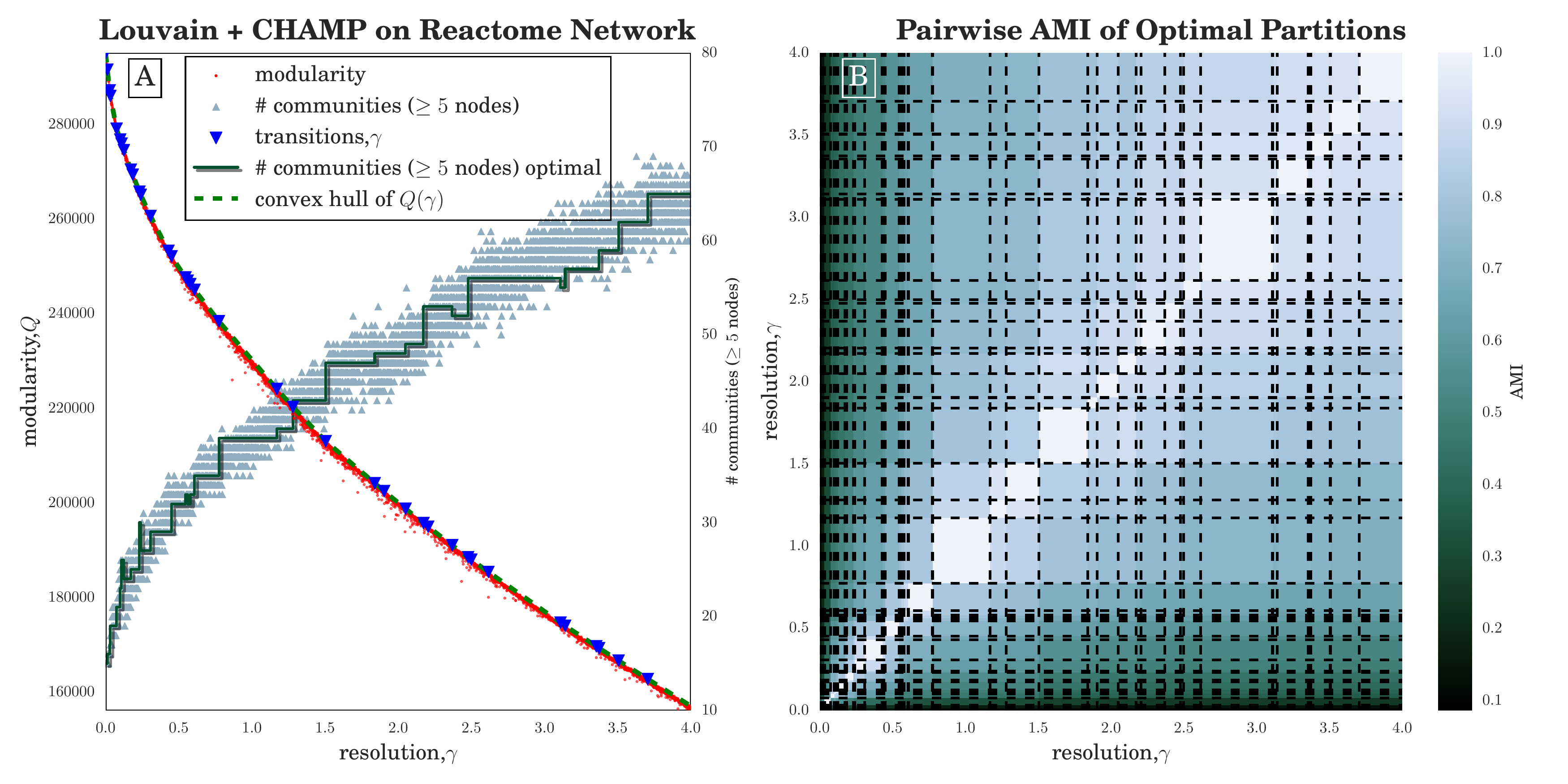}
\caption{(\textbf{A}) Modularity $Q(\gamma)$ given by Equation~\eqref{eq:mod} v.\ resolution parameter $\gamma$ for $20,000$ runs ($25\%$ of results shown) of Louvain \cite{Blondel:2008vn,traag:louvain} on the Human {Protein} Reactome {n}etwork \cite{JoshiTope:2005eh}. Small, grey triangles indicate the number of communities that include $\geq 5$ nodes in each run, while the dark green step function shows the number in the optimal partition in each domain. The dashed green curve is the piecewise-linear modularity function for the optimal partitions, with the transition values marked by blue triangles;  (\textbf{B})~Pairwise AMI between all partitions in the admissible subset identified by CHAMP, arranged by their corresponding $\gamma$-domains of optimality.}
\label{modReact}
\end{figure}

We employed CHAMP to map the domains of modularity optimization for a larger example: the undirected (single-layer) network representation of the Human Protein Reactome \cite{JoshiTope:2005eh,Kunegis:2013ih}, with 6327 nodes representing human proteins and 147,547 edges signifying common biological reactions. We ran the Louvain heuristic 20,000 times on this network with $\gamma\in [0,4]$ uniformly spaced, generating 19,980 unique partitions.   {For this example, each run of Louvain required an average of 2.6 s, generating the input set of partitions in approximately 140 CPU hours.} CHAMP pruned this input set of partitions down to 39 admissible partitions in the convex hull over the  original parameter search space ($\gamma\in[0,4]$). Similar to the figures of the previous example, Figure~\ref{modReact}A shows the spread in the modularities and the numbers of communities identified across all instances of the heuristic, along with the domains of optimization and the number of communities for the admissible subset (see the red step function).

Contrasting Figures~\ref{modAF}A and \ref{modReact}A, we observe in the latter that the red step function decreases with increasing $\gamma$ at some points. Importantly, these decreases are not because of our choice to plot the number of communities that contain at least 5 nodes. The numbers of communities is provably monotonically non-decreasing with increasing resolution parameter in the special case where the null model $P_{ij}=\gamma$ is a constant independent of $i$ and $j$ \cite{Traag:2013}, but we are unaware of any similarly rigorous condition for the Newman-Girvan null model used in Equation~\eqref{eq:mod}. Nevertheless, one typically observes the number of communities to be non-decreasing with increasing $\gamma$, so the results here may indicate values of the resolution parameter near which additional runs of the heuristic might be more likely to identify higher quality partitions.  

\begin{figure}[H]
\centering
\includegraphics[width=0.9\linewidth]{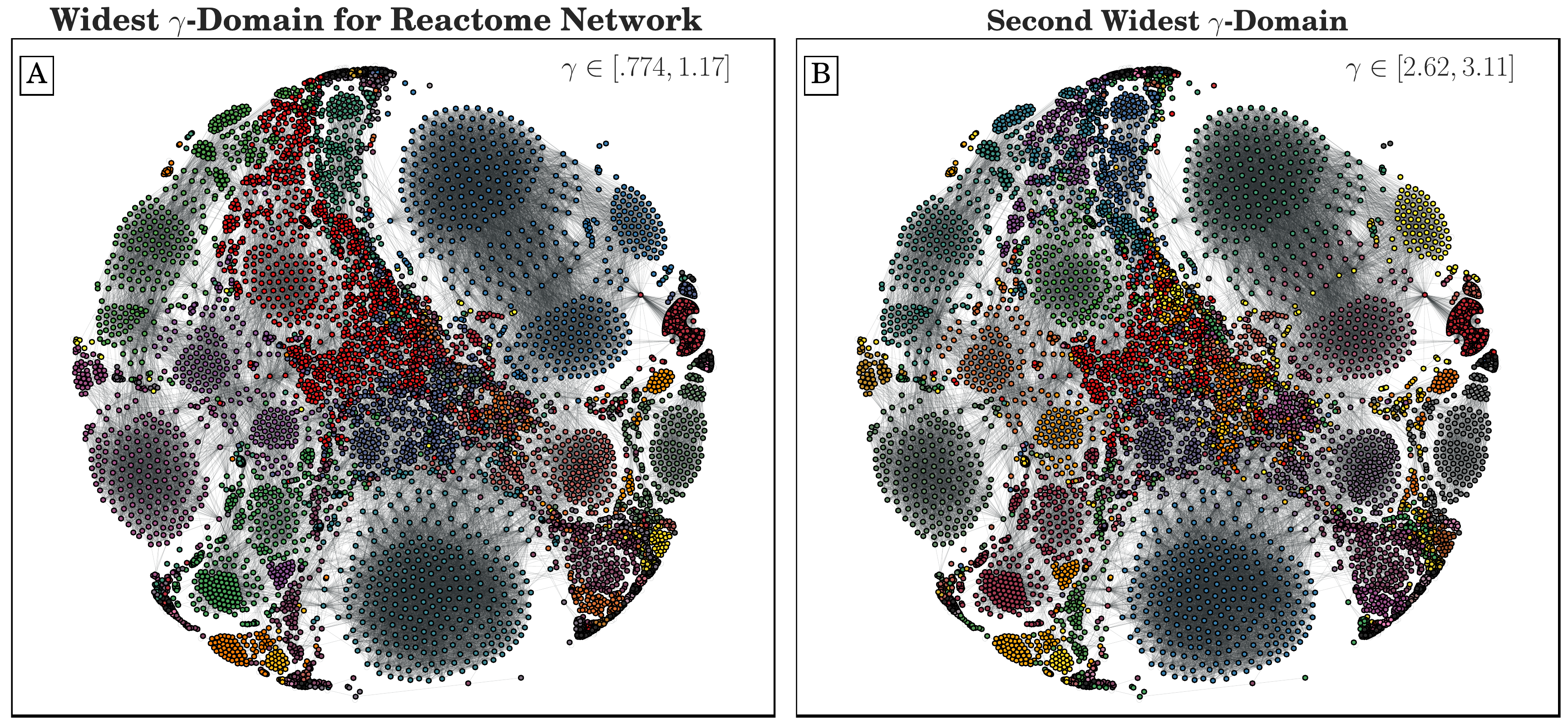}
\caption{ForceAtlas2 layout \cite{jacomy_et_al}, created with \cite{peixoto_graph-tool_2014}, of the Human Reactome Network (edges downsampled to 50,000), colored according to the partitions with the two widest $\gamma$-domains of optimization identified by CHAMP from $20,000$ runs of Louvain.}
\label{reactlayout}
\end{figure}

The number of communities in the initial set of partitions is highly variable, even for small adjustments in $\gamma$, as shown by the yellow triangles in Figure~\ref{modReact}A.  It would be difficult to extract any range of stability from such a plot.  However, when we consider the admissible subset of partitions, we see a few wide domains of optimality in the figure, the two most prominent being $\gamma \in [0.77,1.17]$ and $\gamma \in [2.62,3.11]$.  Layouts of the network colored according to the partitions of these two broadest domains are shown in Figure~\ref{reactlayout}. The pairwise AMI of the admissible partitions are shown in Figure~\ref{modReact}B.  Unlike the college football network, where pairwise AMI appears to indicate two well separated groups of highly similar partitions, the communities here appear to be diffusely similar throughout.  Partitions of adjacent domains are fairly similar but there is no clear divide into groups of partitions.

\subsection{Caltech Facebook Network}
\label{subsec:caltech}
As a final single-layer example, we considered the undirected network of Facebook friendships for students at Caltech in September of 2005 \cite{Traud:2011}, the largest connected component of which includes 762 nodes representing Facebook users and 16,651 unweighted edges representing reciprocal friendships.

\begin{figure}[H]
\centering
\includegraphics[width=0.9\linewidth]{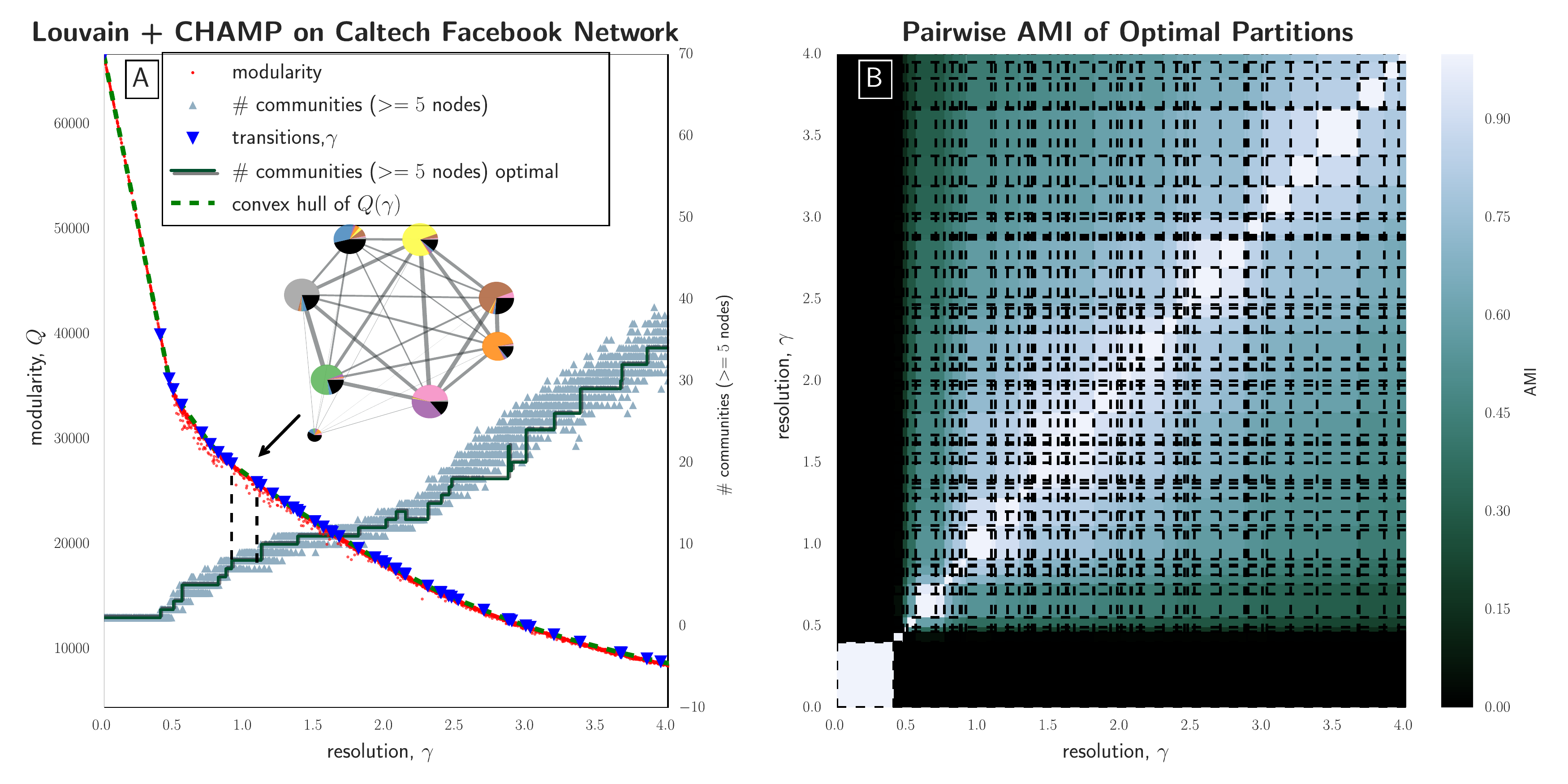}
\caption{(\textbf{A}) Modularity $Q(\gamma)$ v.\ $\gamma$ for $100,000$ runs ($5\%$ of results shown) of Louvain \cite{Blondel:2008vn,traag:louvain} on the Caltech Facebook network \cite{Traud:2011}. Orange triangles indicate the number of communities that include~$\geq 5$ nodes in each run, while the red step function shows the number in the optimal partition in each domain. The dashed green curve is the piecewise-linear modularity function for the optimal partitions, with the transition values marked by blue triangles. The condensed layout of communities (created with \cite{peixoto_graph-tool_2014}) here visualizes the optimal partition found for $\gamma\in [0.908,1.09]$, with each pie-chart corresponding to a community, fractionally colored according to the House membership of the nodes in the community. The AMI between this partition and House labels (including the missing label) is~0.513; (\textbf{B}) Pairwise AMI between all partitions in the admissible subset identified by CHAMP, arranged by their corresponding $\gamma$-domains of optimality.}
\label{caltechfig}
\end{figure}

We used the Louvain algorithm $100,000$ times on $\gamma\in [0,4]$ uniformly spaced, generating $91,080$ unique partitions. CHAMP pruned this set down to $51$ partitions with associated $\gamma$-domains of optimality in the original parameter search space ($\gamma\in[0,4]$).  {That is, the number of partitions in the pruned subset is $1785$ times smaller than that in the set of unique partitions found by our Louvain runs that were input into CHAMP.  Each run of Louvain on the Caltech Facebook network required around 0.8 s with all runs representing approximately 20 CPU hours.}  This output from CHAMP, visualized in Figure~\ref{caltechfig}A, does not indicate the same wide domains of optimality for the community structures in this network as with the previous two examples. The pie-chart visualization within Figure~\ref{caltechfig}A corresponds to one of the wider domains here narrowly straddling the default $\gamma=1$ value. This community structure is reasonably well aligned with the House System at Caltech (see also the associated discussion in \cite{Traud:2011}). At higher values of $\gamma$, we expect that the scales of the communities will be subgroups within the Houses. We observe that some of the wider plateaus in the numbers of communities in the figure correspond to multiple different partitions with the same numbers of communities (note the transition values indicated by blue triangles).

\subsection{U.S. Senate Roll Call Voting Network}
\label{subsec:rollcall}
We demonstrate the use of CHAMP to explore the parameter space for a multilayer network using the roll-call-voting similarity network for the U.S. Senate from 1789 to 2008 (Congresses 1 to 110) as defined in \cite{Waugh:2009vz} and studied with multilayer modularity in \cite{Mucha:2010vk,MuchaPorter:2010}. This data represents the similarities of voting patterns within each two-year Congress between the 1884 distinct U.S. Senators who served across the first 110 Congresses.  {As in \cite{Mucha:2010vk,MuchaPorter:2010}, each two-year Congress starting in the early January following the biennial Congressional elections is represented as a layer, with interlayer connections only between the multiple appearances of each Senator when they appear in nearest-neighbor layers; as~such, multilayer modularity directly handles additions and removals of Senators over time. Self-loops within each layer are zeroed out, since these  only represent perfect agreement of a Senator with herself during the same two-year period. This representation of the voting data is useful for describing legislative voting activity because the community structures typically group together Senators who vote similarly, providing relatively accessible and intuitive examples of communities that are related to the underlying political alignments as expressed by the Senators through voting, independent of their nominally declared party affiliations. The temporal extents of the communities found by multilayer modularity can then indicate different periods of stability in these political alignments.}

\begin{figure}[H]
\centering
\includegraphics[width=0.9\linewidth]{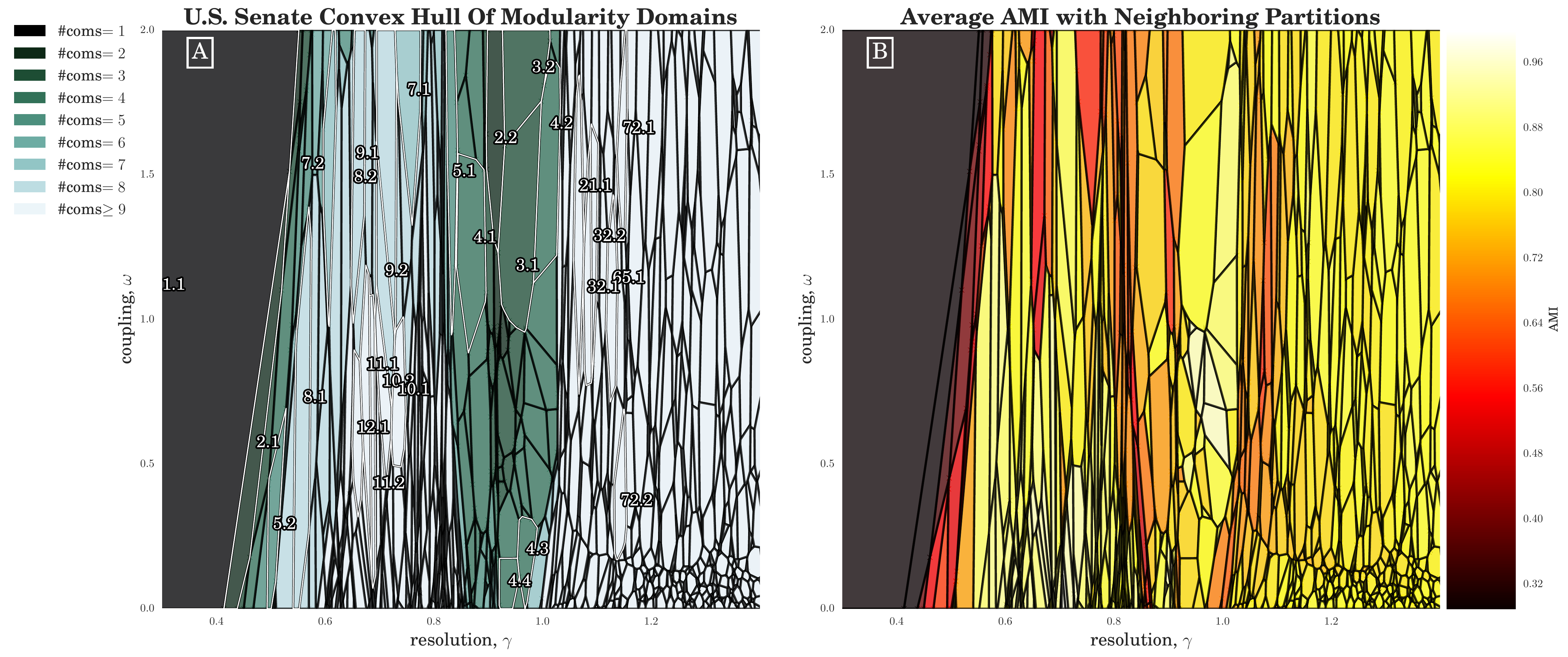}
\caption{(\textbf{A}) Domains of optimization for the pruned set of partitions, colored by the number of communities within each partition. The set of partitions was generated from $240,000$ runs of GenLouvain \cite{GenLouvain} on a $600\times 400$ uniform grid over $[0.3,2]\times[0,2]$ in $(\gamma,\omega)$. The largest partitions are labeled ``$X.Y$'' with $X$ the number of communities with $\geq$ 5 nodes and $Y$ the rank of the domain area (that is, in terms of size) for that given number of communities (e.g., ``5.2'' is the second-largest domain corresponding to 5-community partitions).  The partitions of each labeled domain are visualized in Appendix \ref{app:figures}; (\textbf{B})~Weighted-average AMI of each partition with its neighboring domains' partitions, weighted by the length of the borders between neighboring domains.} 
\label{DODs}
\end{figure}

We ran the GenLouvain \cite{GenLouvain} heuristic 240,000 times, on a 600-by-400 uniform grid over $[0.3,2]~\times~[0,2]$ in $(\gamma,\omega)$, generating 197,879 unique partitions of the network.   {Each run of GenLouvain required approximately 5 s for a total of 340 hours of CPU time.}  CHAMP pruned this set to 1447 partitions admissible in the convex hull of modularity. We note that there were 267 additional partitions with corresponding domains of optimality that were completely outside the selected parameter range $[0.3,2]\times[0,2]$.  In Figure~\ref{DODs} we visualize the $(\gamma,\omega)$-domains of optimality within this region of parameter space. 
In Figure~\ref{DODs}A, a domain's color indicates the numbers of communities for its corresponding optimal partition, whereas in Figure~\ref{DODs}B domain color indicates the average AMI between the corresponding partition and the neighboring optimal partitions (weighted by the lengths of borders between domains).

The trivial 1-community partition dominates the left of the panels in Figure~\ref{DODs} at small $\gamma$. Increasing $\gamma$ outside of this domain, most of the (non-trivial) domains here appear to be relatively long in the $\omega$ direction and much narrower in $\gamma$. Interestingly, we observe a range of $\gamma$ from roughly 0.8 to just above 1 where the domains visually widen in the $\gamma$ direction while also corresponding to a smaller number of communities than partitions below $\gamma\approx 0.8$. Near $\omega=1$, the widths in $\gamma$ of the domains appear larger than those at smaller $\omega$, suggesting perhaps that the stability of identified communities is being enhanced by coupling between the layers. As $\gamma$ increases only slightly past 1, the number of communities in each partition rapidly increases, with the majority of partitions past $\gamma=1.2$ having over 100 communities. At the lower right corner we see the domains are small and highly fragmented in both the $\gamma$ and $\omega$ directions.   

We also aim to identify parameter regions corresponding to similar partitions.  For single-layer networks, we  directly visualized the whole set of pairwise AMI's ordered by $\gamma$. Given two parameters here, we calculate the weighted average AMI of each partition with its neighbors, with weight proportional to the length of the border with the neighboring domain along which the two partitions have the same value of multilayer modularity. The resulting neighbor-averaged AMI of each partition is shown by color in Figure~\ref{DODs}B. We again observe at least three distinct regions of high pairwise similarity, separated by much lower neighbor-averaged AMI, aligned with the different regions in Figure~\ref{DODs}A discussed above: (1) the region below $\gamma \approx 0.8$; (2) the region just below $\gamma=1$, with~particularly high neighbor-averaged AMI for $\omega\in [0.6,0.9]$; and (3) the many-community partitions for $\gamma>1.2$.

\begin{figure}[H]
\centering
\includegraphics[width=0.8\linewidth]{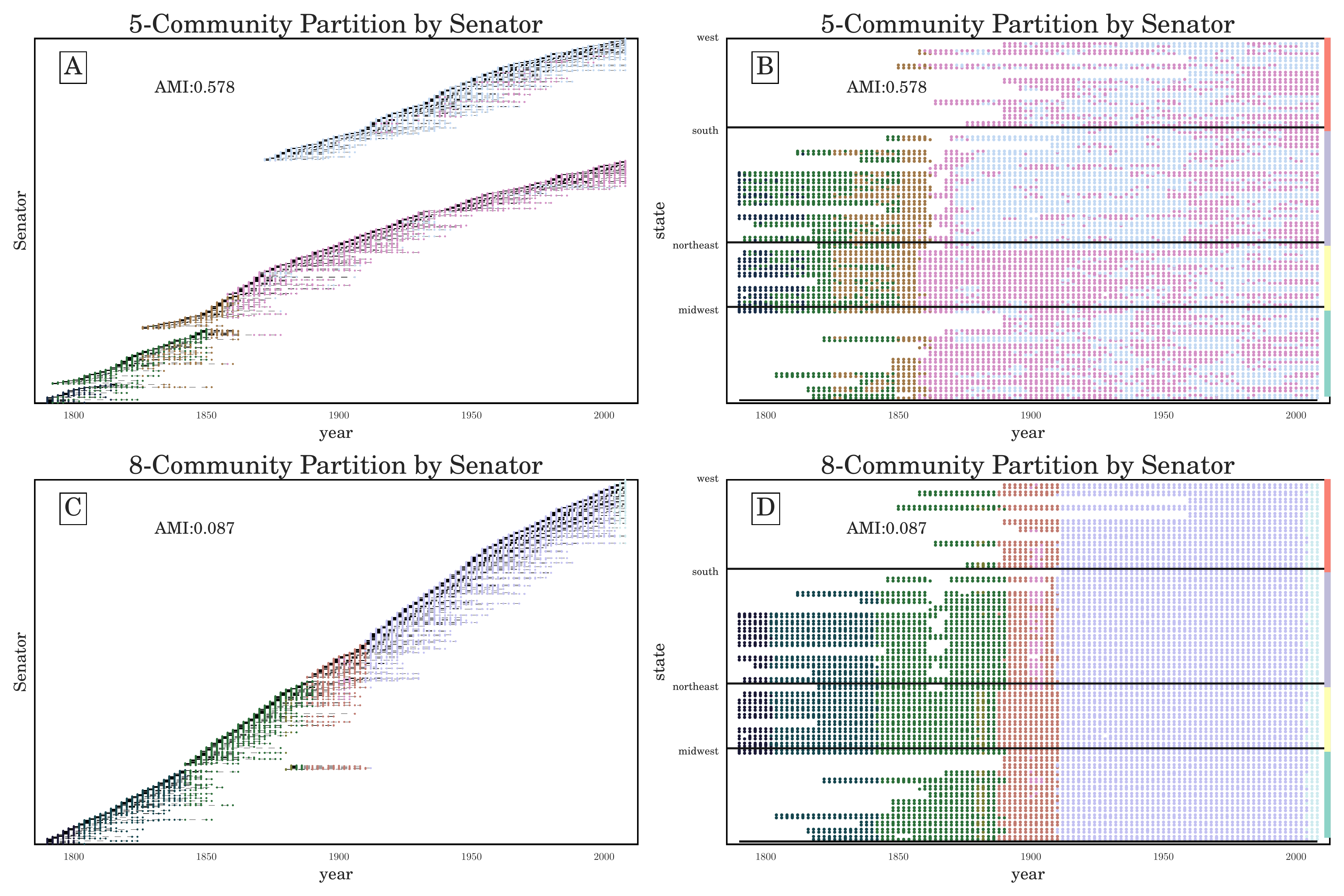}
\caption{
Time-varying community structure for the U.S. Senate from 1789 to 2008 according to the (\textbf{A},\textbf{B}) 5-community  and (\textbf{C},\textbf{D}) 8-community partitions with widest domains of optimality (see labels~$5.1$ and $8.1$ in Figure~\ref{DODs}A); (\textbf{A},\textbf{C})~The vertical axis indicates individual Senators, sorted by community label and time. The AMI reported here is the average over layers (Congresses) of the AMIs in each layer between the identified communities in that layer and political party labels. (This layer-averaged AMI is shown for all partitions in the convex hull over the originally searched parameter range in Figure~\ref{senparty}.) (\textbf{B},\textbf{D})~The vertical axis indicates the state of a Senator, sorted according to geographic region, and the horizontal axis represents time (two-year Congresses). 
}
\label{senlay}
\end{figure}

Indeed, we see a shift in the types of partitions with increasing $\gamma$ across this $\gamma \approx 0.8$ transition boundary. The qualitative difference in community structure between these regions is demonstrated in Figure~\ref{senlay}, highlighting in Figure~\ref{DODs}A the two partitions labeled 5.1 (Figure~\ref{senlay}A) and 8.1 (Figure~\ref{senlay}C). Recall, that these are the partitions with the largest domains of optimality with 5 and with 8 communities, respectively. Most of the Congress layers in the 8-community partition include only a single community label per Congress (see Figure~\ref{senlay}D). In contrast, the 5-community partition divides the Senators both across time and within each Congress, typically into 2 communities in each Congress. These intralayer divisions that extend across time are additionally highlighted by the individual Senator layout in Figure~\ref{senlay}A showing distinct branches, because the Senators have been sorted here first by community label and then, within each community by time. Layouts for the other domains labeled in white in Figure~\ref{DODs}A further demonstrate qualitatively similar patterns, as shown in Appendix \ref{app:figures}.

\begin{figure}[H]
\centering
\includegraphics[width=0.5\linewidth]{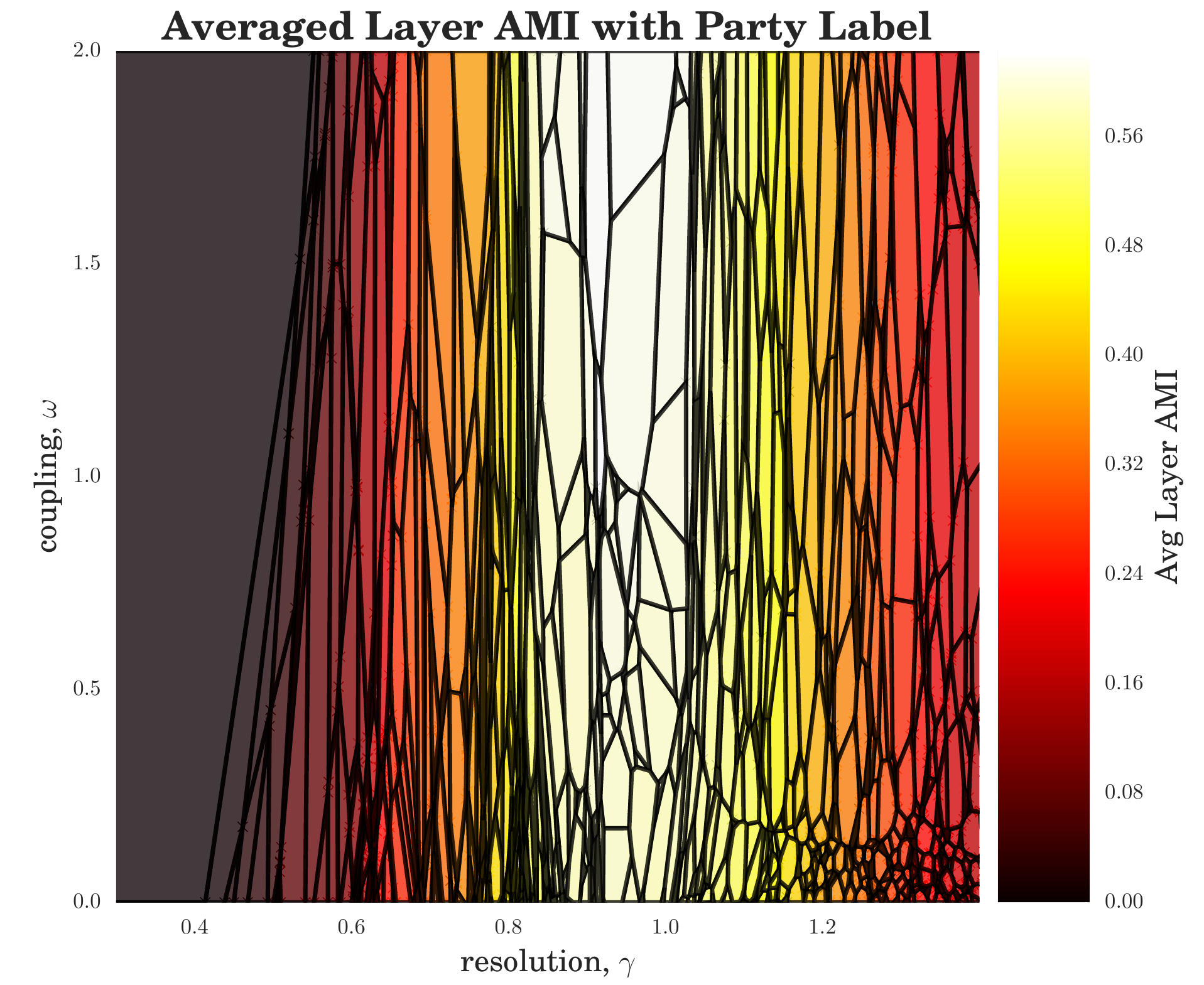}
\caption{The domains of optimality for the time-varying U.S. Senate roll-call similarity network (as in Figure~\ref{DODs}), colored by the layer-averaged AMI between the political-party affiliations of Senators and the community labels $\{c_{i\sigma}\}$ for that layer. 
}
\label{senparty}
\end{figure}

In Figure~\ref{senparty}, we again visualize the domains of optimality in the $(\gamma,\omega)$ parameter space, now color-coded by the layer-averaged AMI between each partition and the known political affiliations of Senators. Specifically,  we compute for each layer the AMI between the community labels $\{c_{i\sigma}\}$ and the Senators' party affiliations, and then we average the AMIs across layers (i.e., across Congresses). The~central, broadest domains have the highest AMI with the mostly 2-party system seen throughout the different session of Congress, consistent with our observations above.  For the most part, partitions with neighboring domains have fairly similar structure within the layers.  There are a few places in the Figure where a darker border represents a transition in the qualitative features of the community structure, such as the transition region around $\gamma\approx 0.8$ discussed above.

\section{Discussion}
\label{sec:discussion}
There are a number of features of CHAMP that make it a useful tool for community detection, as we have demonstrated by way of a variety of examples. By eliminating partitions that are non-admissible to the convex hull, CHAMP can greatly reduce the number of partitions remaining for consideration. By assessing the sizes of the domains of optimality of the partitions in the pruned admissible subset, and through direct pairwise comparisons of partitions in the admissible subset, CHAMP provides a framework for identifying stable parameter domains that signal robust community structures in the network. 

The set of input partitions can be obtained as a result of a community-detection method across a range of parameter choices (as we explored here) or from the comparison of different community-detection methods. Ideally the input set contains near-optimal partitions with relevance for the application at hand. Because each partition is allowed to compete across the whole space of resolution and coupling parameters, CHAMP can surmount some of the pathologies associated with modularity-based community detection heuristics. For example, CHAMP has uncovered several cases where there is a parameter range over which Louvain consistently identifies suboptimal partitions compared to partitions that Louvain itself identifies at other parameter values. In our study of the Human Protein Reactome network (see section \ref{subsec:HPR}), we have seen that the stochasticity over multiple runs of the heuristic makes finding a plateau in the number of communities challenging; nevertheless, CHAMP is able to identify regions where a single partition is intrinsically stable, regardless of how frequently a particular detection algorithm uncovers such a partition. By identifying a manageable-sized and organized subset of admissible partitions with CHAMP, one can then apply a pairwise measure of similarity such as AMI to adjacent partitions to identify shifts in the landscape of optimal community structure. 

We in no way claim that CHAMP resolves all of the problems with modularity-based methods\linebreak (see, e.g., the discussion in \cite{Fortunato:2016}).  {And CHAMP is certainly not the only way to try to process different results across various resolution parameters (see again the Introduction). However, by~taking advantage of the underlying properties of modularity, including the fact that each partition defines a linear function for $Q$ in terms of the resolution and interlayer coupling parameters, CHAMP provides a principled method built directly on the definition of modularity} to make better sense of the parameter space when modularity methods are employed.  {In particular, many of the various other proposed approaches assess each partition at the particular parameter value input into the community detection heuristic that found the partition, that is treating each partition as a single point in $(\gamma,Q)$. In contrast, CHAMP returns to the underlying definition of modularity with a resolution parameter to recognize that each partition here is more completely represented as a line in $(\gamma,Q)$ [in the multilayer case, as a plane in $(\gamma,\omega,Q)$]. The single point is on that line but does not completely explore the potential of that partition to compete against the other identified partitions. By using the full linear subspace associated with each partition, CHAMP prunes away the vast majority of partitions in practice.}

Importantly, CHAMP itself is not a method for partitioning a network, and as such its ability to highlight partitions is limited by the set of partitions given as input to the algorithm. Given the many available heuristics, the computational complexity of maximizing modularity \cite{Brandes:2008}, and the potentially large number of near-optimal partitions \cite{Good:2010}, it is possible that interesting and important community features may be missing from the provided input set. CHAMP as developed here is restricted to processing hard partitions of nodes into community labels, whereas overlapping communities and background nodes (those not belonging to any community) can be important for some applications. One may also reasonably worry about the potential value of partitions in the input set that are near-optimal over a wide domain of the parameters but yet never achieve admission to the convex hull itself and are thus discarded by the algorithm. 

With the introduction of CHAMP presented here, we have left open many other possible uses of this general approach that may be worth exploring.	Although we apply Louvain to discover partitions, CHAMP is agnostic to the detection method used to generate the set of partitions. The partitions input into CHAMP do not even need to be generated by modularity-maximizing heuristics;  {for example, one may also include new partitions as generated by ensemble learning \cite{Ovelgonne:2012} or consensus clustering~\cite{Lancichinetti:2012,Bassett:2013}}. By comparing the results between sets of partitions generated by different methods, CHAMP might be useful as an additional method for making comparisons between these methods. 

Of course, even with a resolution parameter, modularity may not be a good measure for what constitutes a good ``community'' in some networks, and one could investigate whether other quality functions with parameters might be explored with an analogous approach. Even within the consideration of modularity, it would be interesting to generalize the approach of CHAMP to exploring different scales as resolved with different self-loop weights as proposed in \cite{Arenas:2008} (see also \cite{Granell:2011} for an application of this approach). Unlike the resolution and coupling parameters used here, changing the self-loop weight makes a nonlinear change to modularity. Nevertheless, we believe it may be possible to extend CHAMP to the self-loop method for resolving different scales. It would also be useful to extend CHAMP to methods for community structures with overlap and with background nodes.

In further developing CHAMP, it is important to recognize the inability of many community-detection algorithms to assess the reliability of identified communities versus apparent structures arising in random network models. The particular value of modularity, for example, does not immediately indicate whether an identified partition is significant; in fact, the modularities of many classes of random networks such as trees of fixed degree can be quite high in the asymptotic limit~\cite{DeMontgolfier:2011jp,Bagrow:2012hy}. 
Thus, it may be interesting to use CHAMP to further explore and characterize the domains of optimization for partitions of such random networks,  to determine the extent to which leveraging such partition stability information can address questions about detected structures and random noise.

Additionally, it would be interesting to study the consistency of optimality domains output from the application of CHAMP to different input sets of partitions in order to possibly provide insight about how quickly the convex intersection of half-spaces shrinks to the underlying true but unknown upper envelope as the set of input partitions grows. For the networks tested here, the numbers of admissible partitions remaining in the pruned subset were only a very small fraction of the numbers in the input sets. In our experience, the numbers of partitions in the final pruned admissible subset appeared to increase slowly as the size of the input set was increased, but the position of the larger domains appeared to remain relatively consistent  {in practice.} The number of initial partitions needed to get a good mapping of the parameter space undoubtedly depends on the structure of the network and the computational heuristics used. It may also be possible to use a variant of CHAMP to iteratively steer the parameters at which additional partitions might be sought. For instance, input parameters that consistently give rise to dominant partitions with broad domains could be targeted for more runs in an iterative fashion.

In summary, we have presented the CHAMP algorithm as a post-processing tool for pruning a set of network partitions down to the admissible subset in the convex hull that optimizes modularity at different parameters. We have demonstrated the utility of CHAMP on various single-layer networks and on a multilayer network, identifying partitions and their associated domains of optimality in the parameter space. Further research may focus on how the sizes of these domains and the comparisons between domains can be best used to ascertain confidence in identified community structures, to explore subgraphs of a network, and to further process the admissible subset for consensus clustering, as well as other uses of the pruned subset identified by CHAMP.
\vspace{6pt} 


\acknowledgments{This project was supported by the James S. McDonnell Foundation 21st Century Science Initiative -Complex Systems Scholar Award grant \#220020315, by the National Institutes of Health through Award Numbers R01HD075712, R56DK111930 and T32GM067553, and by the CDC Prevention Epicenter Program. The content is solely the responsibility of the authors and does not necessarily represent the official views of the funding agencies.}

\authorcontributions{W.H.W, S.E., R.G., D.T. and P.J.M. each contributed different details of the algorithm. W.H.W and P.J.M coded the algorithm and designed the example experiments. W.H.W performed the examples and analyzed the results. W.H.W, D.T. and P.J.M. wrote the paper.}

\conflictsofinterest{The authors declare no conflict of interest.} 



\appendixtitles{yes} 
\appendixsections{multiple} 
\appendix

\section{Additional Figures}
\label{app:figures}

\begin{figure}[H]
\centering
\includegraphics[width=.24\linewidth]{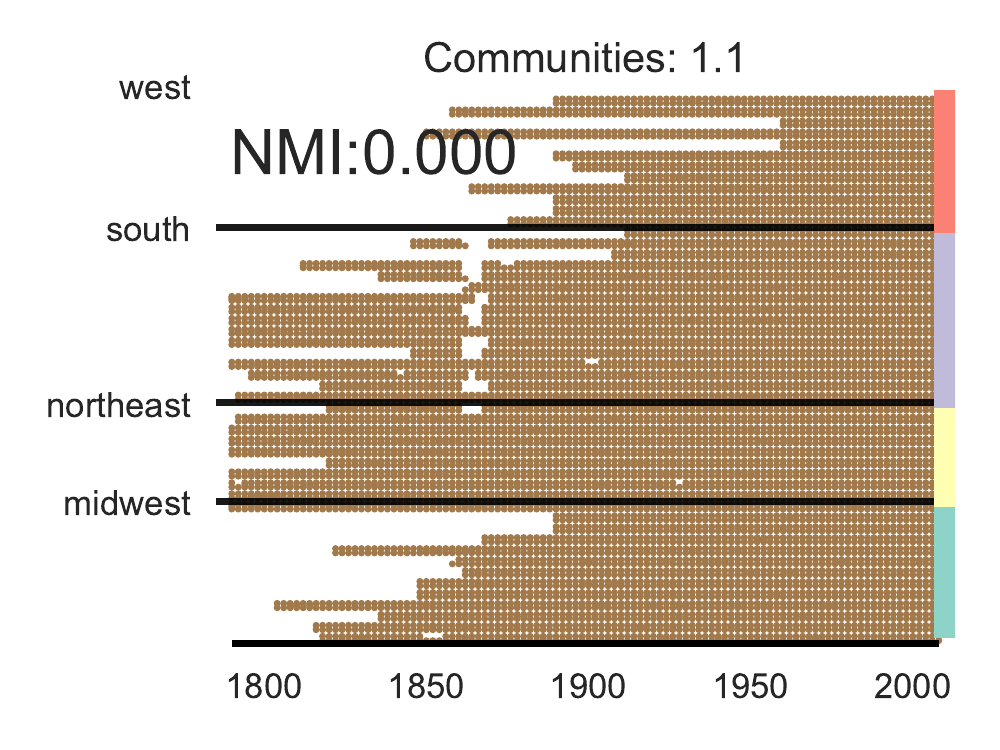}
\includegraphics[width=.24\linewidth]{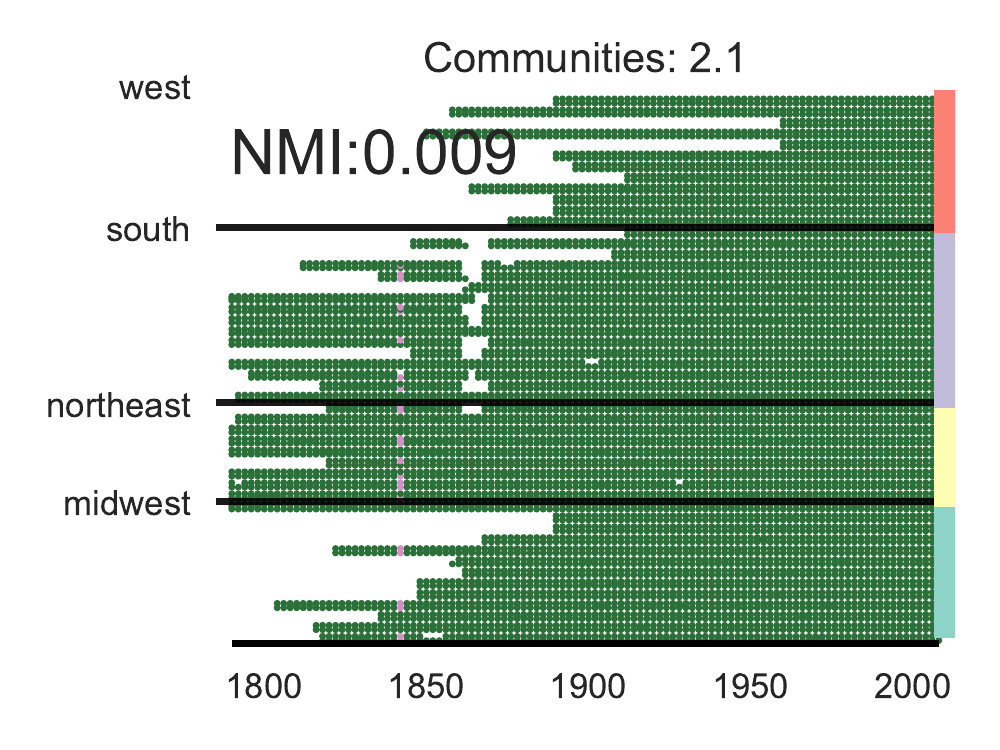}
\includegraphics[width=.24\linewidth]{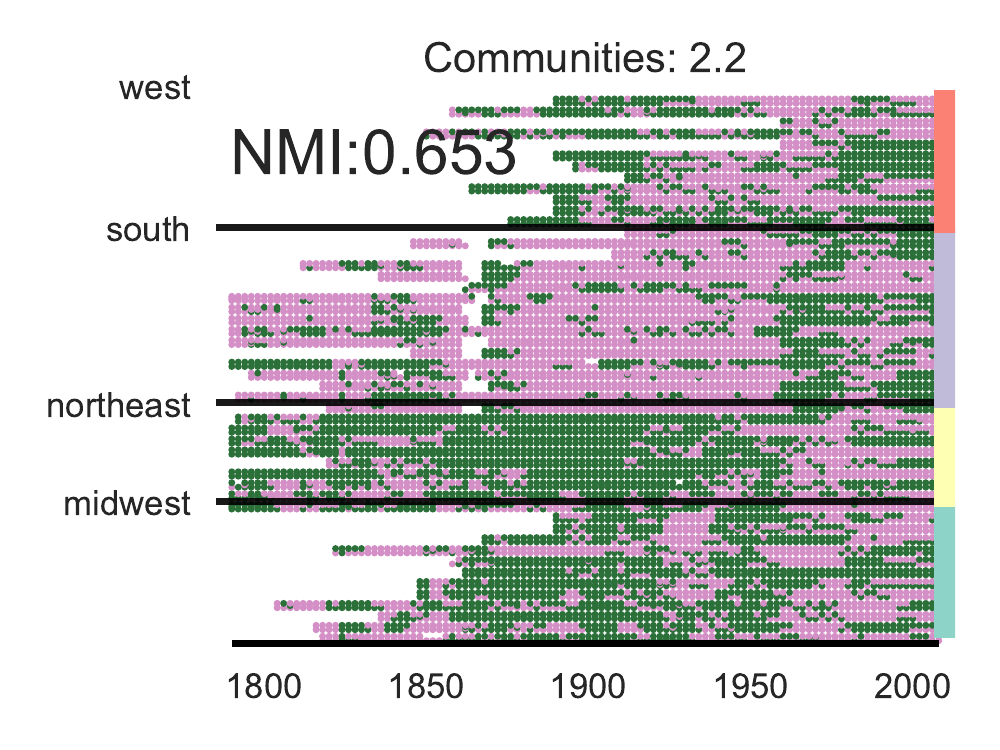}
\includegraphics[width=.24\linewidth]{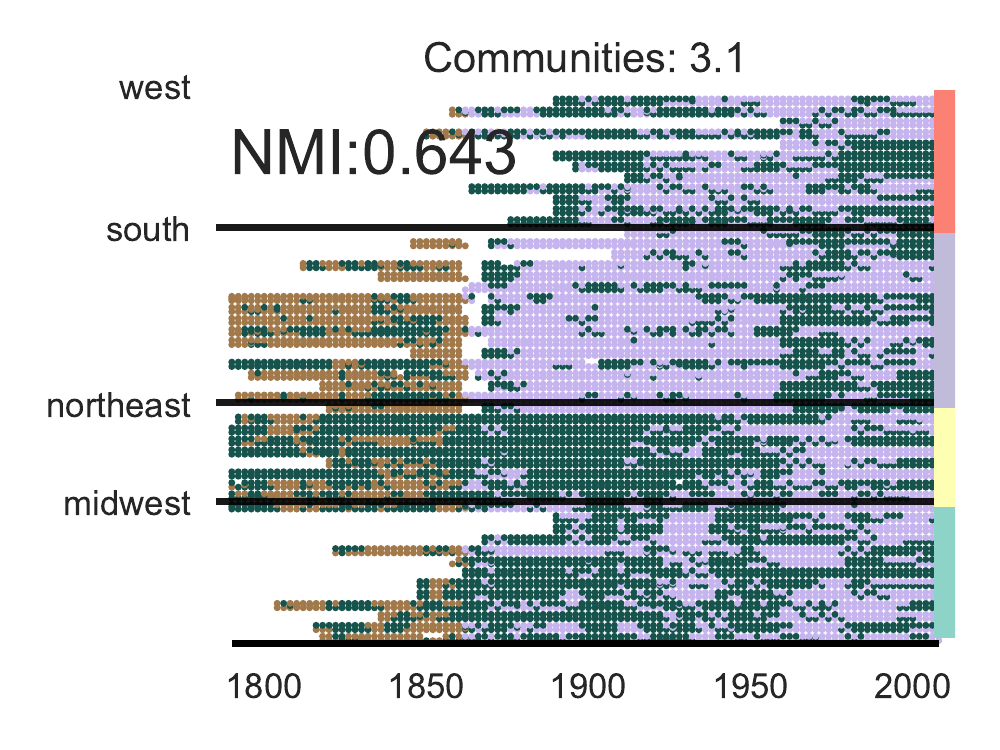}\\
\includegraphics[width=.24\linewidth]{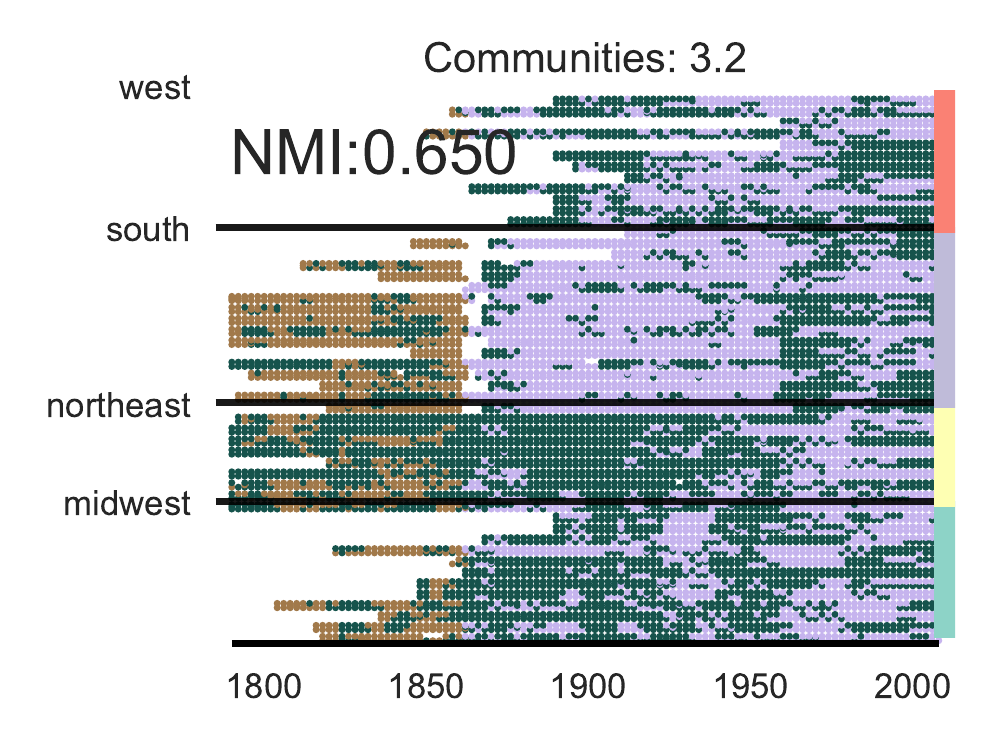}
\includegraphics[width=.24\linewidth]{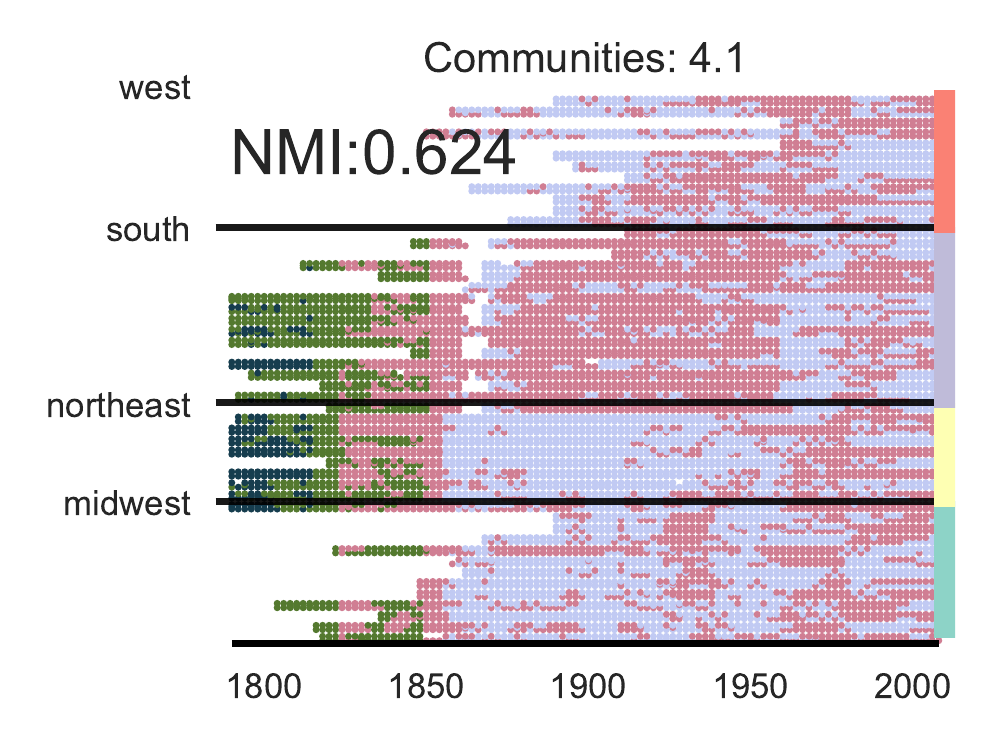}
\includegraphics[width=.24\linewidth]{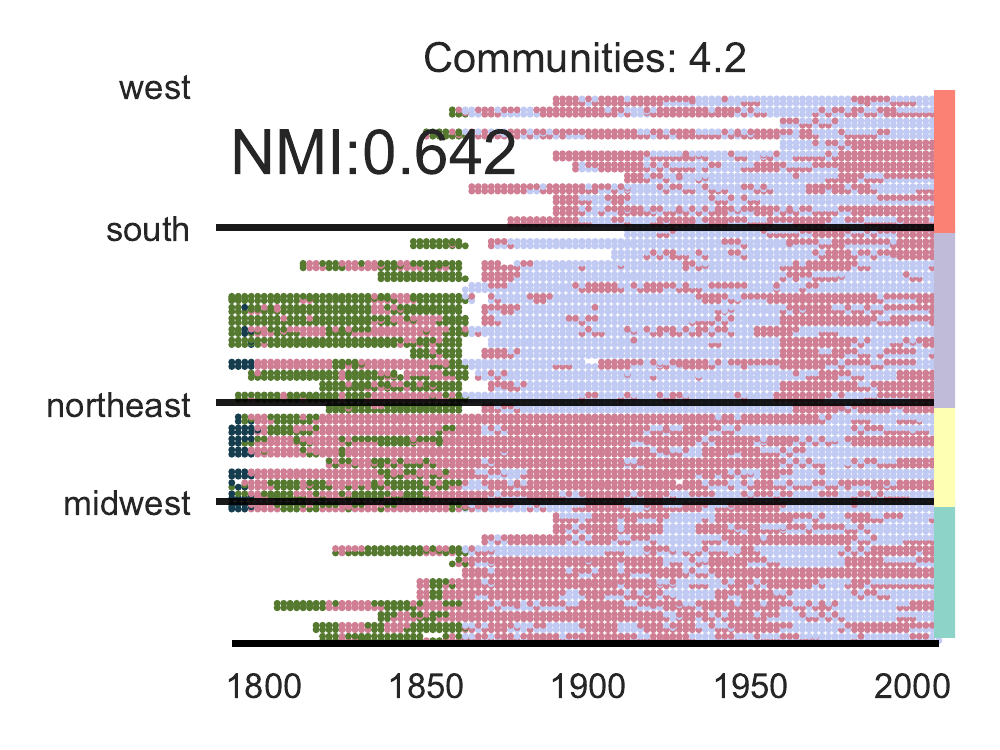}
\includegraphics[width=.24\linewidth]{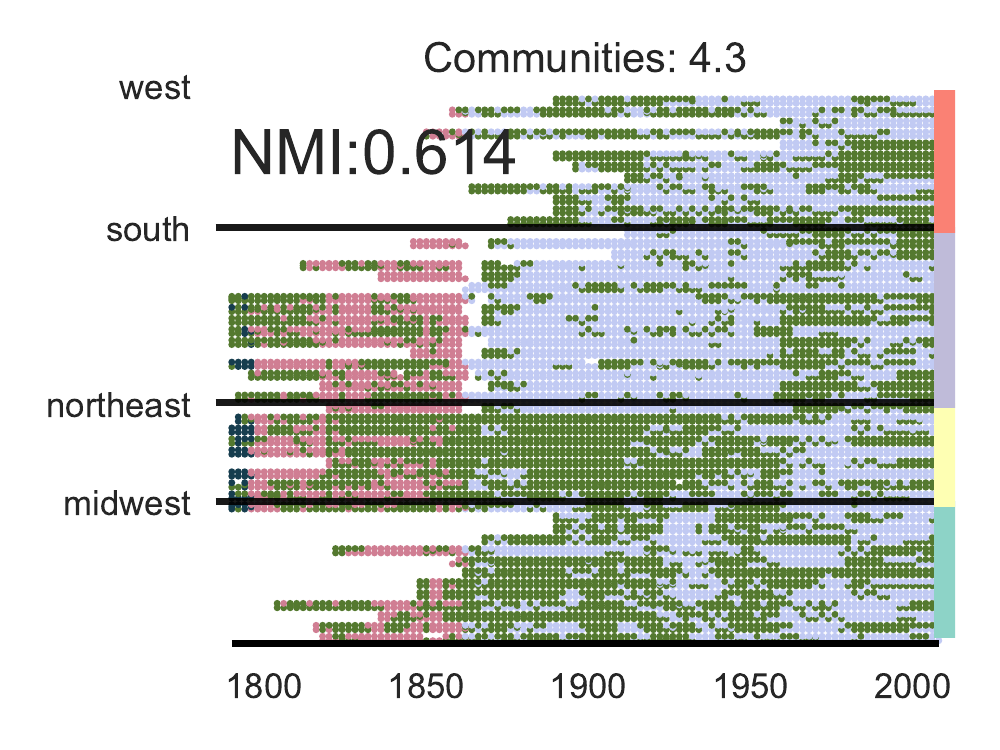}\\
\includegraphics[width=.24\linewidth]{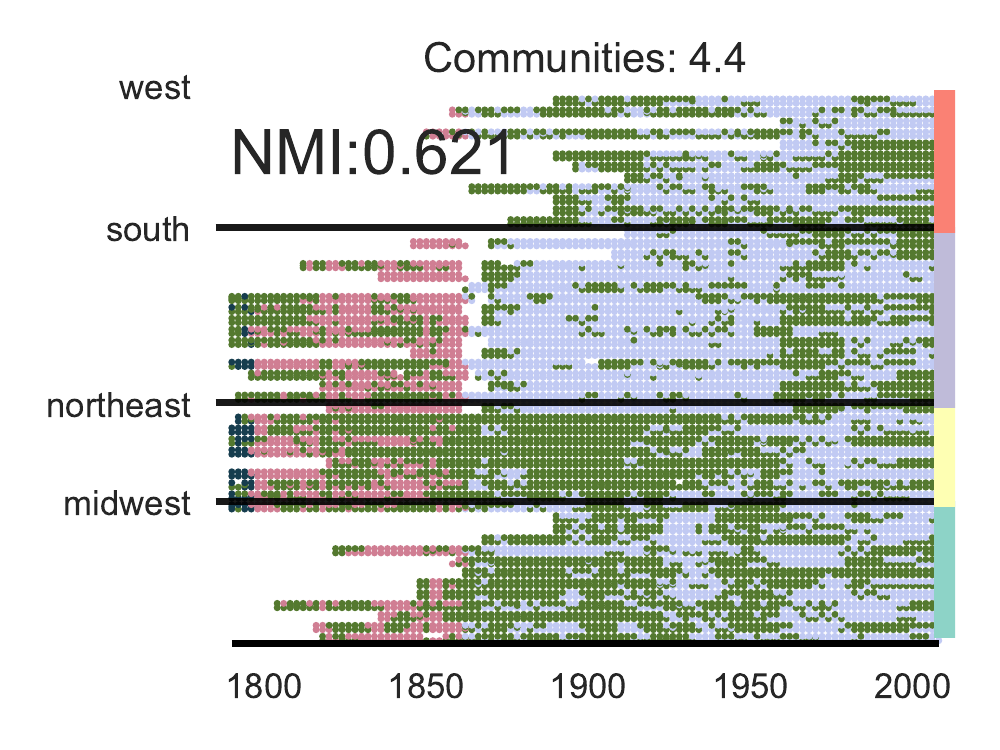}
\includegraphics[width=.24\linewidth]{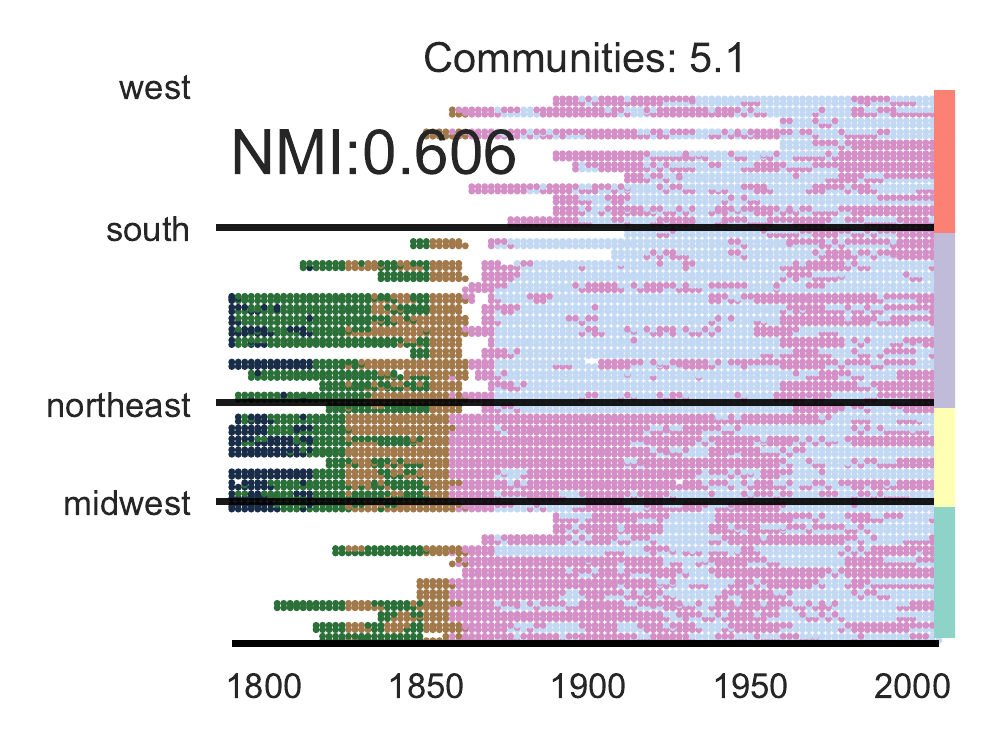}
\includegraphics[width=.24\linewidth]{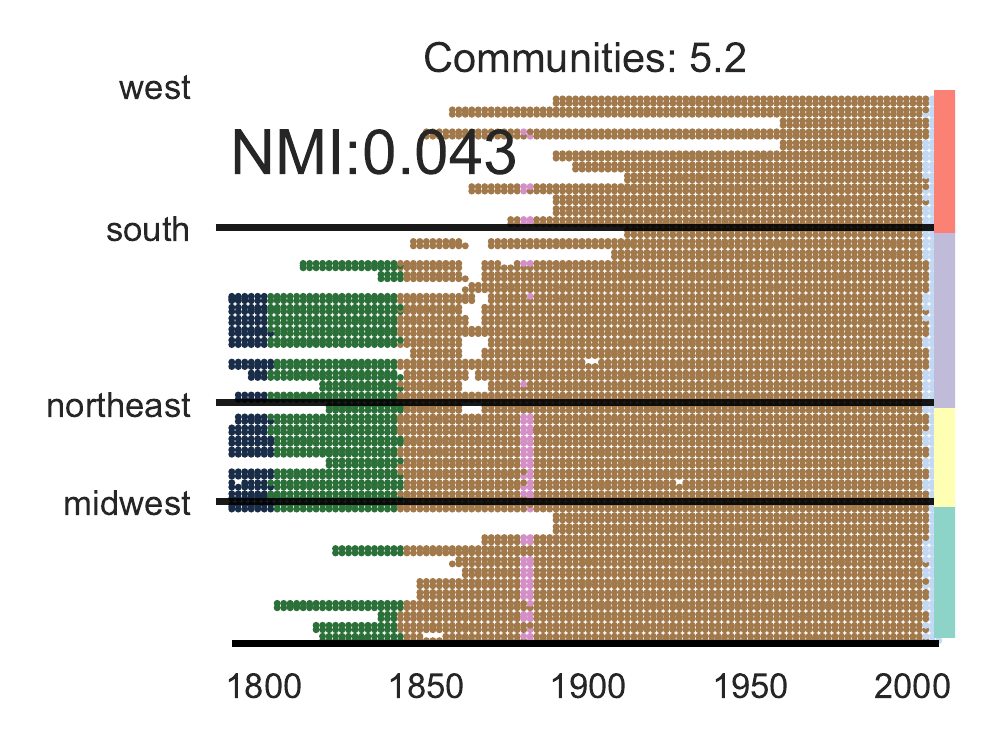}
\includegraphics[width=.24\linewidth]{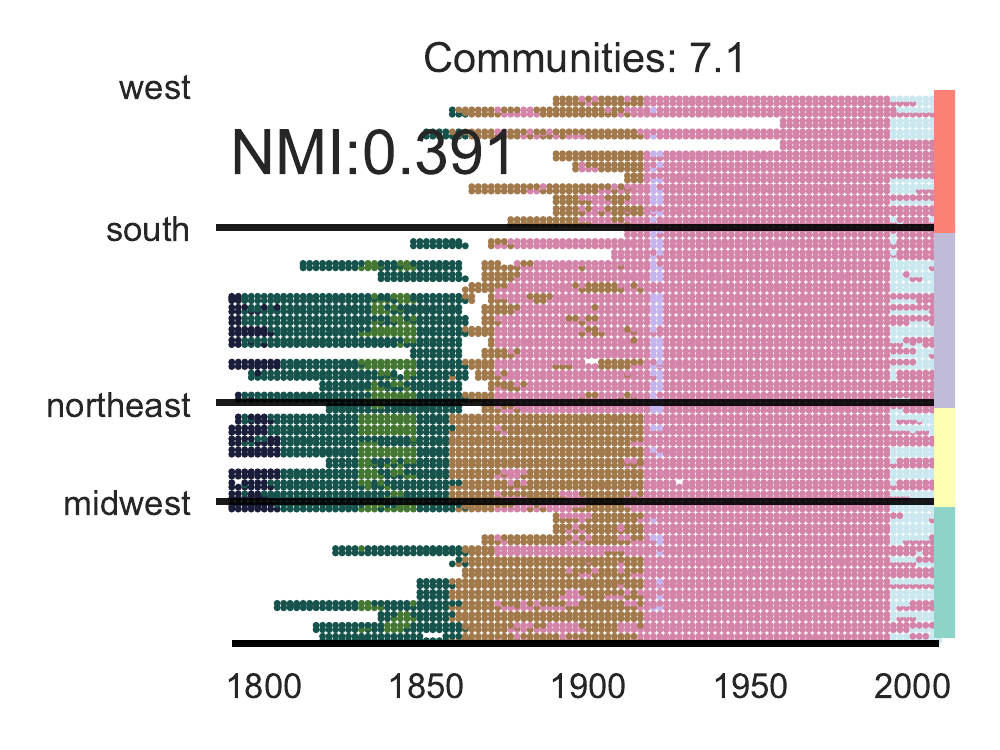}\\
\includegraphics[width=.24\linewidth]{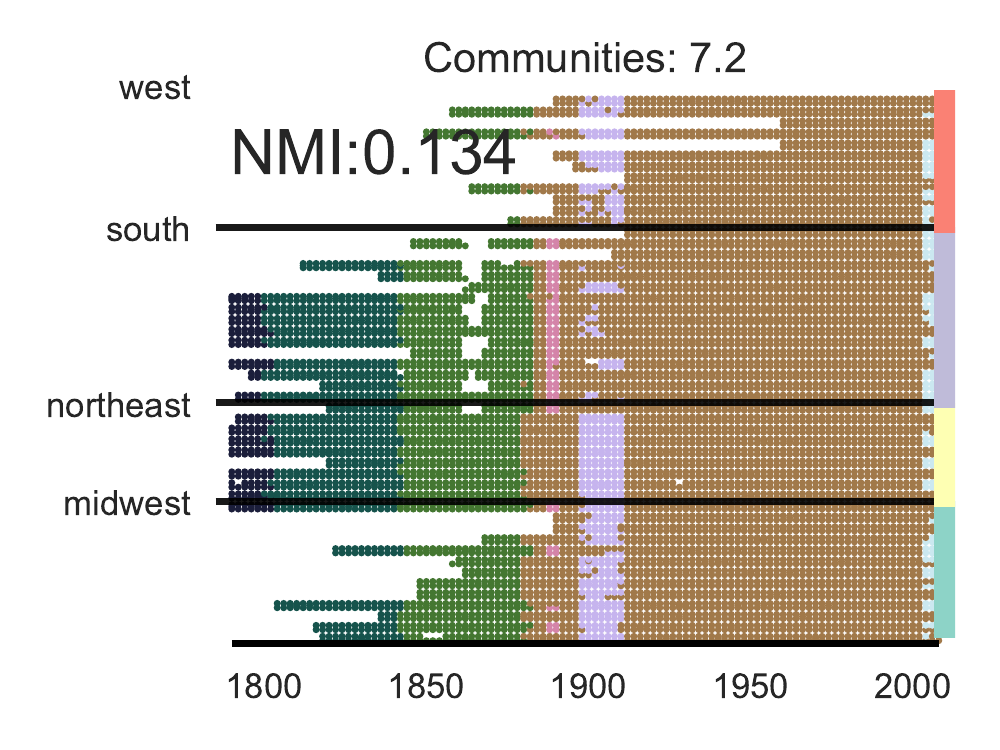}
\includegraphics[width=.24\linewidth]{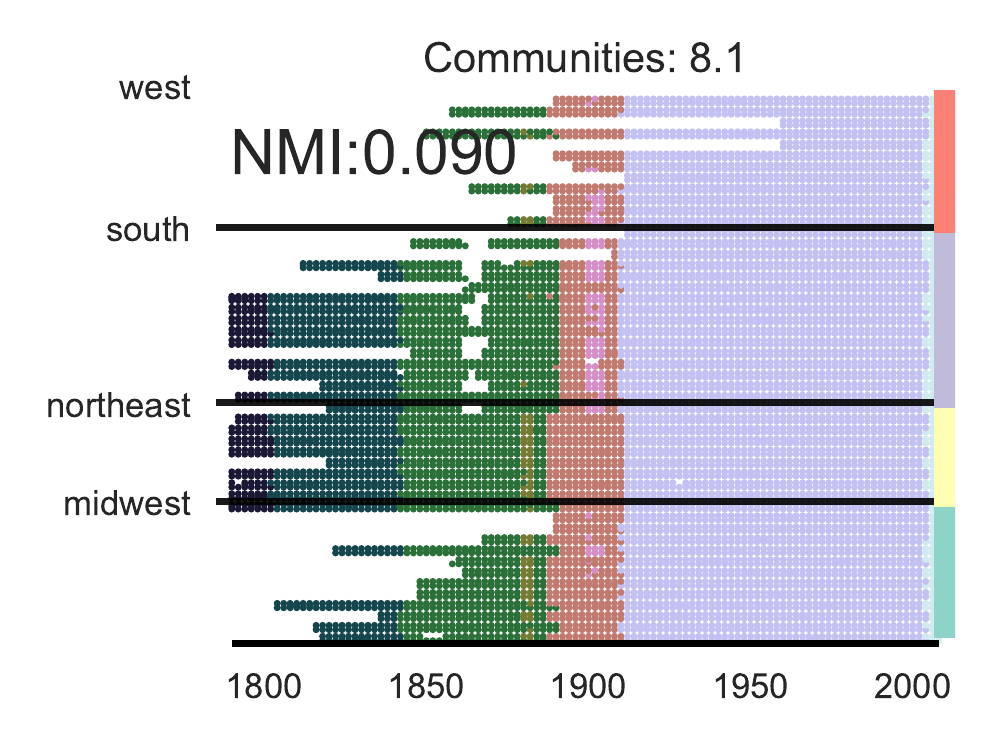}
\includegraphics[width=.24\linewidth]{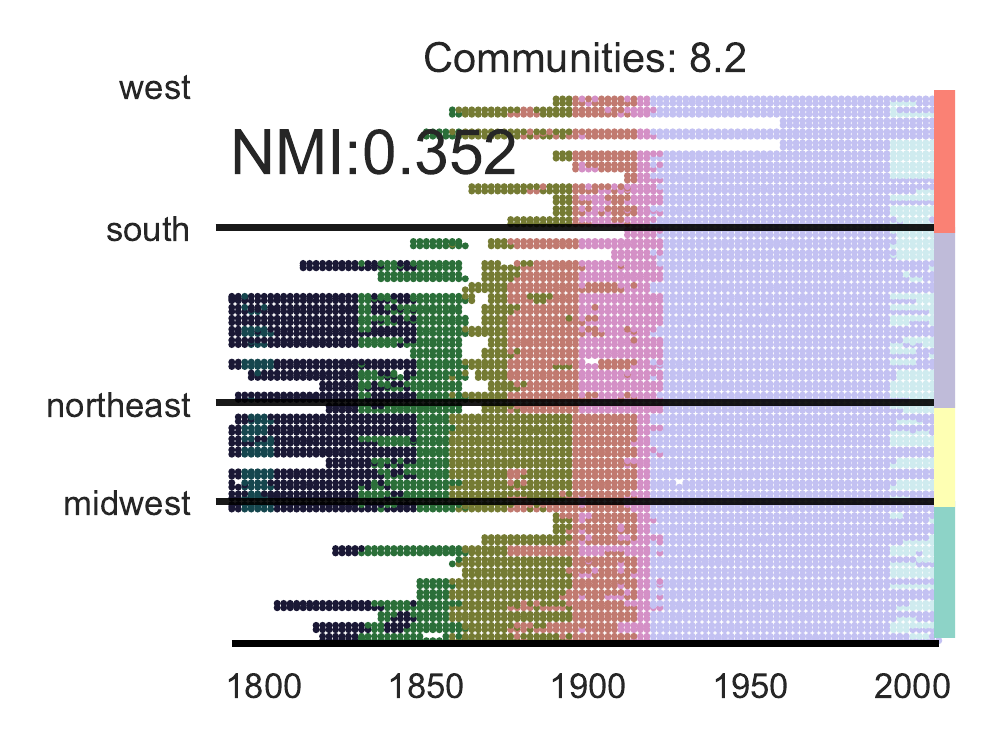}
\includegraphics[width=.24\linewidth]{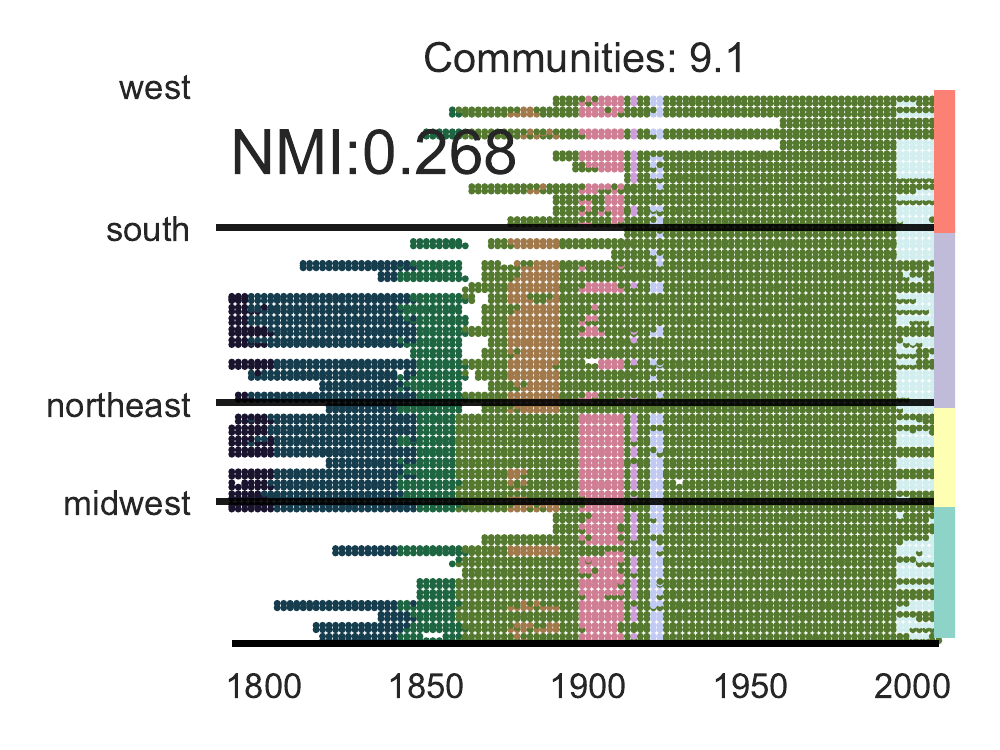}\\
\includegraphics[width=.24\linewidth]{senate_layouts/state_layout_9_1.pdf}
\includegraphics[width=.24\linewidth]{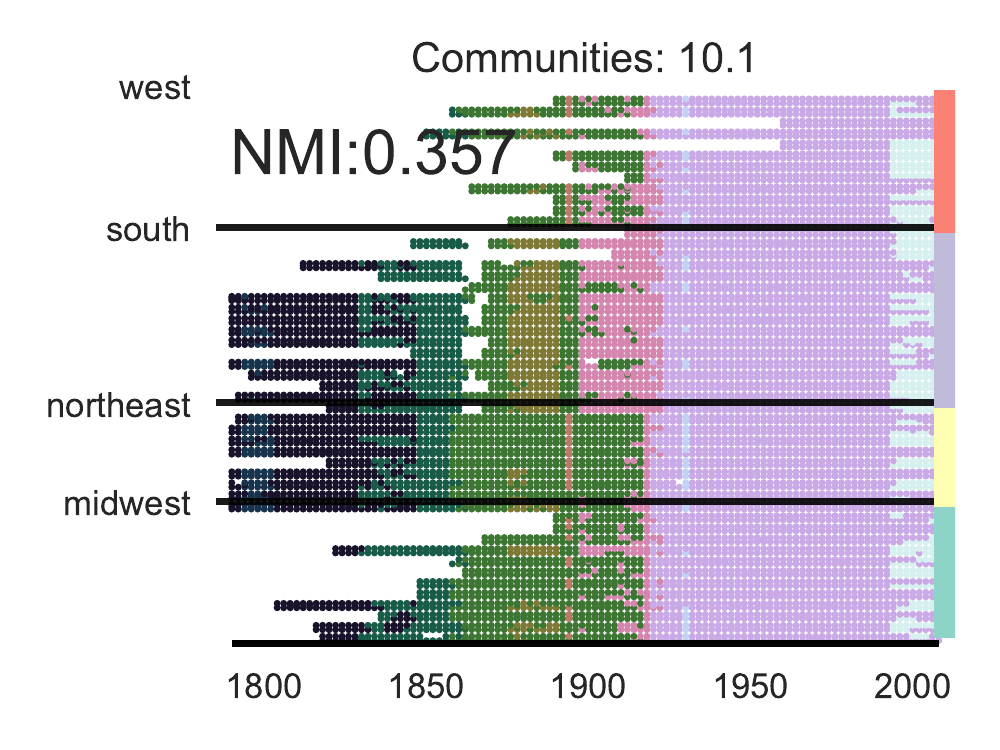}
\includegraphics[width=.24\linewidth]{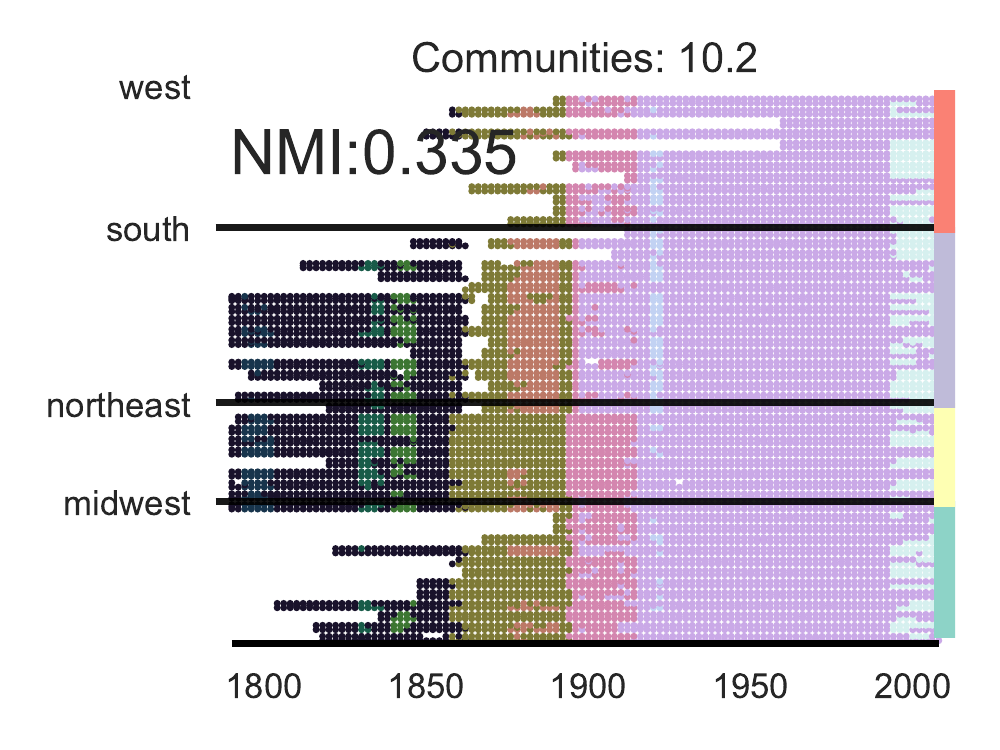}
\includegraphics[width=.24\linewidth]{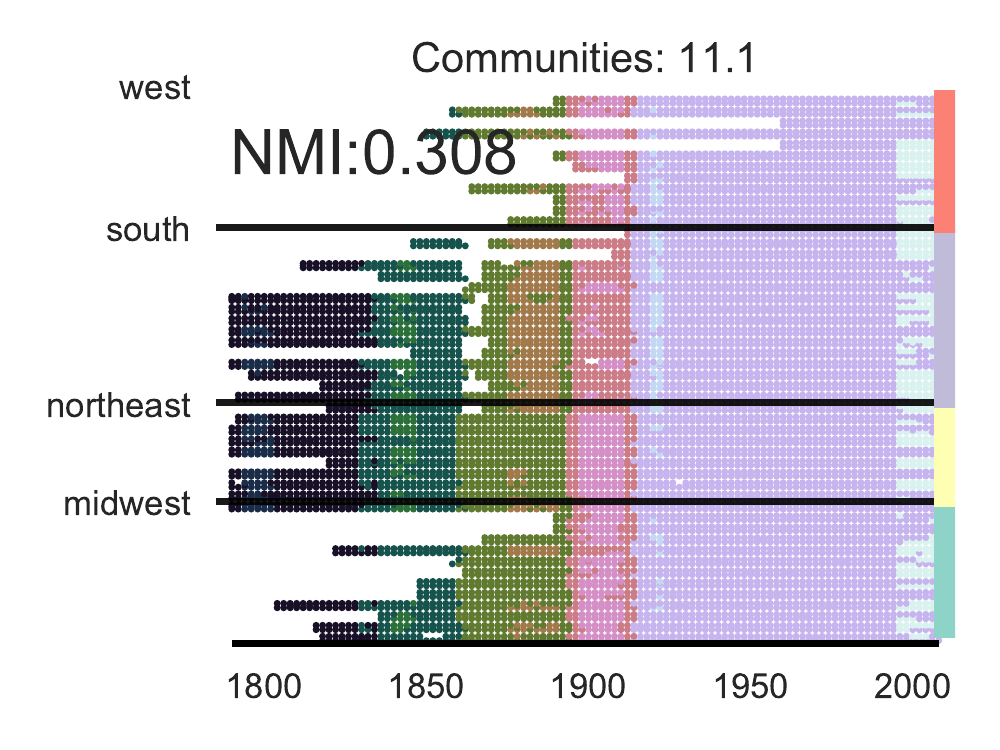}\\
\includegraphics[width=.24\linewidth]{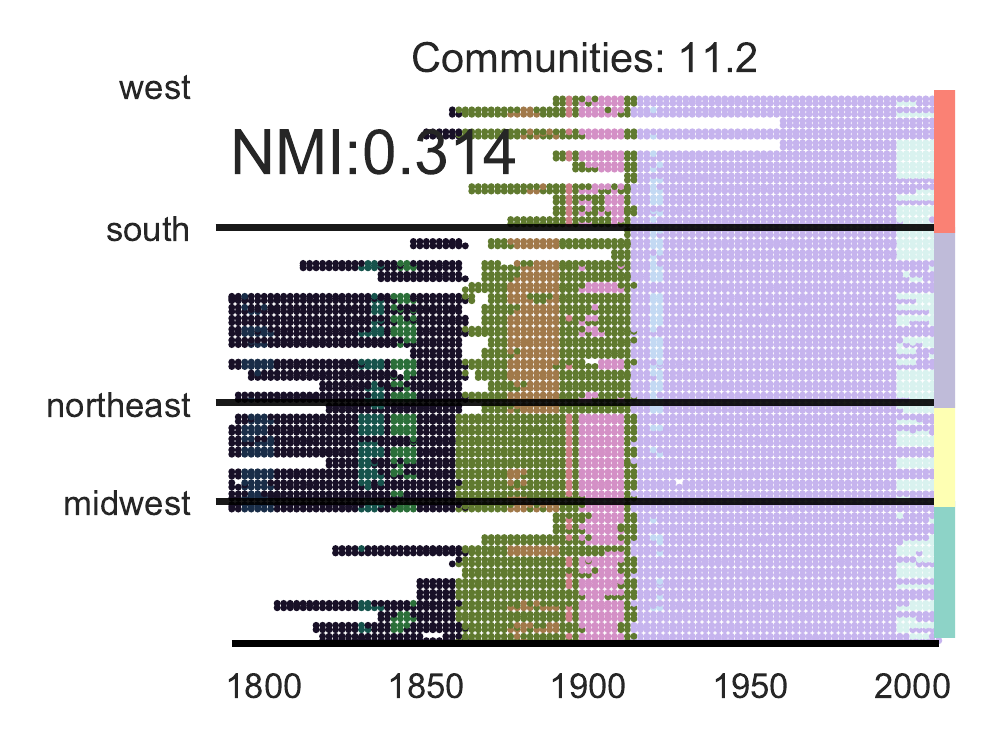}
\includegraphics[width=.24\linewidth]{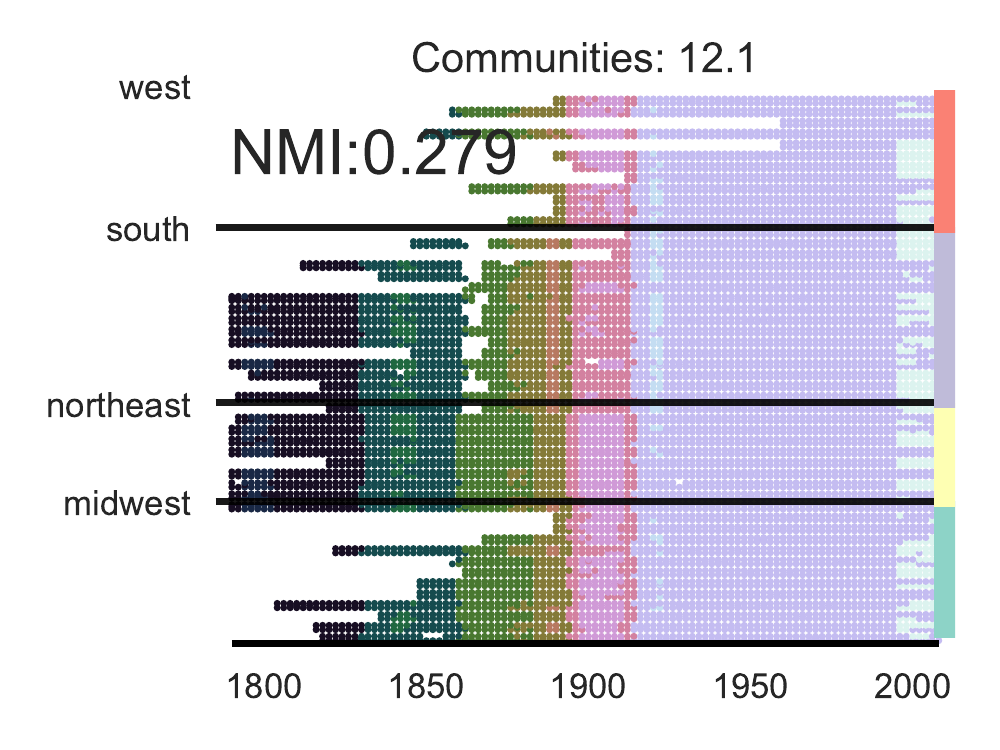}
\includegraphics[width=.24\linewidth]{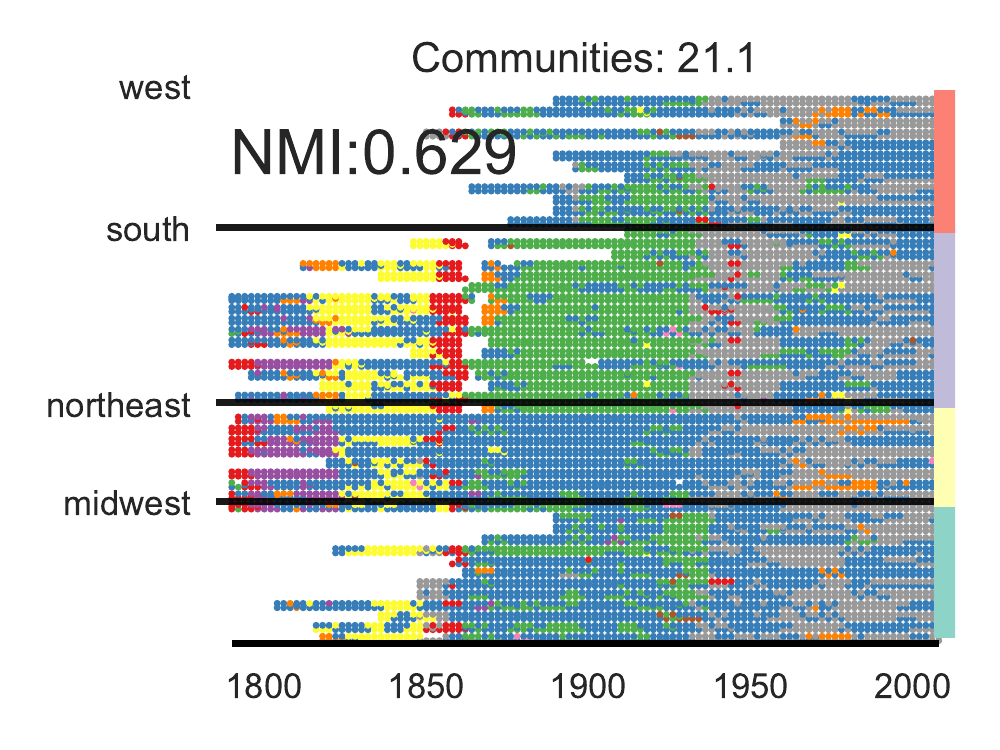}
\includegraphics[width=.24\linewidth]{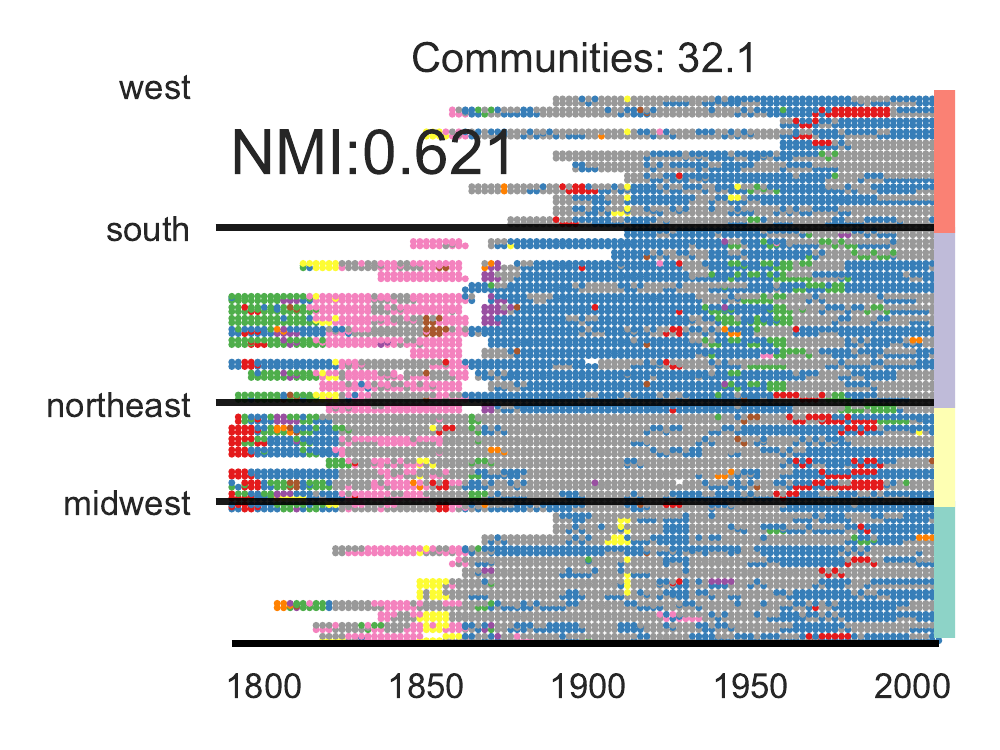}\\
\includegraphics[width=.24\linewidth]{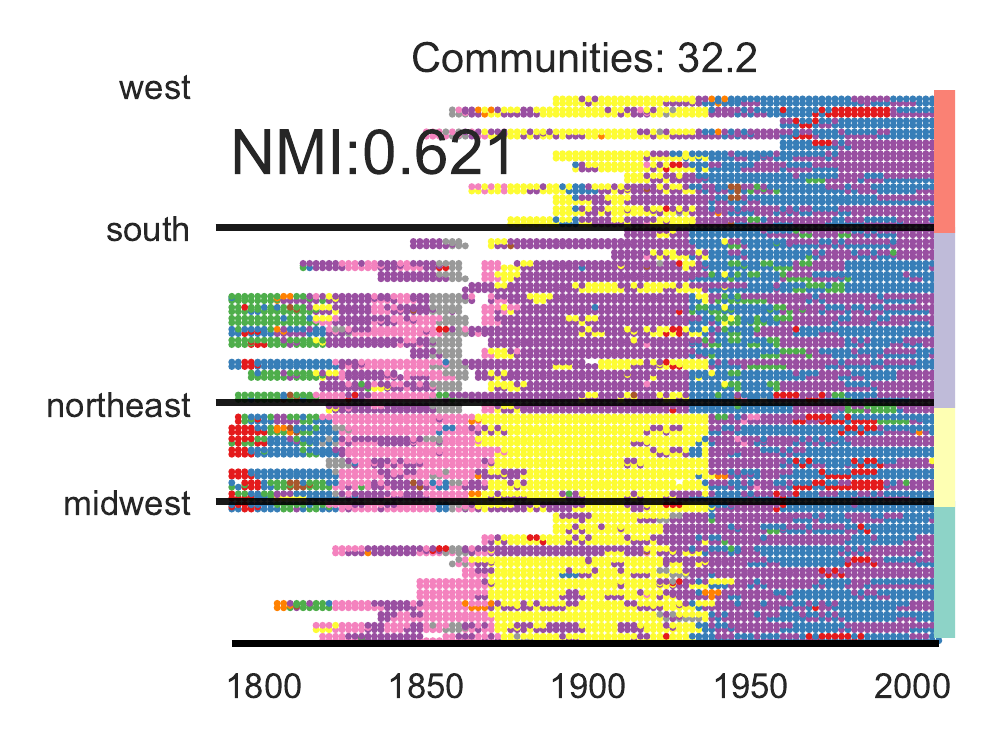}
\includegraphics[width=.24\linewidth]{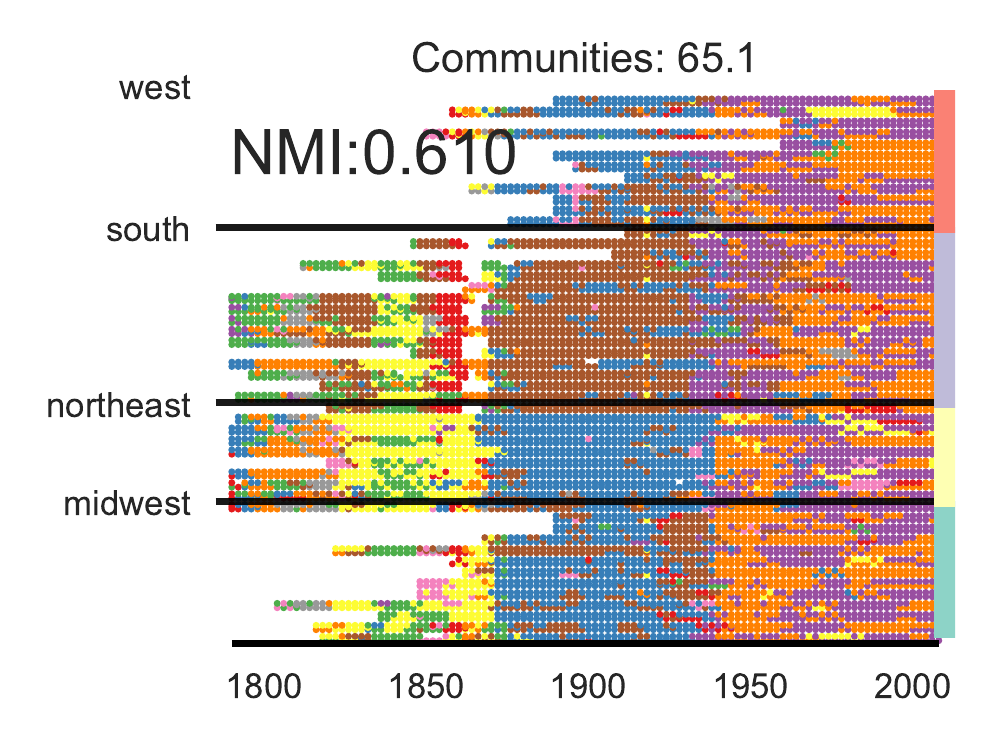}
\includegraphics[width=.24\linewidth]{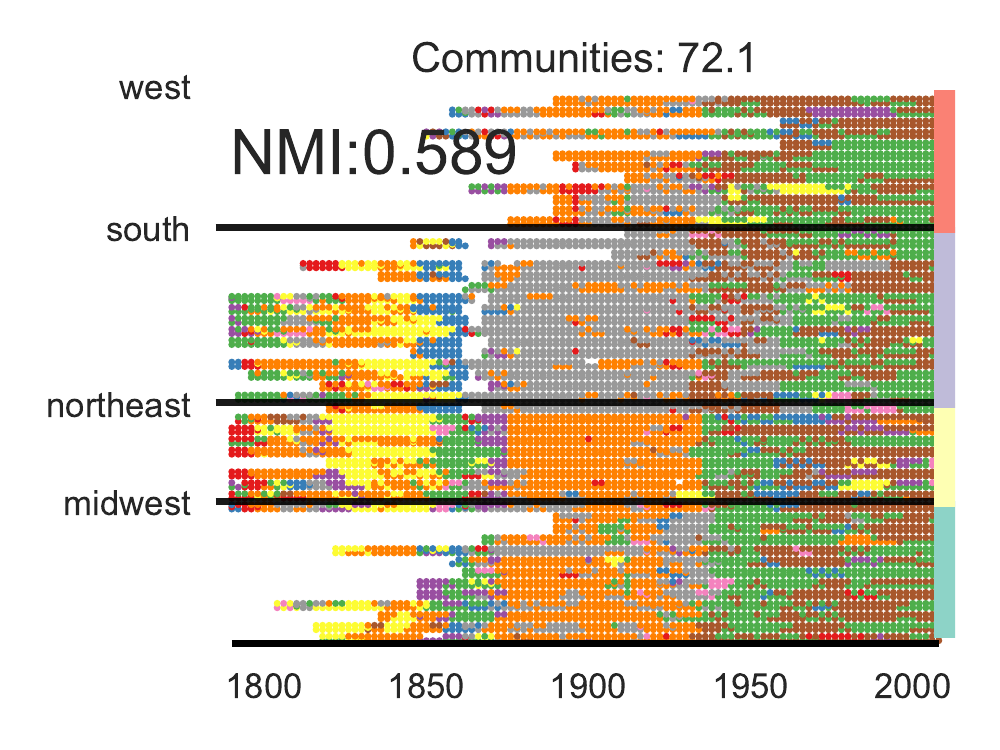}
\includegraphics[width=.24\linewidth]{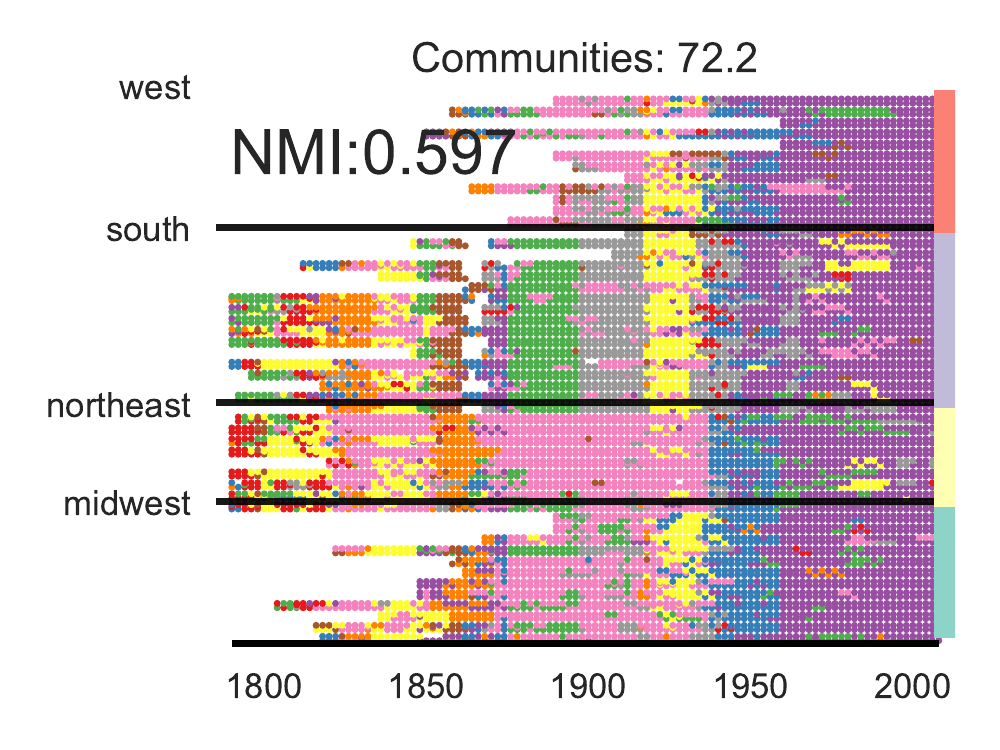}

\caption{Visualizations of partitions labeled in white in Figure~\ref{DODs}A, with Senators grouped according to their state. The listed AMI is the average over layers of the AMI in each layer (Congress) between the communities and political party affiliations for that Congress. Partitions are labeled ``$X.Y$'' with $X$ the number of communities with $\geq$ 5 nodes and $Y$ the rank of the domain area for that number of~communities.}
\label{state}
\end{figure}

\begin{figure}[H]
\centering
  \includegraphics[width=.24\linewidth]{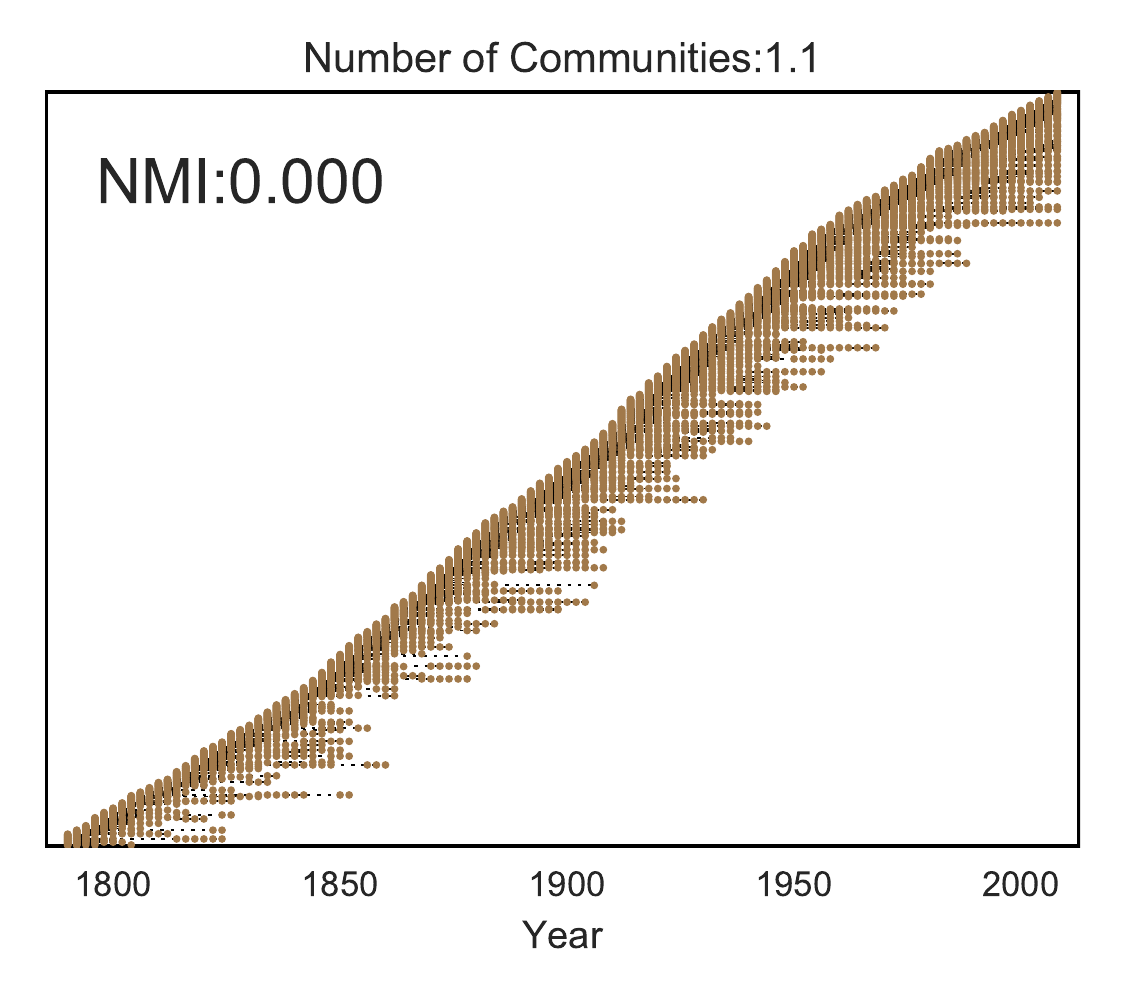}
  \includegraphics[width=.24\linewidth]{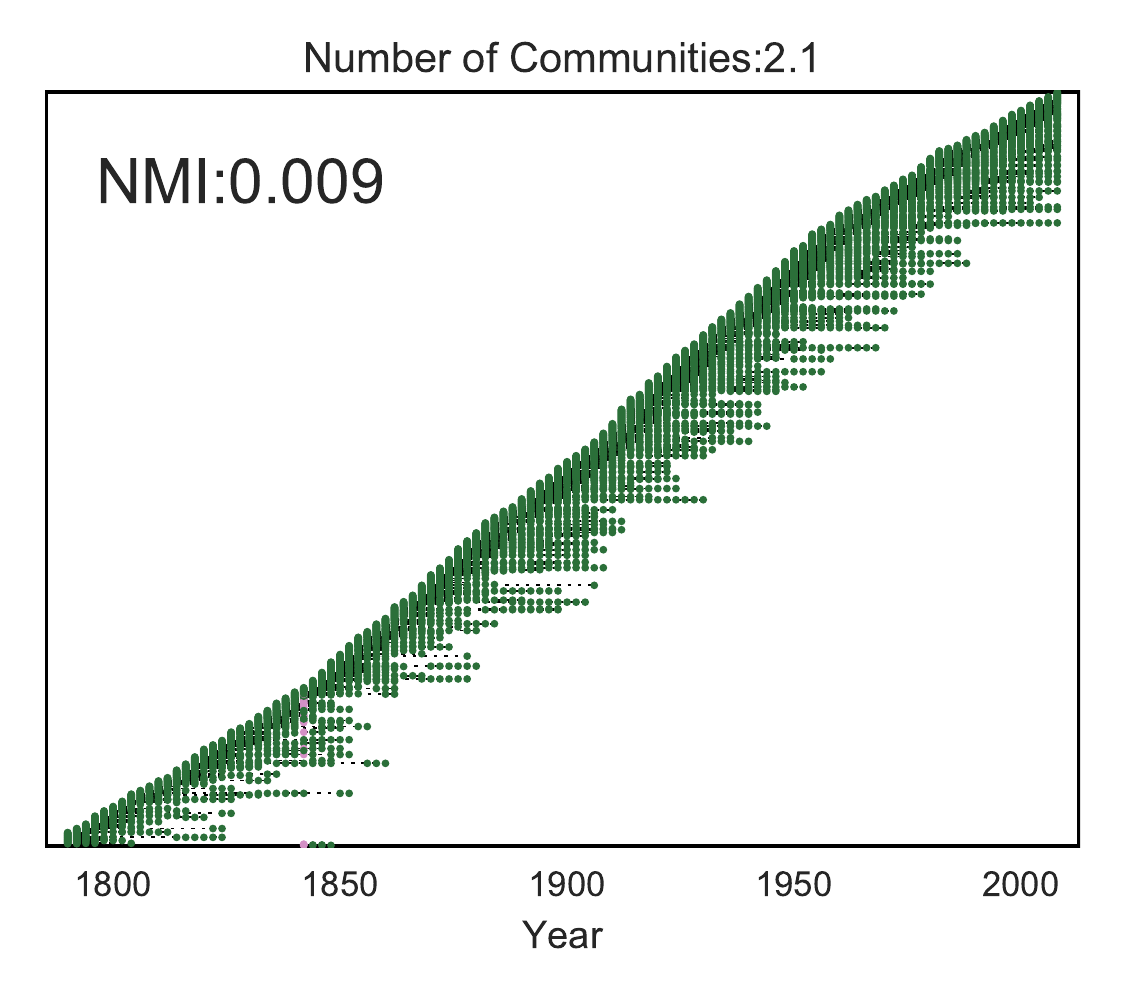}
  \includegraphics[width=.24\linewidth]{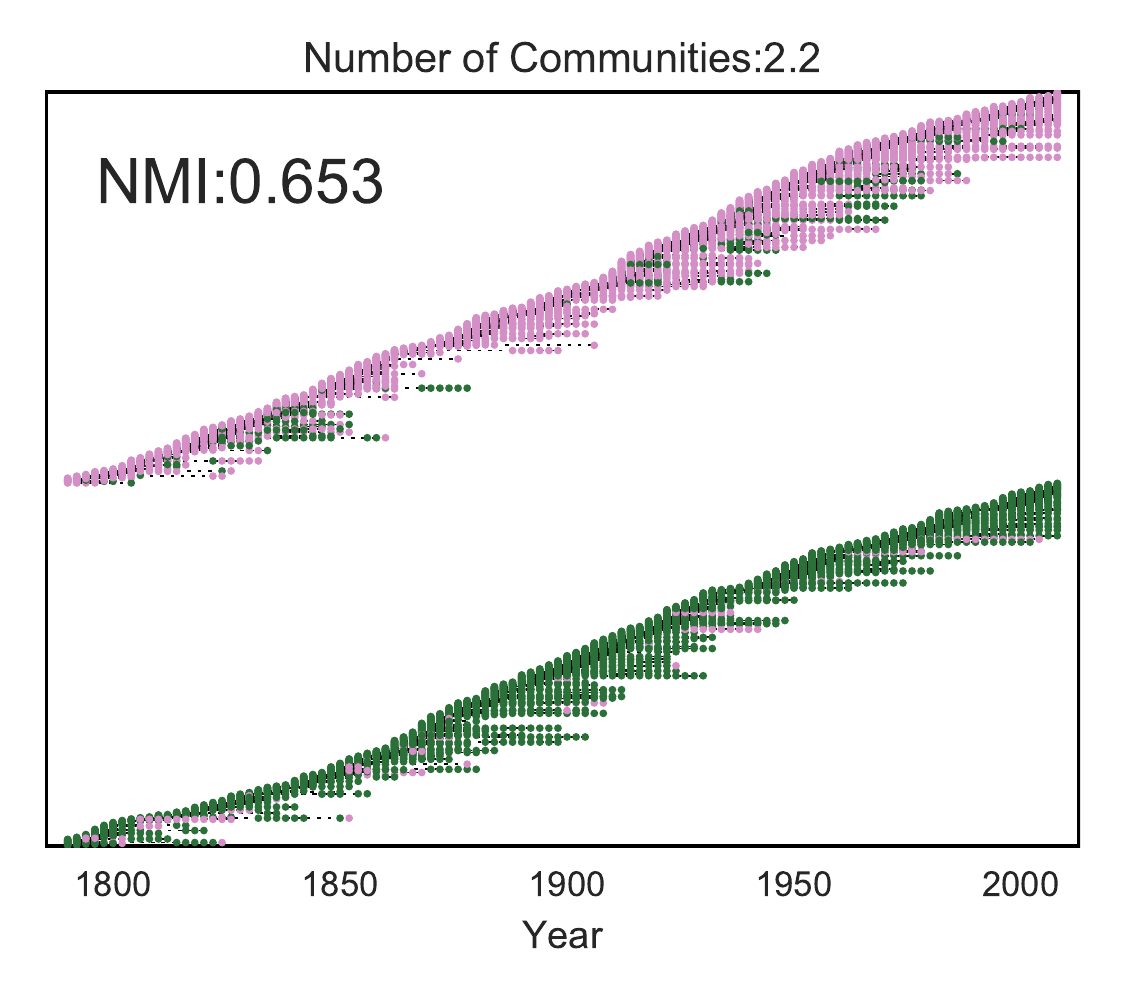}
  \includegraphics[width=.24\linewidth]{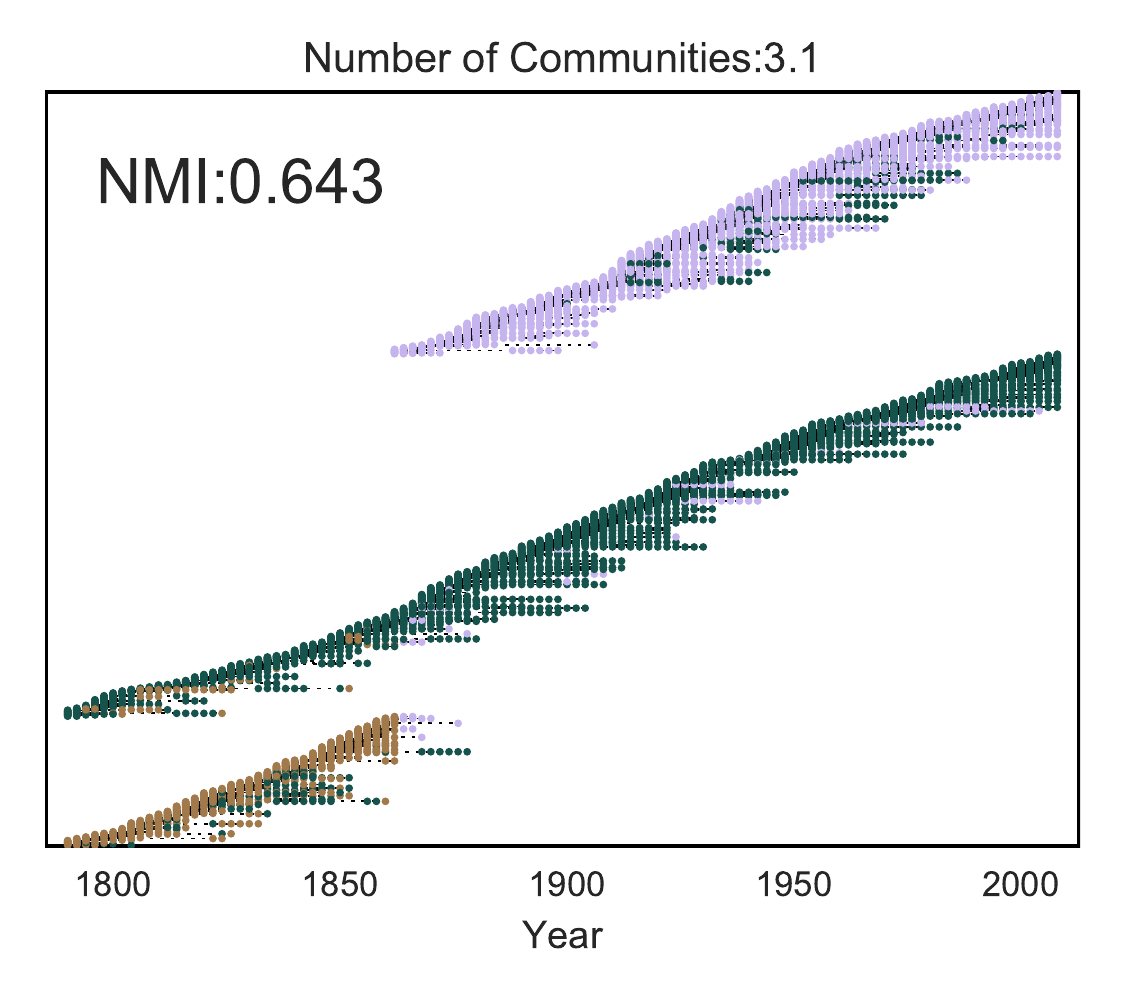}\\
  \includegraphics[width=.24\linewidth]{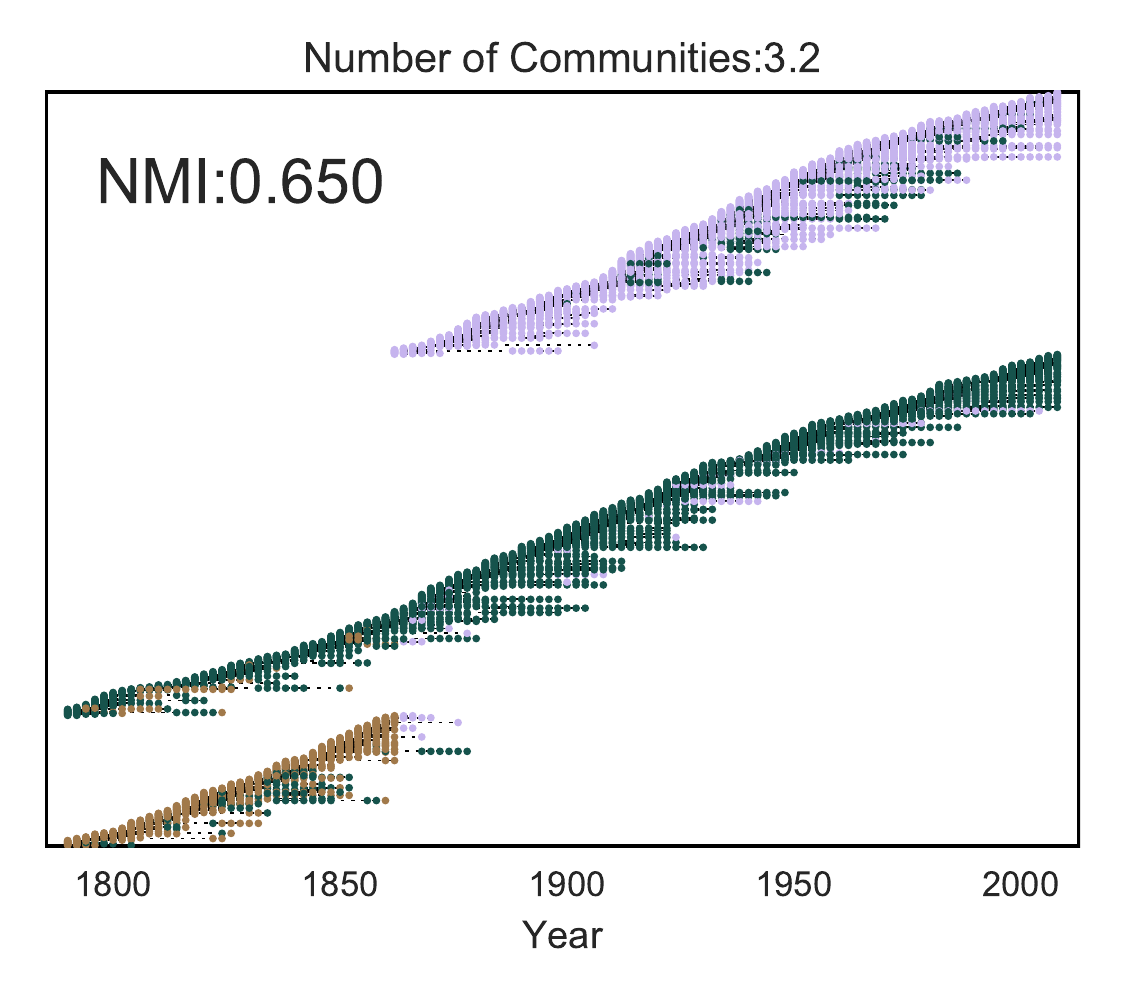}
  \includegraphics[width=.24\linewidth]{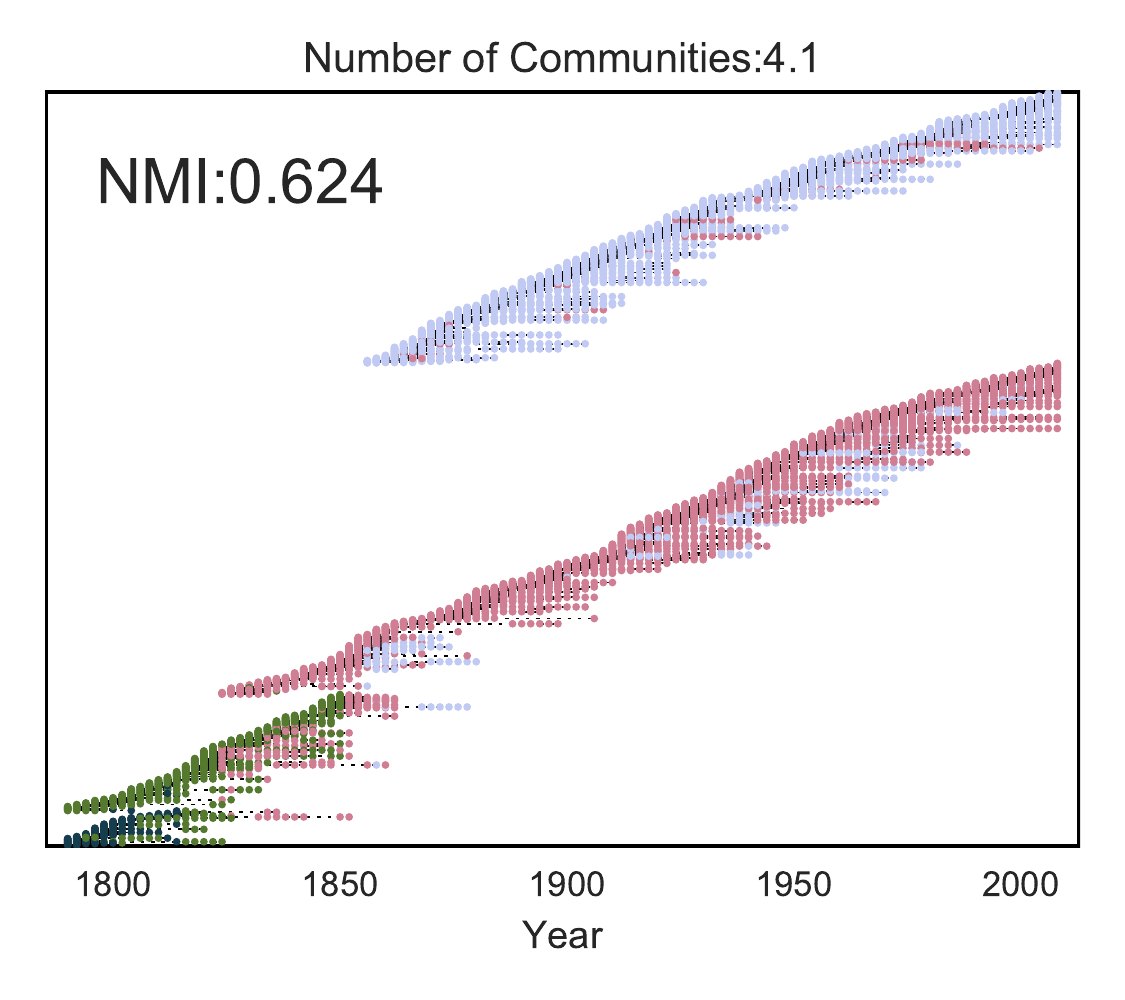}
  \includegraphics[width=.24\linewidth]{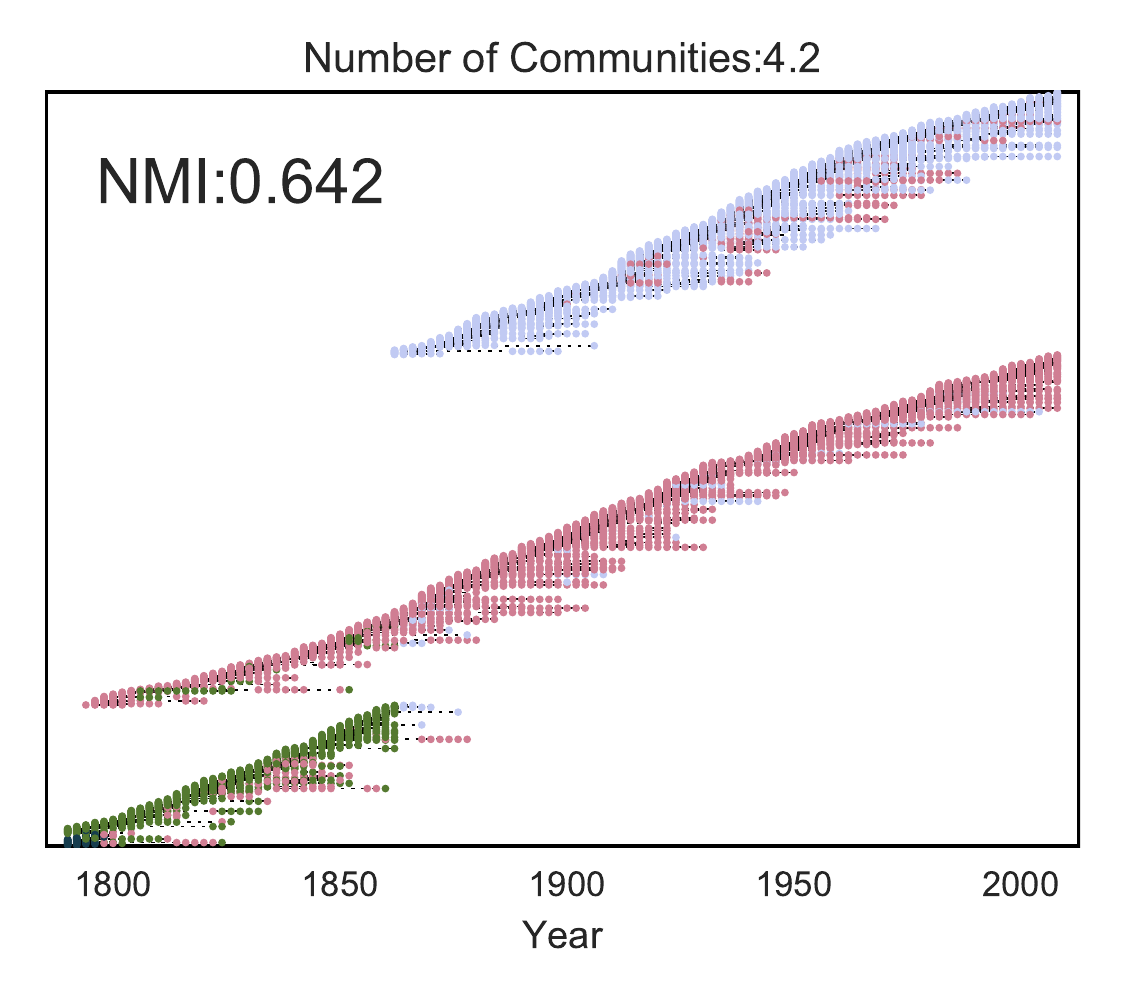}
  \includegraphics[width=.24\linewidth]{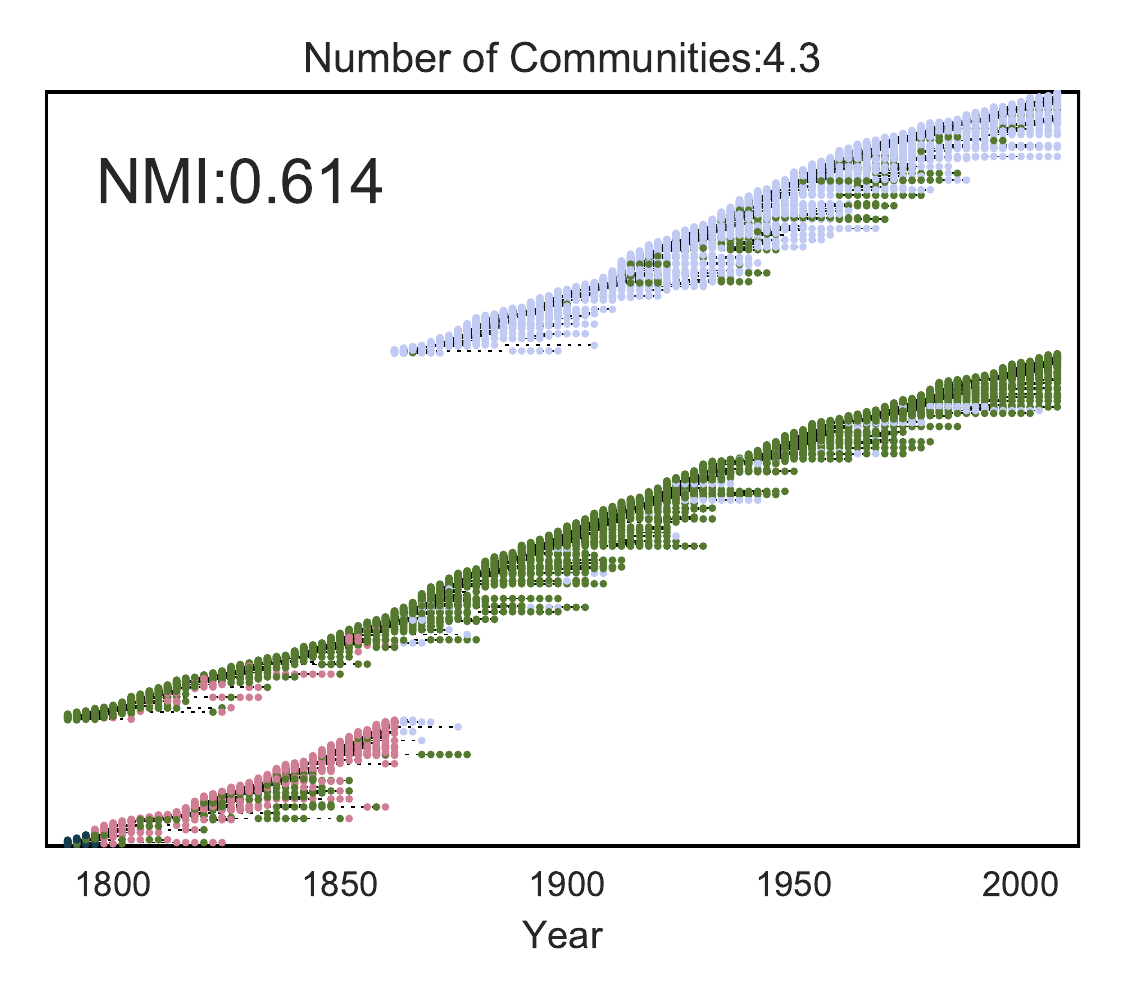}\\
  \includegraphics[width=.24\linewidth]{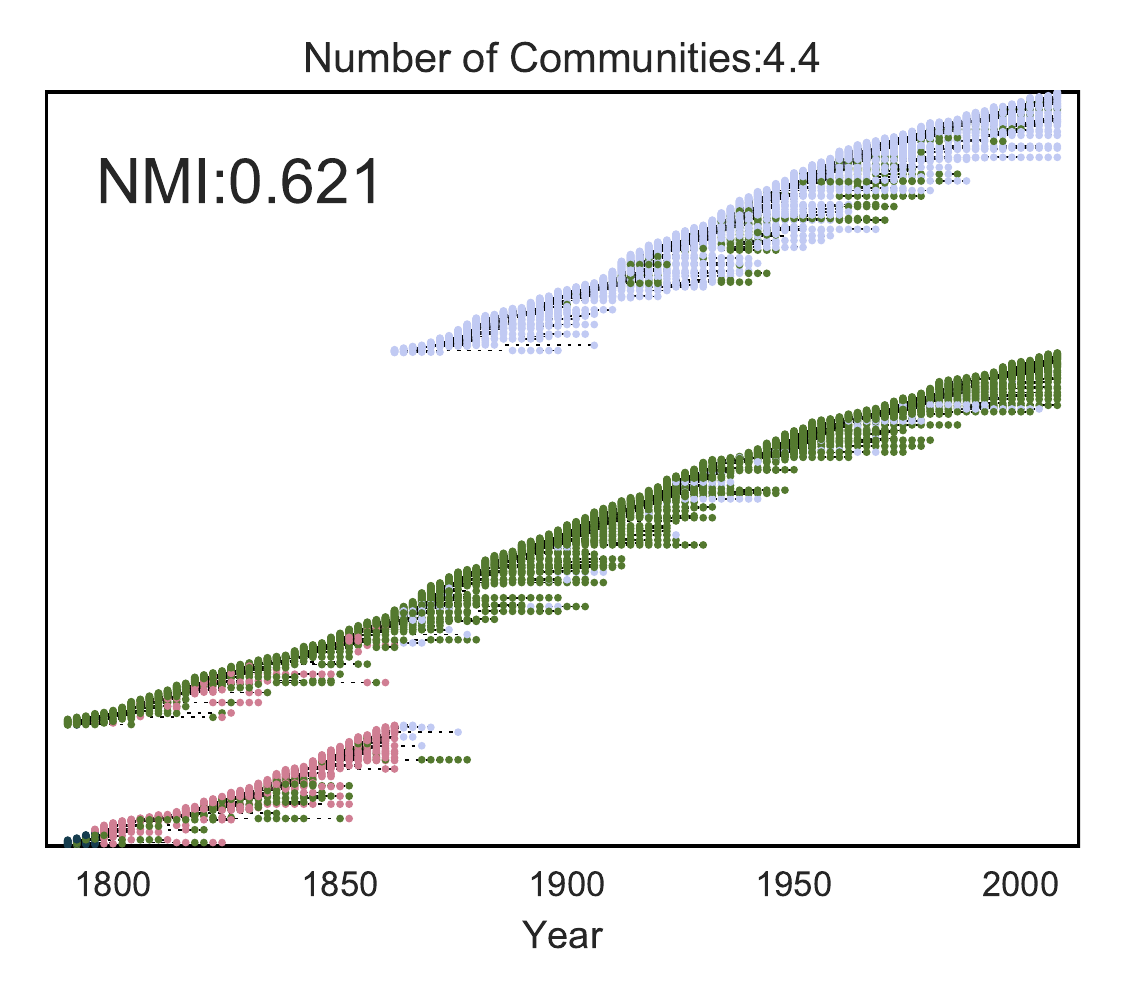}
  \includegraphics[width=.24\linewidth]{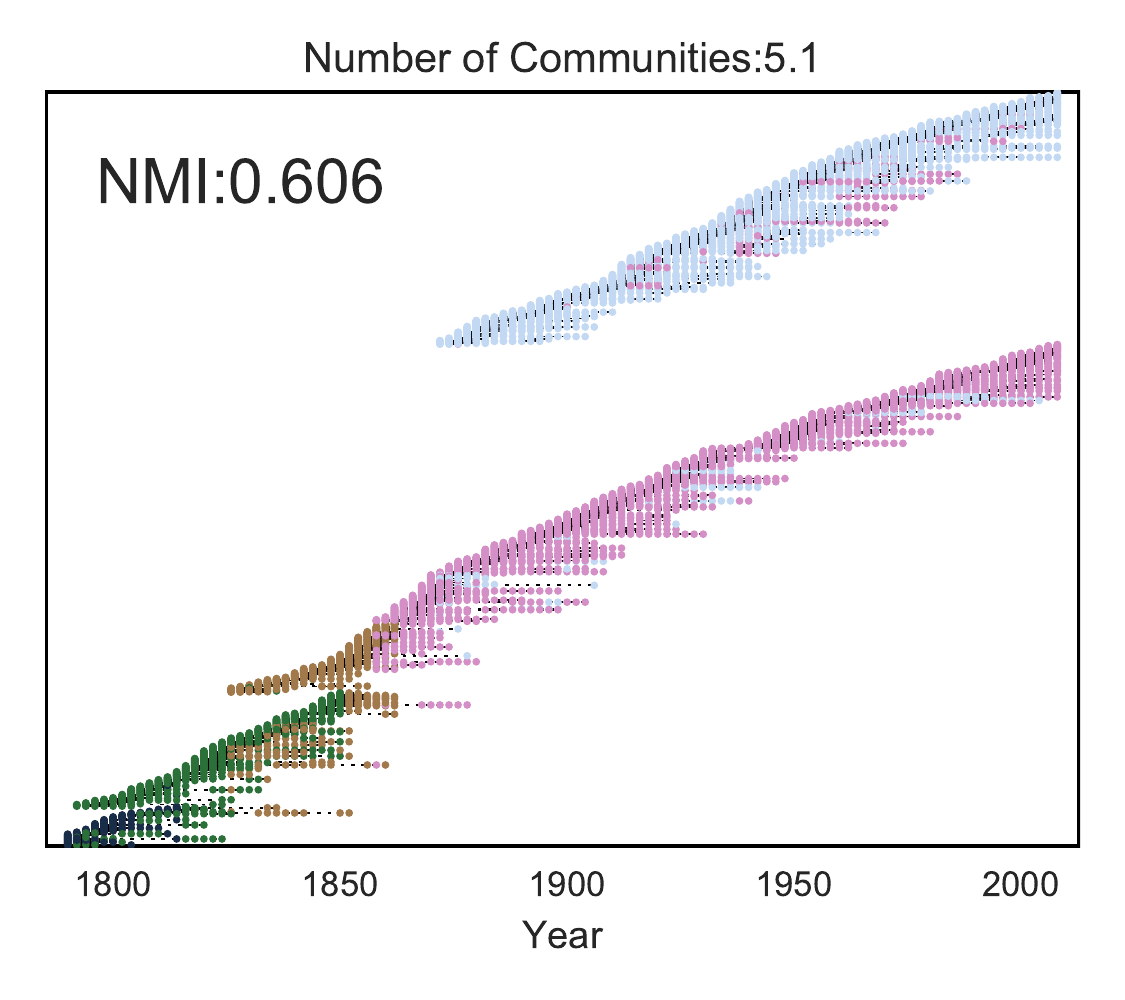}
  \includegraphics[width=.24\linewidth]{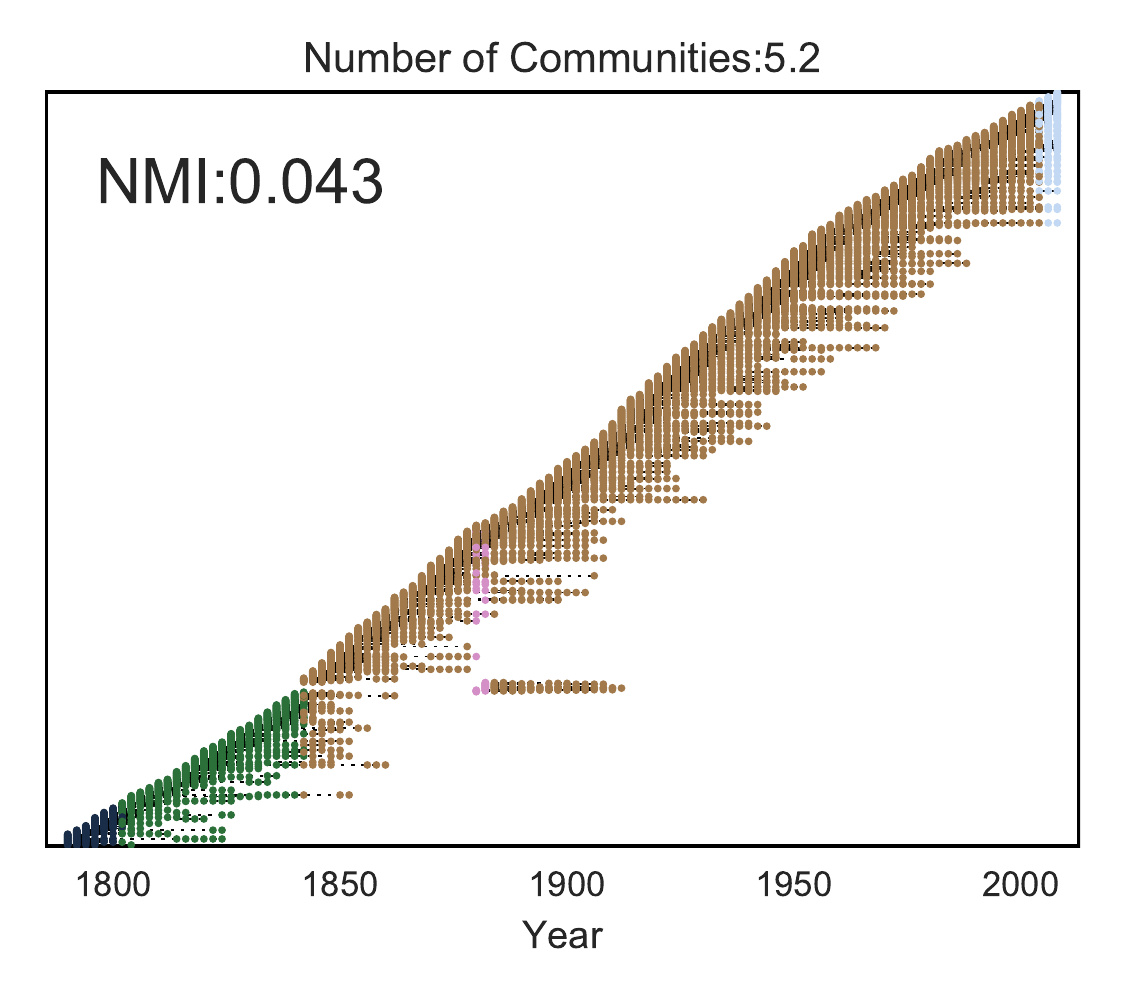}
  \includegraphics[width=.24\linewidth]{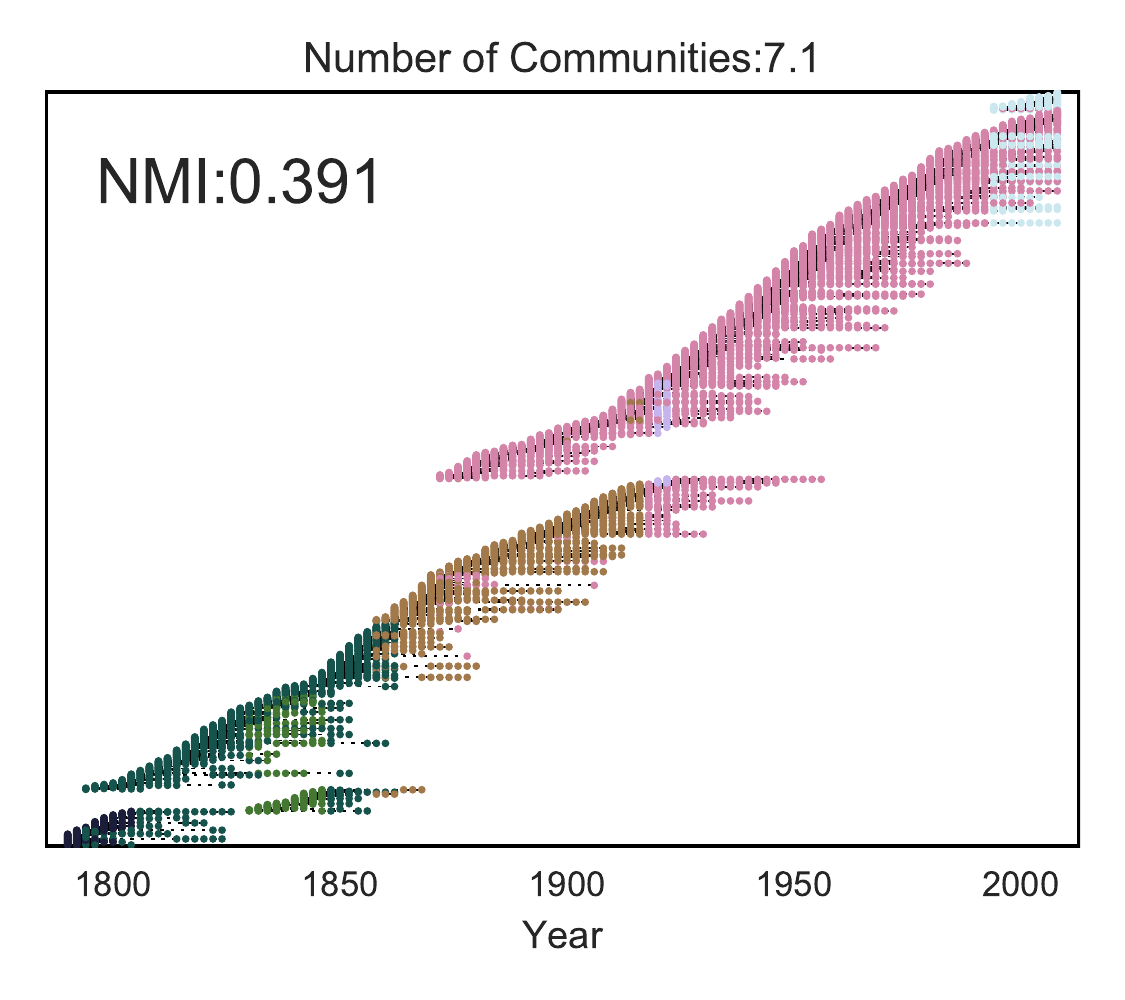}\\
  \includegraphics[width=.24\linewidth]{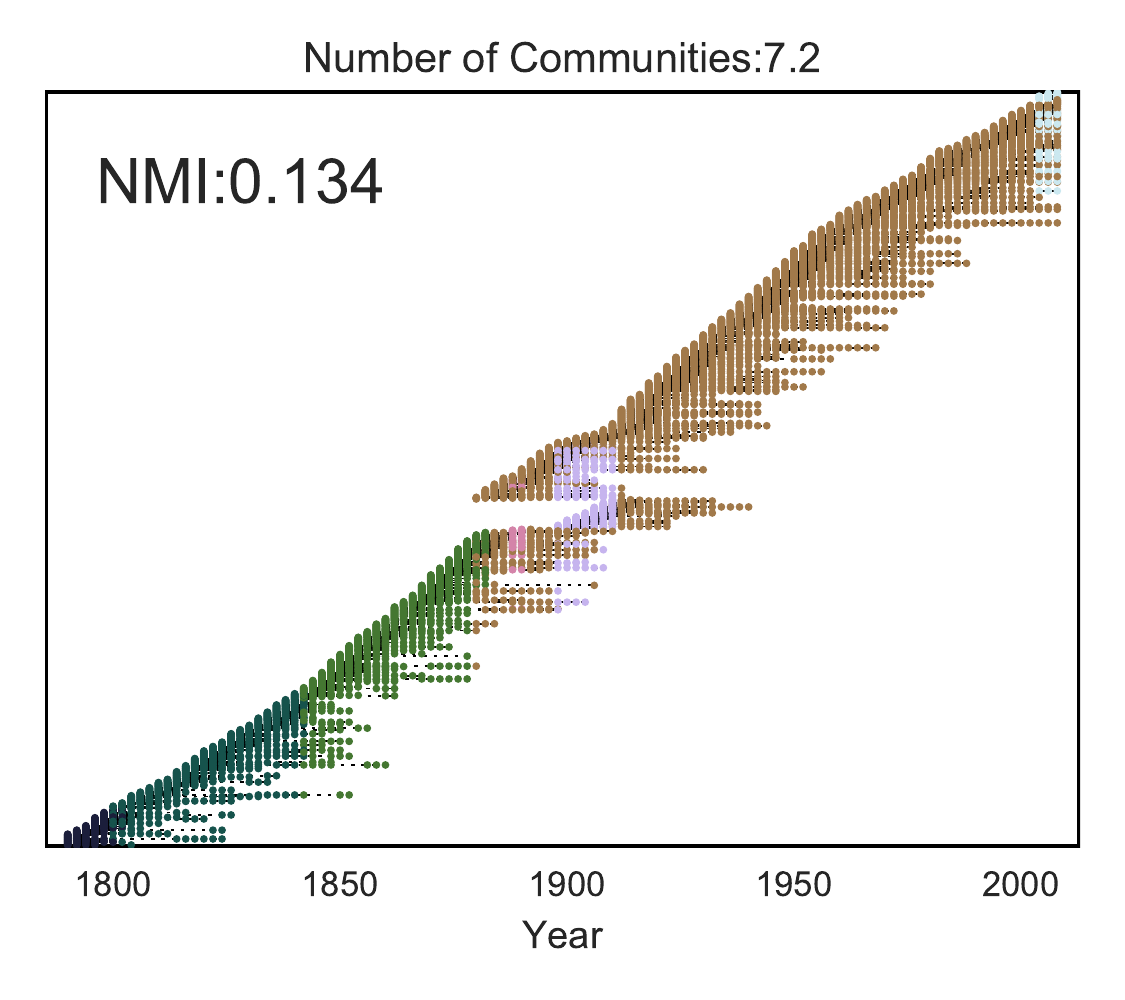}
  \includegraphics[width=.24\linewidth]{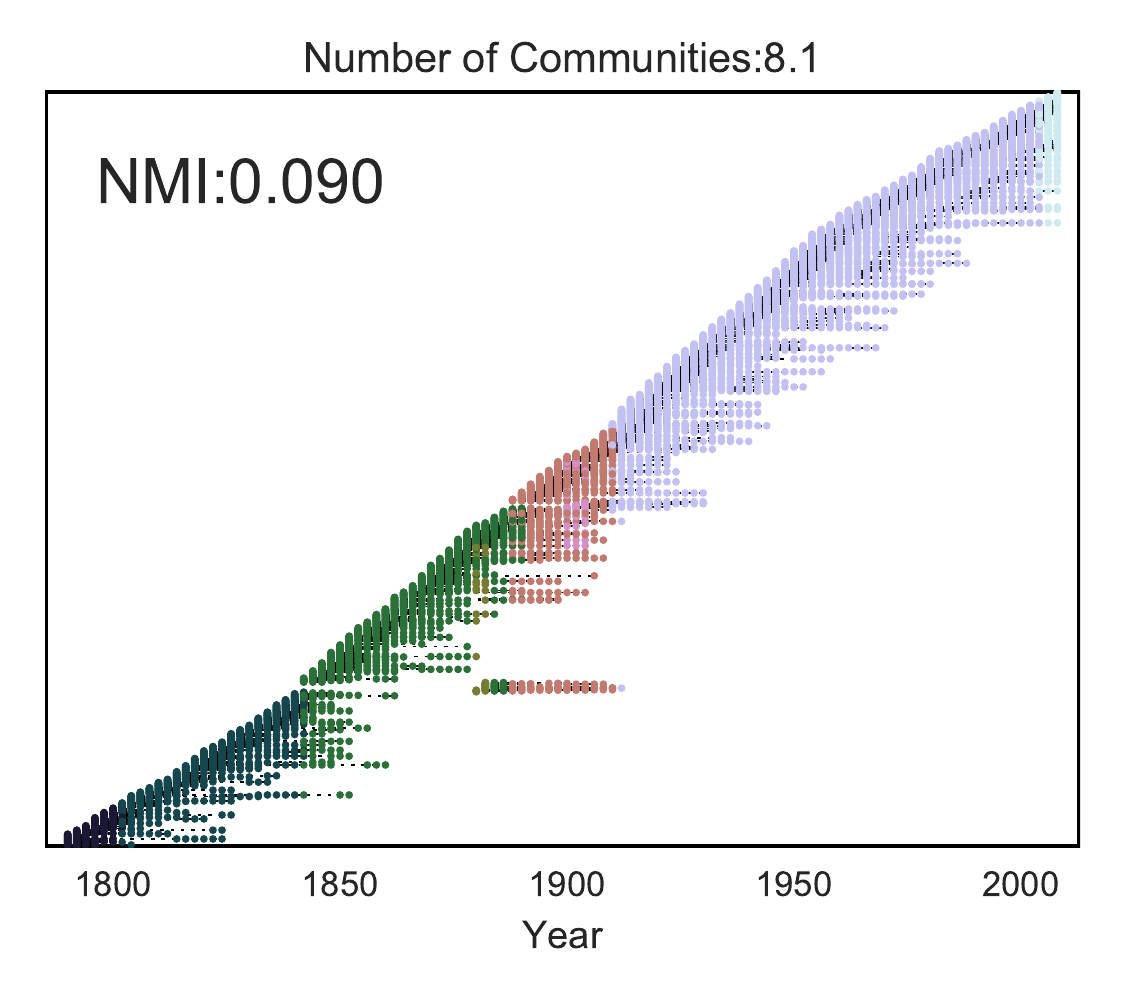}
  \includegraphics[width=.24\linewidth]{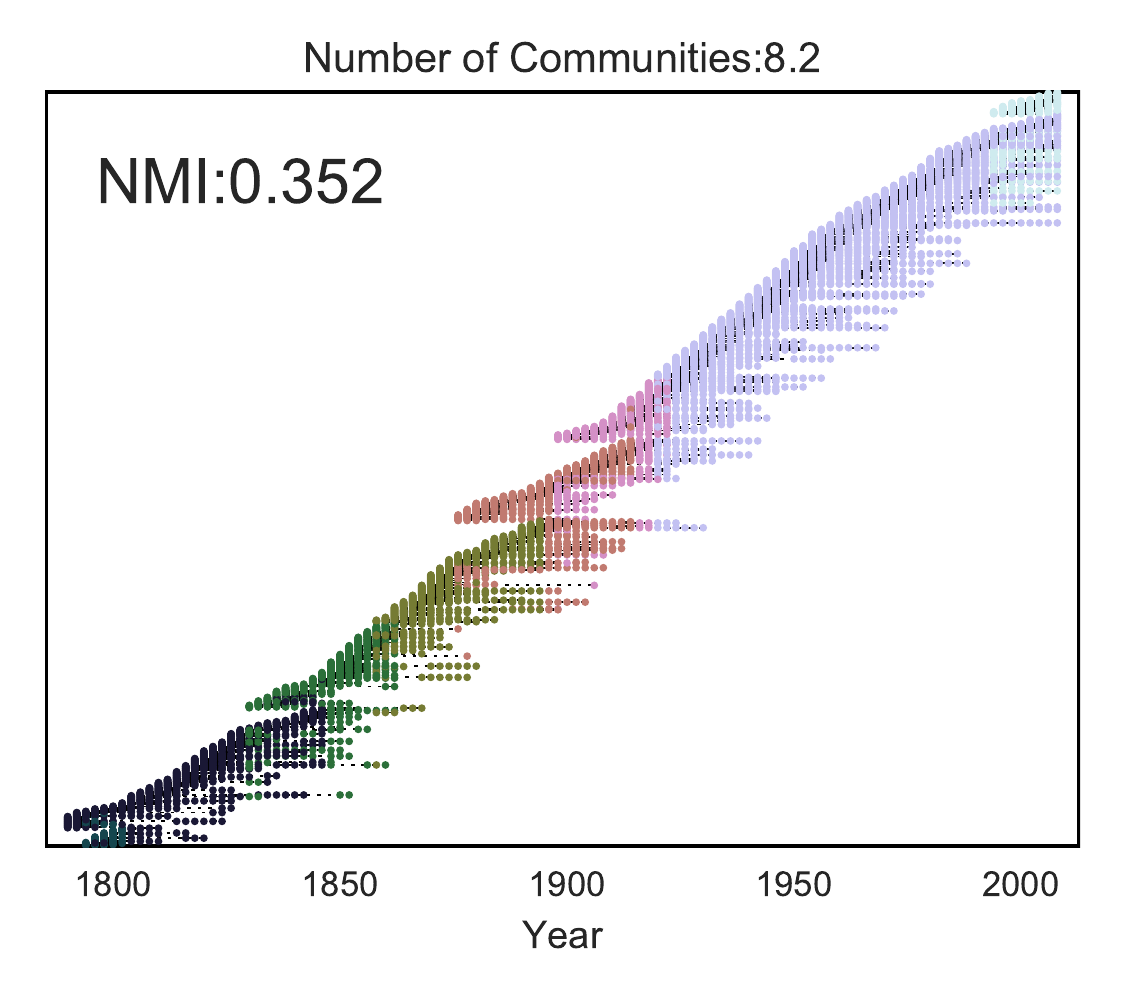}
  \includegraphics[width=.24\linewidth]{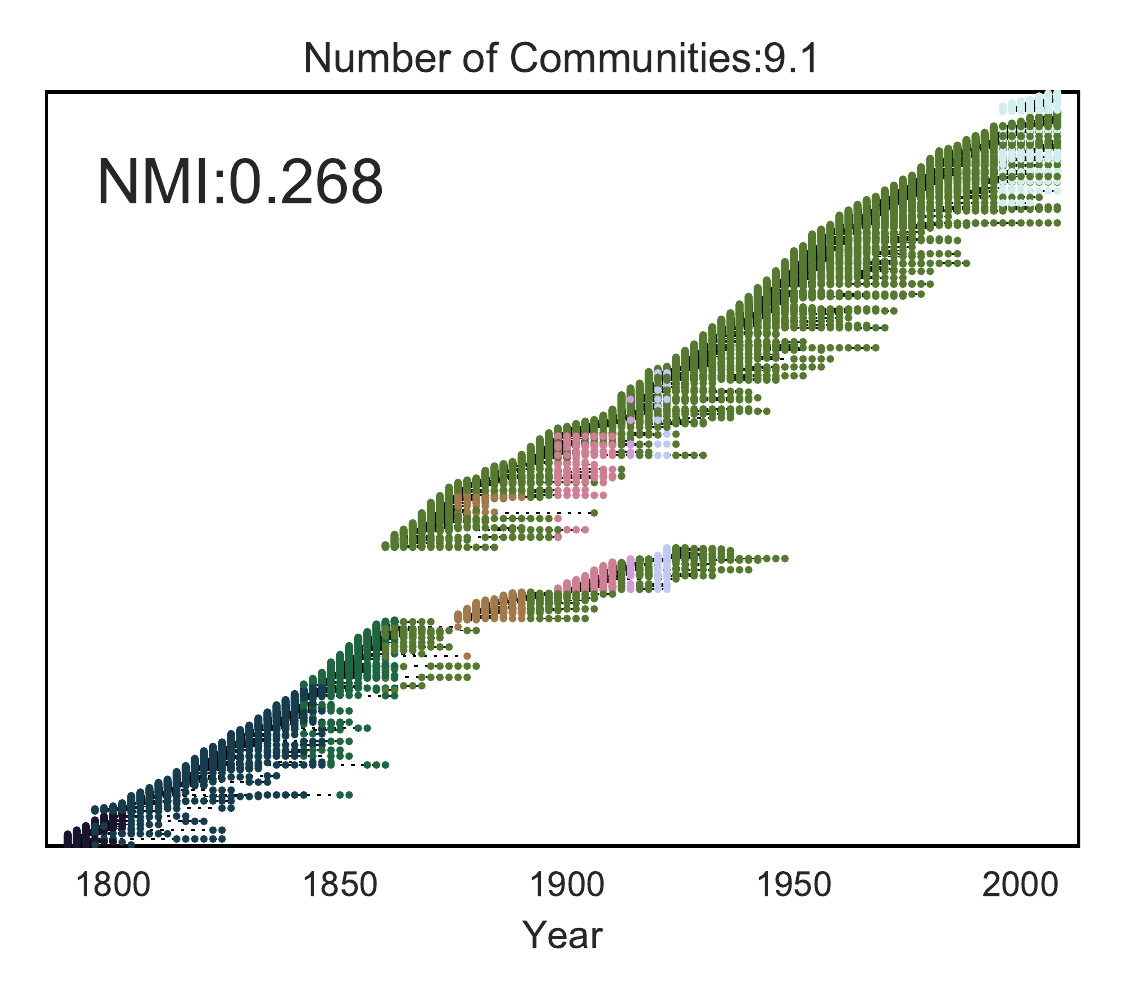}\\
  \includegraphics[width=.24\linewidth]{senate_layouts/individ_layout_9_1.pdf}
  \includegraphics[width=.24\linewidth]{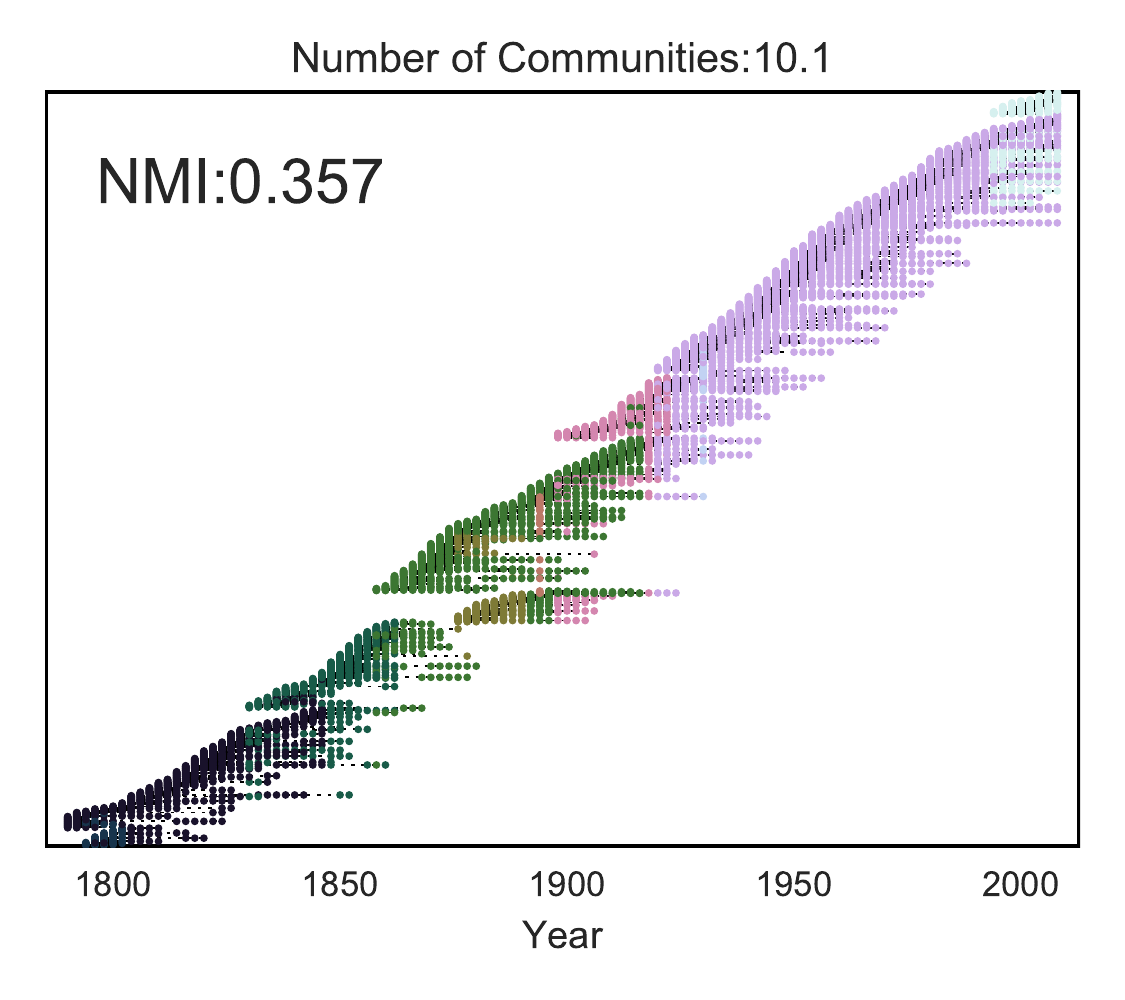}
  \includegraphics[width=.24\linewidth]{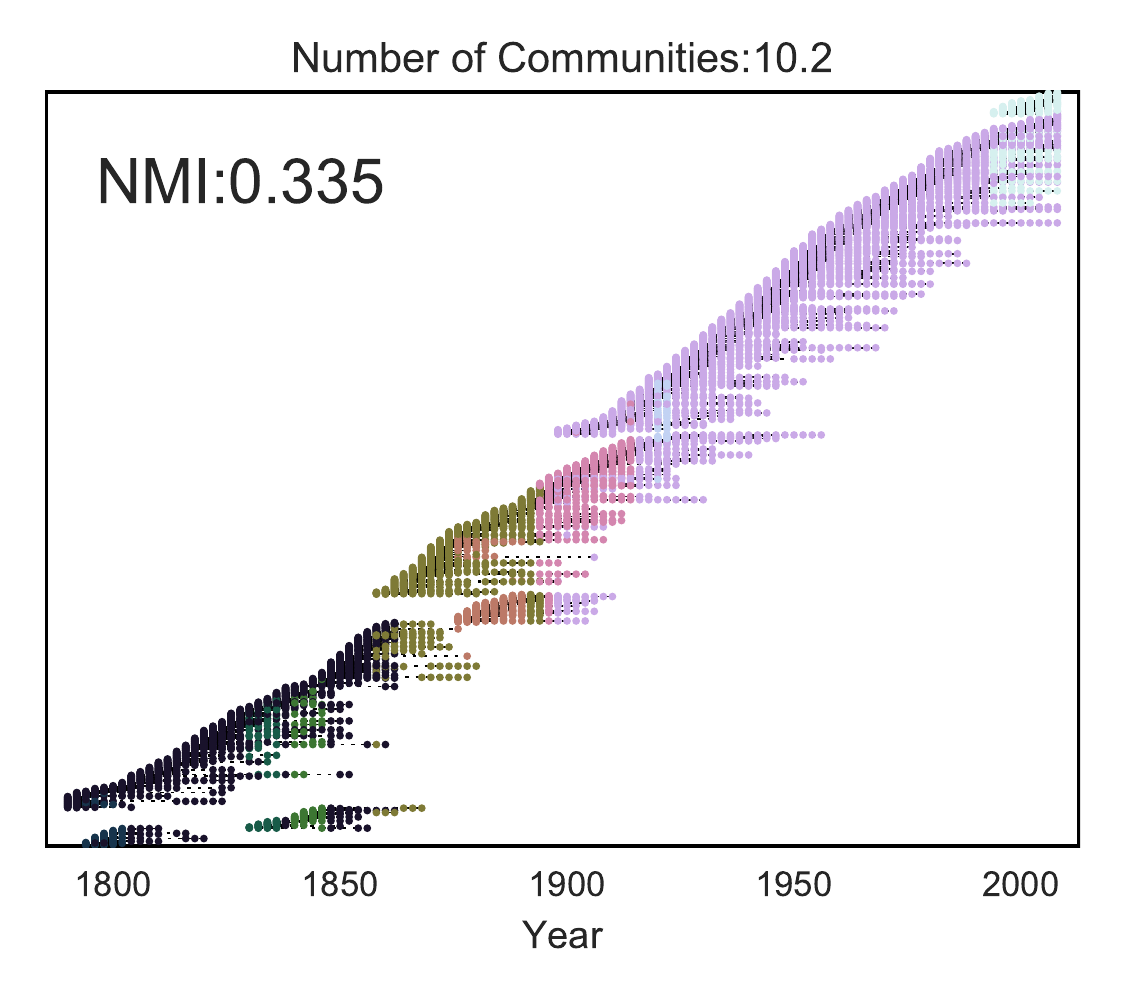}
  \includegraphics[width=.24\linewidth]{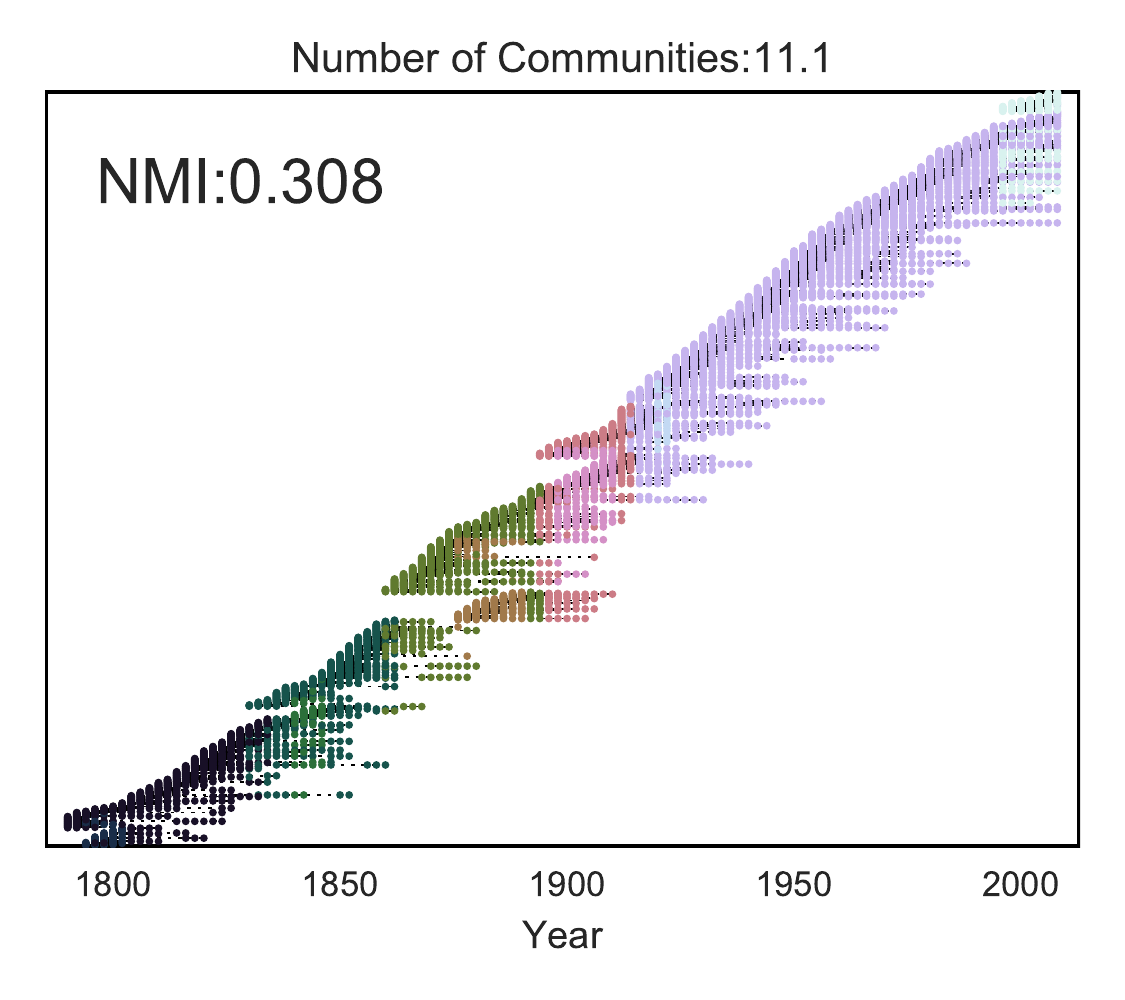}\\
   \includegraphics[width=.24\linewidth]{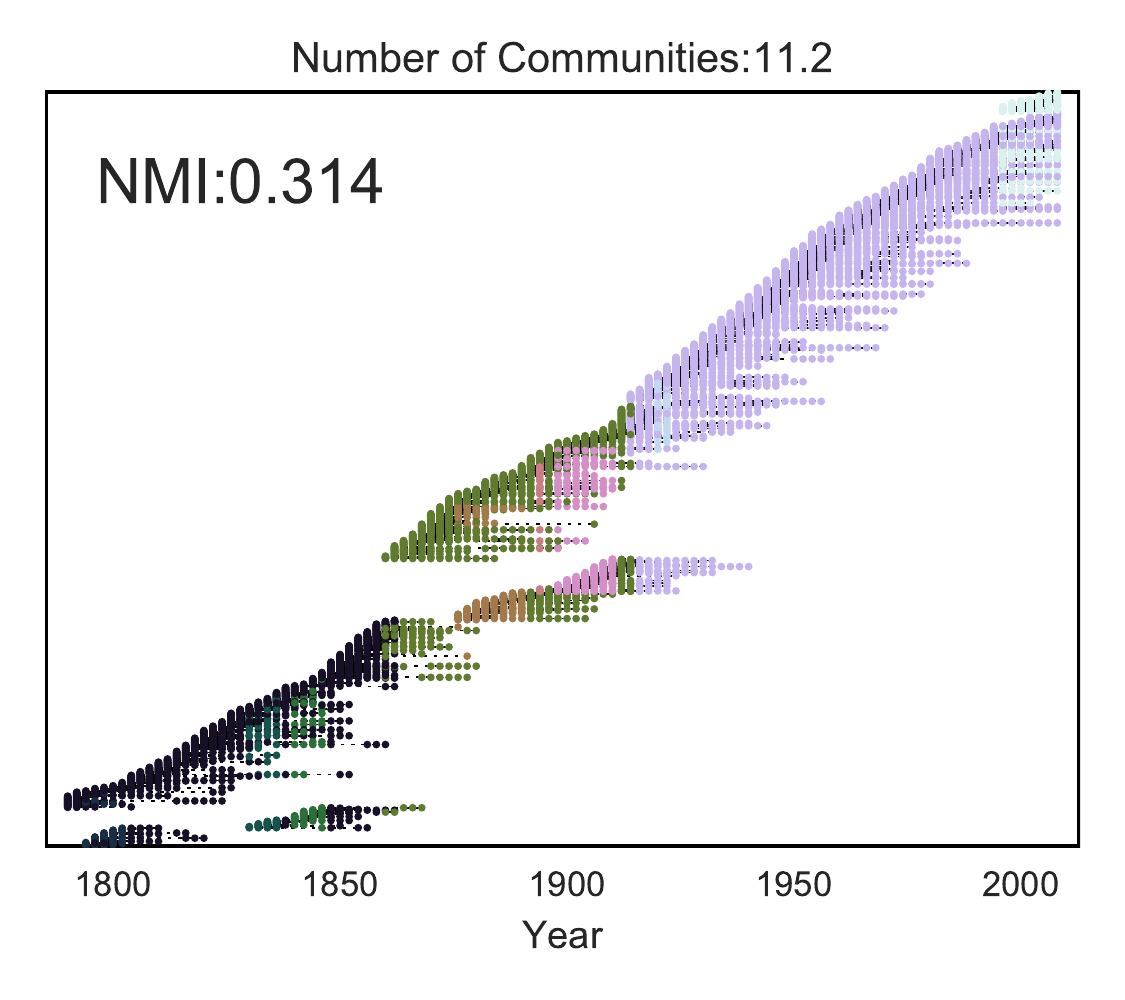}
  \includegraphics[width=.24\linewidth]{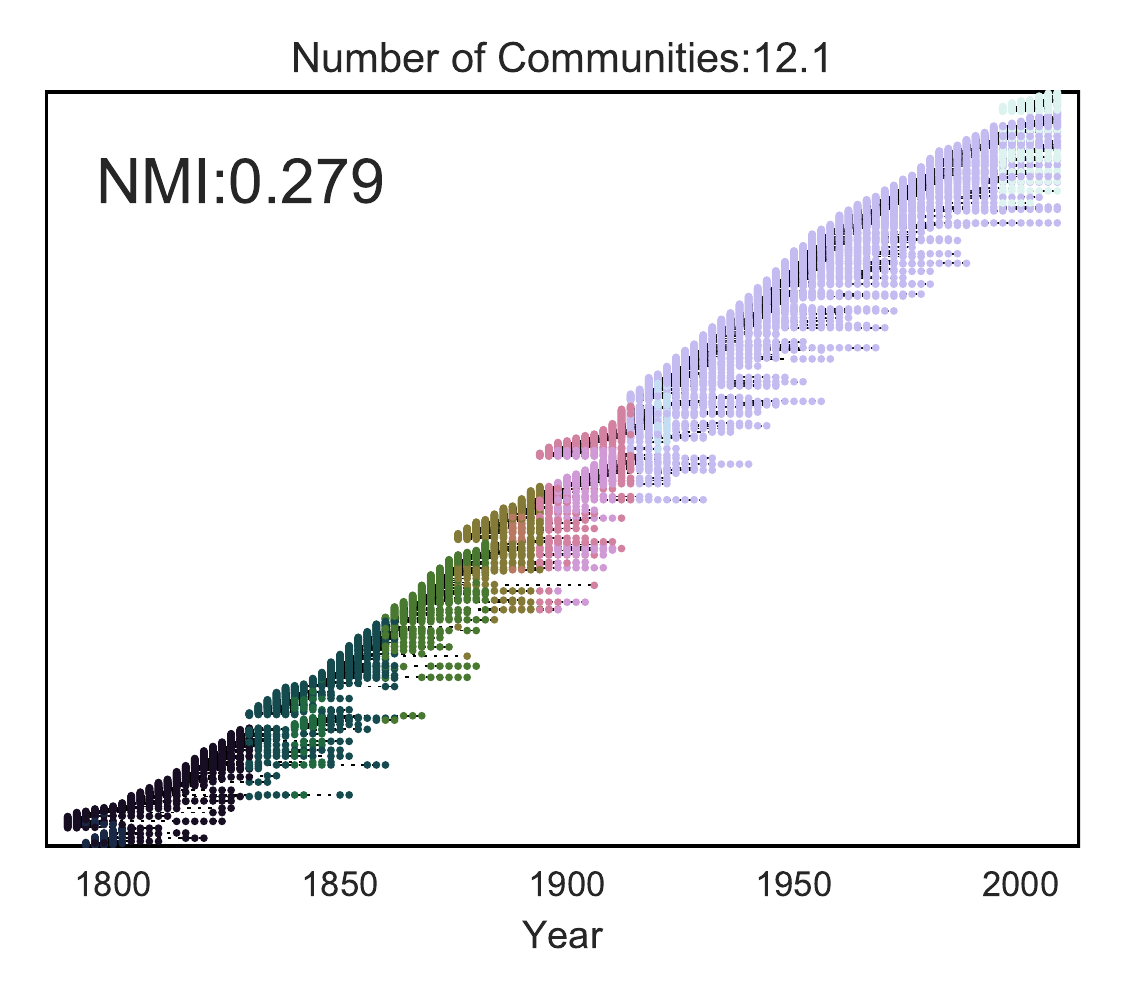}
  \includegraphics[width=.24\linewidth]{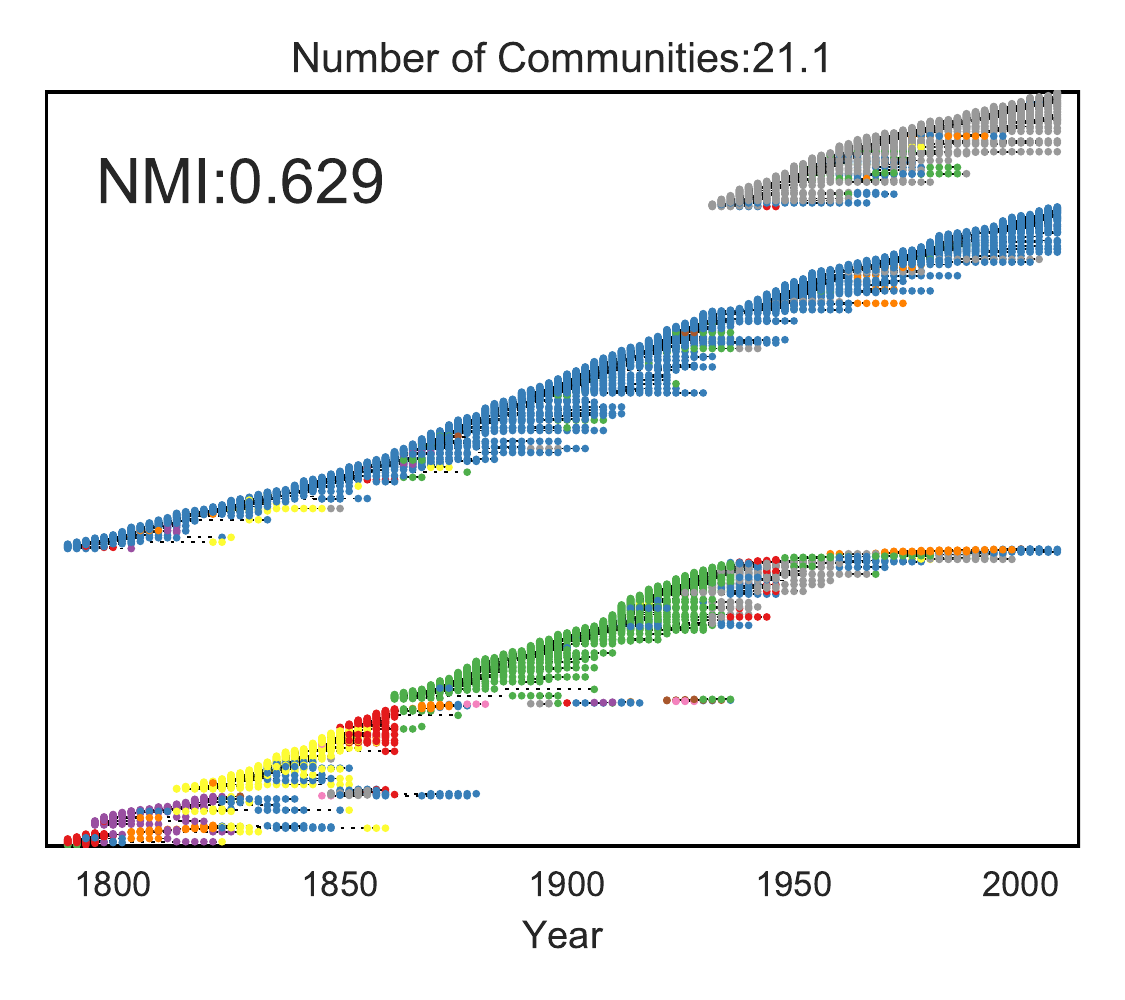}
  \includegraphics[width=.24\linewidth]{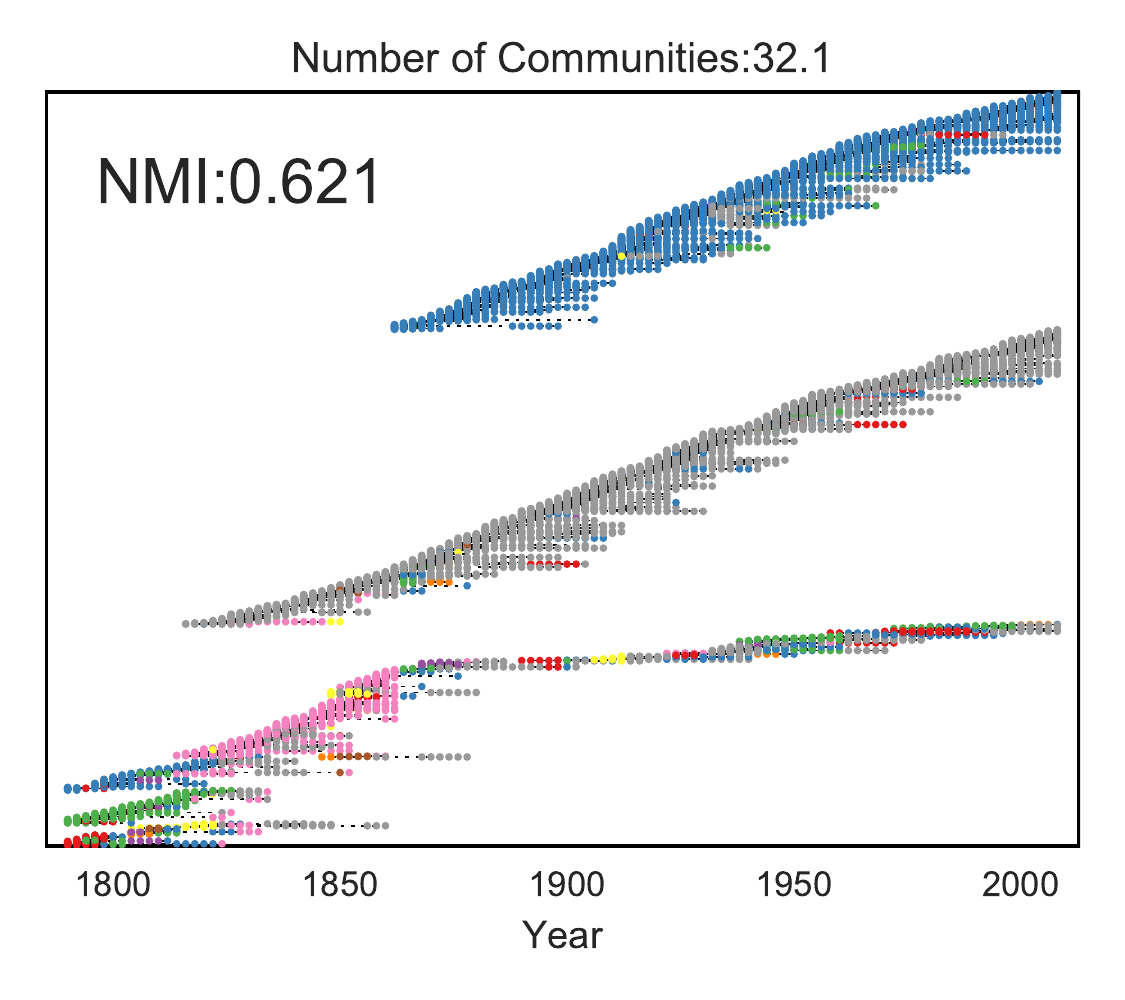}\\
  \includegraphics[width=.24\linewidth]{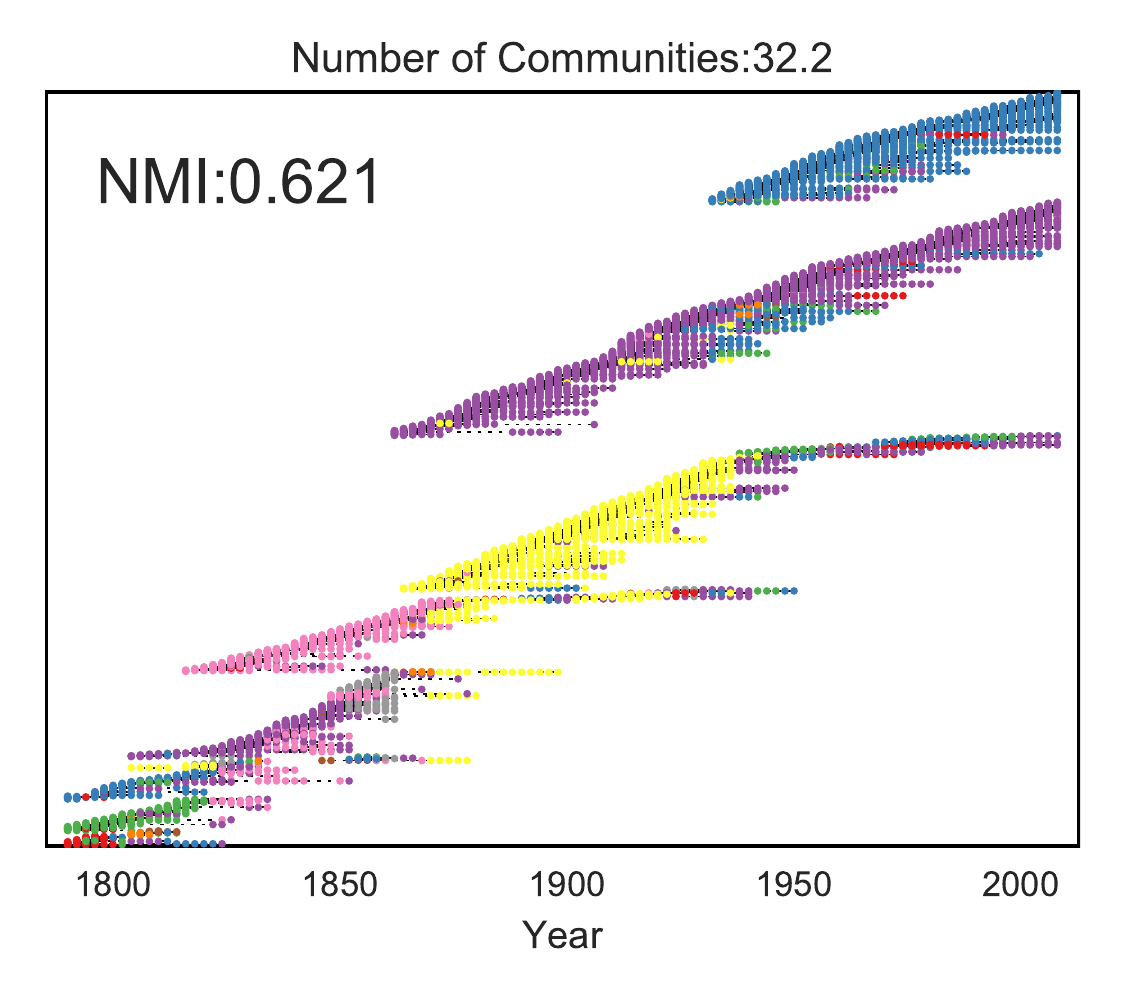}
  \includegraphics[width=.24\linewidth]{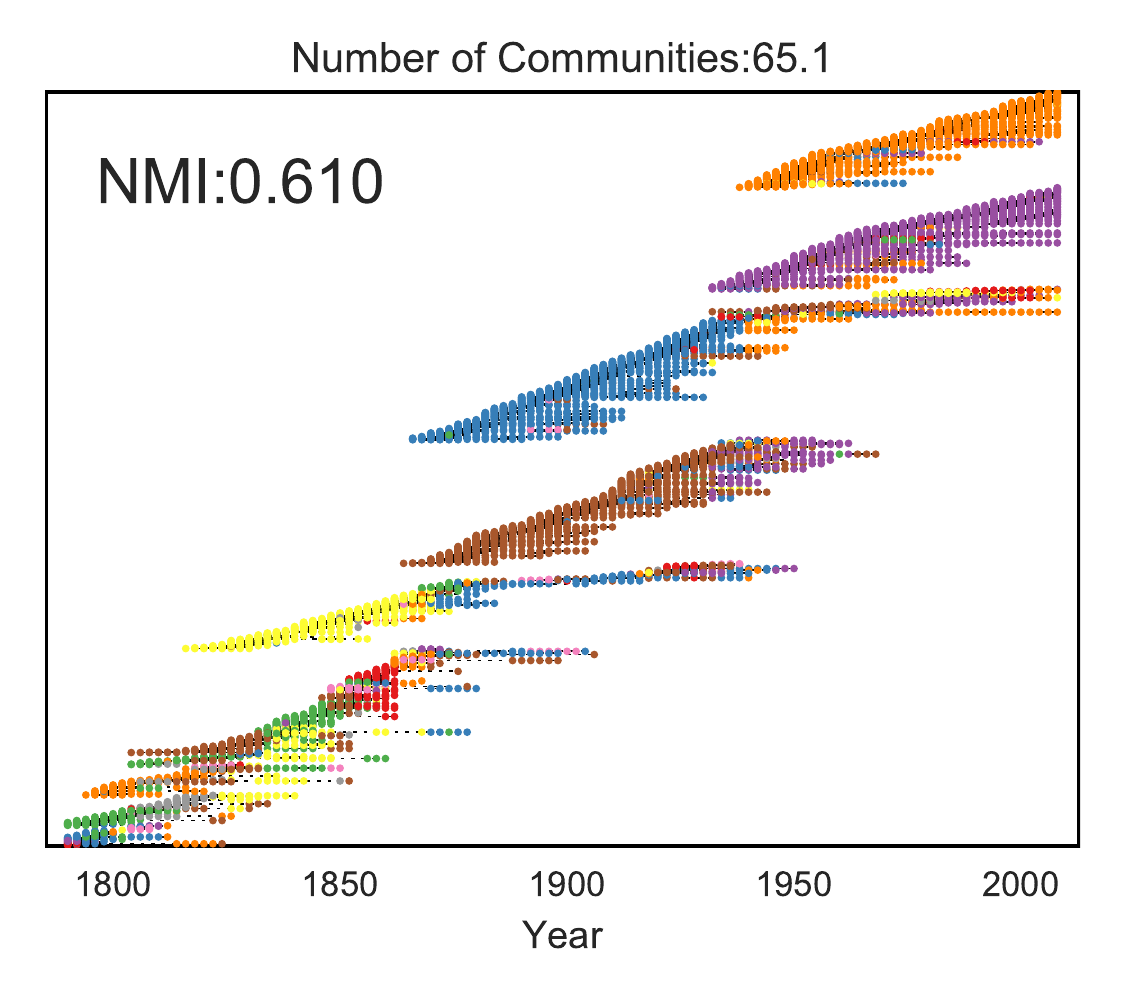}
  \includegraphics[width=.24\linewidth]{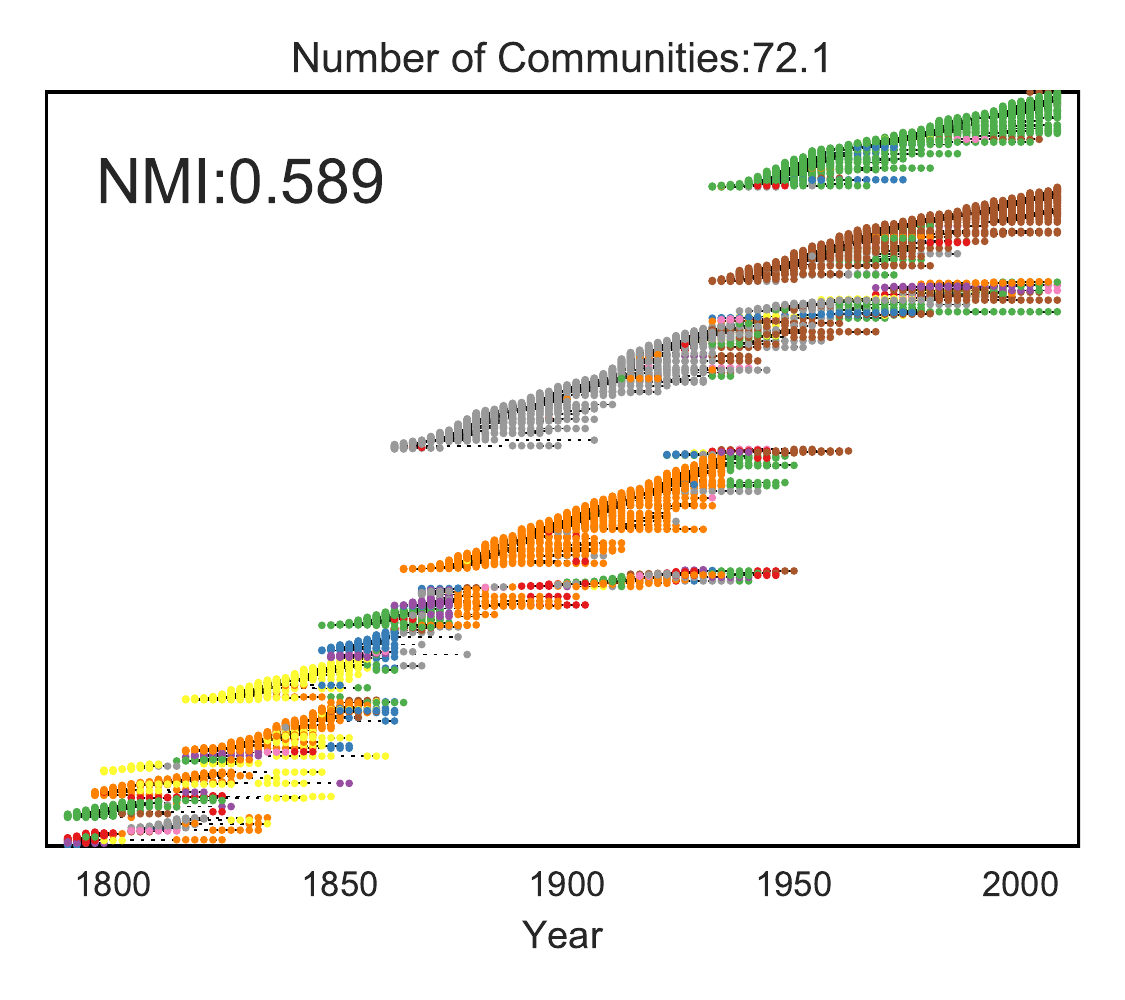}
  \includegraphics[width=.24\linewidth]{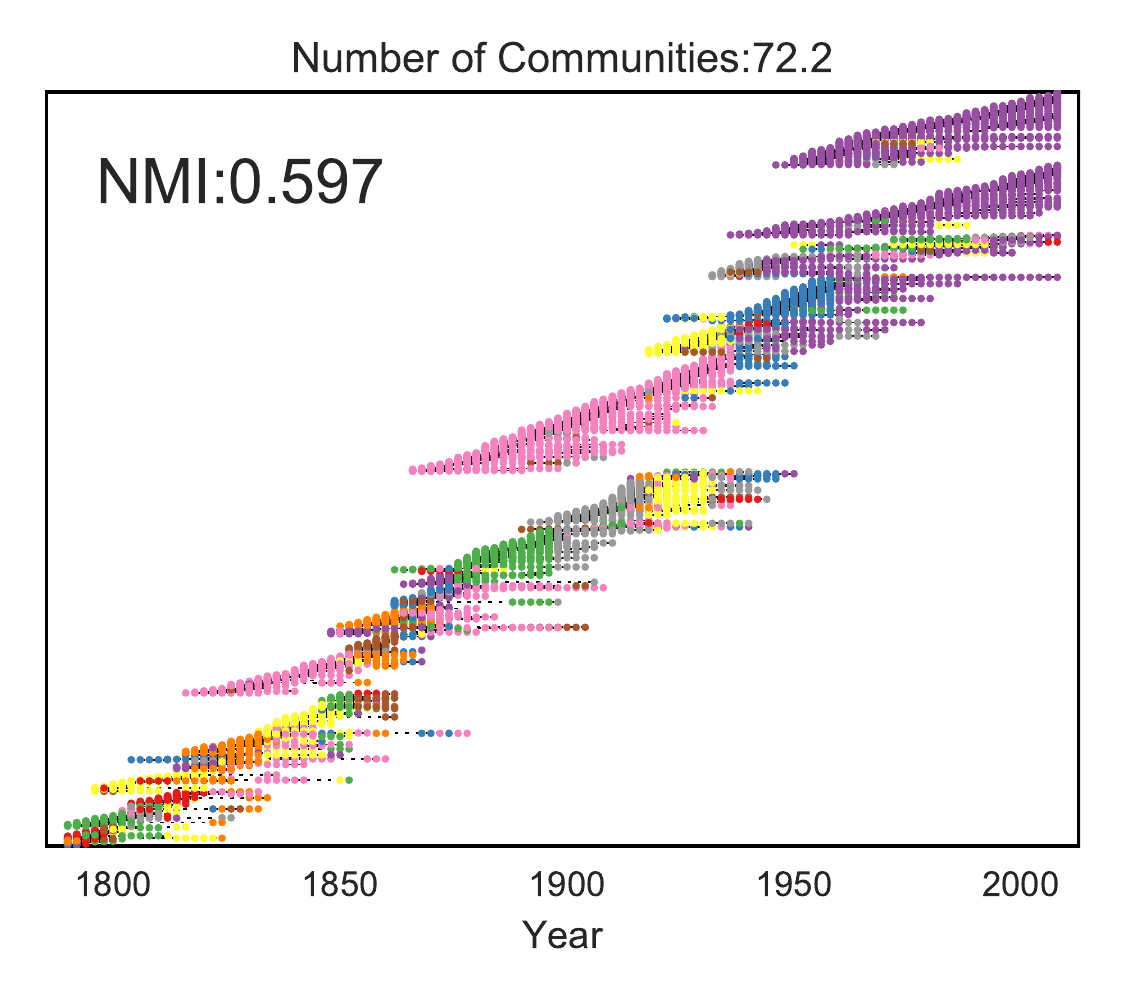}
\caption{Visualizations of  partitions labeled in white in Figure~\ref{DODs}A, with Senators sorted by their most frequent community label (with the labels sorted by last appearance in time), and within communities by first appearance. The listed AMI is the average over layers of the AMI in each layer (Congress) between the communities and political party affiliations in that Congress. }
\label{individs}
\end{figure}

In this Appendix, we include visualizations of each of the partitions with domains of optimization labeled in white text in Figure~\ref{DODs}A. In Figure~\ref{state}, the Senators are plotted according to their states. 
In~Figure~\ref{individs}, the individual Senators have been sorted according to community assignment and, within communities, time of first appearance in the Senate. 

We call particular attention to the qualitative difference between the community structures with domains above and below the transition around $\gamma\approx 0.8$. Below $\gamma\approx 0.8$, each Congress layer has only a single community, with the communities broken up across time. In the region just above this transition, the typical Congress layer has two communities, with the community structure corresponding to an evolving two-party system over time. 

\externalbibliography{yes}


\begin{thebibliography}{-------}
\providecommand{\natexlab}[1]{#1}

\end{thebibliography}


\begin{thebibliography}{-------}
\providecommand{\natexlab}[1]{#1}

\bibitem[Porter \em{et~al.}(2009)Porter, Onnela, and Mucha]{Porter:2009}
Porter, M.A.; Onnela, J.P.; Mucha, P.J.
\newblock Communities in networks.
\newblock {\em Not. AMS} {\bf 2009}, {\em 56},~1082--1097,
  1164--1166.

\bibitem[Fortunato(2010)]{Fortunato:2010}
Fortunato, S.
\newblock Community detection in graphs.
\newblock {\em Phys. Rep.} {\bf 2010}, {\em 486},~75--174.

\bibitem[Fortunato and Hric(2016)]{Fortunato:2016}
Fortunato, S.; Hric, D.
\newblock Community detection in networks: A user guide.
\newblock {\em Phys. Rep.} {\bf 2016}, {\em 659},~1 -- 44.

\bibitem[Abbe(2017)]{Abbe:2017}
Abbe, E.
\newblock Community detection and stochastic block models: Recent developments.
\newblock {\em arXiv} {\bf 2017}, arXiv:1703.10146. 

\bibitem[Schaub \em{et~al.}(2017)Schaub, Delvenne, Rosvall, and
  Lambiotte]{Schaub:2017}
Schaub, M.T.; Delvenne, J.C.; Rosvall, M.; Lambiotte, R.
\newblock The many facets of community detection in complex networks.
\newblock {\em Appl. Netw. Sci.} {\bf 2017}, {\em 2},~4.

\bibitem[Shai \em{et~al.}(2017)Shai, Stanley, Granell, Taylor, and
  Mucha]{Shai:2017}
Shai, S.; Stanley, N.; Granell, C.; Taylor, D.; Mucha, P.J.
\newblock Case studies in network community detection.
\newblock {\em arXiv} {\bf 2017}, arXiv:1705.02305.

\bibitem[Newman and Girvan(2004)]{NewmanGirvan:2004}
Newman, M.E.J.; Girvan, M.
\newblock Finding and evaluating community structure in networks.
\newblock {\em Phys. Rev. E} {\bf 2004}, {\em 69},~026113.

\bibitem[Reichardt and Bornholdt(2006)]{Reichardt:2006}
Reichardt, J.; Bornholdt, S.
\newblock Statistical mechanics of community detection.
\newblock {\em Phys. Rev. E} {\bf 2006}, {\em 74},~016110.

\bibitem[Fortunato and Barth\'elemy(2007)]{Fortunato:2007}
Fortunato, S.; Barth\'elemy, M.
\newblock Resolution limit in community detection.
\newblock {\em Proc. Natl. Acad. Sci. USA} {\bf 2007},
  {\em 104},~36--41.

\bibitem[Arenas \em{et~al.}(2008)Arenas, Fern\'andez, and G\'omez]{Arenas:2008}
Arenas, A.; Fern\'andez, A.; G\'omez, S.
\newblock Analysis of the structure of complex networks at different resolution
  levels.
\newblock {\em New J. Phys.} {\bf 2008}, {\em 10}, 053039.

\bibitem[Granell \em{et~al.}(2011)Granell, G\'omez, and Arenas]{Granell:2011}
Granell, C.; G\'omez, S.; Arenas, A.
\newblock Mesoscopic analysis of networks: Applications to exploratory analysis
  and data clustering.
\newblock {\em Chaos} {\bf 2011}, {\em 21}, 016102.

\bibitem[Leicht and Newman(2008)]{Leicht:2008}
Leicht, E.A.; Newman, M.E.J.
\newblock Community Structure in Directed Networks.
\newblock {\em Phys. Rev. Lett.} {\bf 2008}, {\em 100},~118703.

\bibitem[Barber(2007)]{Barber:2007}
Barber, M.J.
\newblock Modularity and community detection in bipartite networks.
\newblock {\em Phys. Rev. E} {\bf 2007}, {\em 76},~066102.

\bibitem[Gomez \em{et~al.}(2009)Gomez, Jensen, and Arenas]{Gomez:2009}
Gomez, S.; Jensen, P.; Arenas, A.
\newblock Analysis of community structure in networks of correlated data. \linebreak
\newblock {\em Phys. Rev. E} {\bf 2009}, {\em 80},~016114.

\bibitem[Traag and Bruggeman(2009)]{Traag:2009}
Traag, V.A.; Bruggeman, J.
\newblock Community detection in networks with positive and negative links.
\newblock {\em Phys. Rev.~E} {\bf 2009}, {\em 80},~036115.

\bibitem[Mucha \em{et~al.}(2010)Mucha, Richardson, Macon, and
  Porter]{Mucha:2010vk}
Mucha, P.J.; Richardson, T.; Macon, K.; Porter, M.A.
\newblock {Community structure in time-dependent, multiscale, and multiplex
  networks}.
\newblock {\em Science} {\bf 2010}, {\em 328}, 876.

\bibitem[Brandes \em{et~al.}(2008)Brandes, Delling, Gaertler, Goerke, Hoefer,
  Nikoloski, and Wagner]{Brandes:2008}
Brandes, U.; Delling, D.; Gaertler, M.; Goerke, R.; Hoefer, M.; Nikoloski, Z.;
  Wagner, D.
\newblock On modularity clustering.
\newblock {\em IEEE Trans. Knowl. Data Eng.} {\bf 2008},
  {\em 20},~172--188.

\bibitem[Good \em{et~al.}(2010)Good, de~Montjoye, and Clauset]{Good:2010}
Good, B.H.; de~Montjoye, Y.A.; Clauset, A.
\newblock Performance of modularity maximization in practical contexts.
\newblock {\em Phys. Rev. E} {\bf 2010}, {\em 81},~046106.

\bibitem[Kivel{\"a} \em{et~al.}(2014)Kivel{\"a}, Arenas, Barthelemy, Gleeson,
  Moreno, and Porter]{Kivela:2014}
Kivel{\"a}, M.; Arenas, A.; Barthelemy, M.; Gleeson, J.P.; Moreno, Y.; Porter,
  M.A.
\newblock Multilayer networks.
\newblock {\em J.~Complex Netw.} {\bf 2014}, {\em 2},~203--271.

\bibitem[De~Domenico \em{et~al.}(2015)De~Domenico, Lancichinetti, Arenas, and
  Rosvall]{Rosvall:2015}
De~Domenico, M.; Lancichinetti, A.; Arenas, A.; Rosvall, M.
\newblock Identifying Modular Flows on Multilayer Networks Reveals Highly
  Overlapping Organization in Interconnected Systems.
\newblock {\em Phys. Rev. X} {\bf 2015}, {\em 5},~011027.

\bibitem[Rosvall and Bergstrom(2008)]{Rosvall:2008}
Rosvall, M.; Bergstrom, C.T.
\newblock Maps of random walks on complex networks reveal community structure.
\newblock {\em Proc. Natl. Acad. Sci. USA} {\bf 2008},
  {\em 105},~1118--1123.

\bibitem[Han \em{et~al.}(2015)Han, Xu, and Airoldi]{Airoldi:2015}
Han, Q.; Xu, K.S.; Airoldi, E.M., Consistent Estimation of Dynamic and
  Multi-layer Block Models.
\newblock In~Proceedings of the 32Nd International Conference on
  International Conference on Machine Learning - Volume 37, Lille, France,  6--11 July  2015; pp. 1511--1520.

\bibitem[Stanley \em{et~al.}(2016)Stanley, Shai, Taylor, and
  Mucha]{Stanley:2016}
Stanley, N.; Shai, S.; Taylor, D.; Mucha, P.J.
\newblock Clustering Network Layers with the Strata Multilayer Stochastic Block
  Model.
\newblock {\em IEEE Trans. Netw. Sci. Eng.} {\bf
  2016}, {\em 3},~95--105.

\bibitem[Taylor \em{et~al.}(2016)Taylor, Shai, Stanley, and Mucha]{Taylor:2016}
Taylor, D.; Shai, S.; Stanley, N.; Mucha, P.J.
\newblock Enhanced Detectability of Community Structure in Multilayer Networks
  through Layer Aggregation.
\newblock {\em Phys. Rev. Lett.} {\bf 2016}, {\em 116},~228301.

\bibitem[Paj()]{Pajek}
Detecting communities ({Pajek} and {PajekXXL}). Available online: \url{http://mrvar.fdv.uni-lj.si/pajek/community/CommunityDrawExample.htm} (accessed on 16 August 2017).

\bibitem[Fenn \em{et~al.}(2009)Fenn, Porter, McDonald, Williams, Johnson, and
  Jones]{Fenn:2009}
Fenn, D.J.; Porter, M.A.; McDonald, M.; Williams, S.; Johnson, N.F.; Jones,
  N.S.
\newblock Dynamic communities in multichannel data: An application to the
  foreign exchange market during the 2007--2008 credit crisis.
\newblock {\em Chaos} {\bf 2009}, {\em 19},~033119.

\bibitem[Fenn \em{et~al.}(2012)Fenn, Porter, Mucha, McDonald, Williams,
  Johnson, and Jones]{Fenn:2012}
Fenn, D.J.; Porter, M.A.; Mucha, P.J.; McDonald, M.; Williams, S.; Johnson,
  N.F.; Jones, N.S.
\newblock Dynamical clustering of exchange rates.
\newblock {\em Quant. Finance} {\bf 2012}, {\em 12},~1493--1520.

\bibitem[Macon \em{et~al.}(2012)Macon, Mucha, and Porter]{Macon:2012}
Macon, K.T.; Mucha, P.J.; Porter, M.A.
\newblock Community structure in the United Nations General Assembly.
\newblock {\em Phys. A} {\bf 2012}, {\em 391},~343--361.

\bibitem[Traag \em{et~al.}(2013)Traag, Krings, and Van~Dooren]{Traag:2013}
Traag, V.A.; Krings, G.; Van~Dooren, P.
\newblock Significant Scales in Community Structure.
\newblock {\em Sci. Rep.} {\bf 2013}, {\em 3}, 2930.

\bibitem[Lewis \em{et~al.}(2010)Lewis, Jones, Porter, and Deane]{Lewis:2010}
Lewis, A.C.; Jones, N.S.; Porter, M.A.; Deane, C.M.
\newblock The function of communities in protein interaction networks at
  multiple scales.
\newblock {\em BMC Syst. Biol.} {\bf 2010}, {\em 4}, doi:10.1186/1752-0509-4-100.

\bibitem[Traud \em{et~al.}(2011)Traud, Kelsic, Mucha, and Porter]{Traud:2011}
Traud, A.L.; Kelsic, E.D.; Mucha, P.J.; Porter, M.A.
\newblock Comparing Community Structure to Characteristics in Online Collegiate
  Social Networks.
\newblock {\em SIAM Rev.} {\bf 2011}, {\em 53},~526--543.

\bibitem[Meil\v{a}(2007)]{Meila:2007}
Meil\v{a}, M.
\newblock Comparing clusterings---An information based distance.
\newblock {\em J. Multivar. Anal.} {\bf 2007}, {\em 98},~873--895.

\bibitem[Fred and Jain(2003)]{FredJain:2003}
Fred, A.L.N.; Jain, A.K. Robust data clustering.
\newblock In  Proceedings of the   2003 IEEE Computer Society Conference on Computer Vision and Pattern Recognition,  Madison, WI, USA, 18--20 June 2003; Volume~2, pp. II--128.

\bibitem[Bassett \em{et~al.}(2013)Bassett, Porter, Wymbs, Grafton, Carlson, and
  Mucha]{Bassett:2013}
Bassett, D.S.; Porter, M.A.; Wymbs, N.F.; Grafton, S.T.; Carlson, J.M.; Mucha,
  P.J.
\newblock Robust detection of dynamic community structure in networks.
\newblock {\em Chaos } {\bf
  2013}, {\em 23},~013142.

\bibitem[Ovelg\"onne and Geyer-Schulz(2012)]{Ovelgonne:2012}
Ovelg\"onne, M.; Geyer-Schulz, A.
\newblock An ensemble learning strategy for graph clustering. {\em Contemp. Math.}
\newblock  2012, {\em 588}, 187--206.

\bibitem[Lancichinetti and Fortunato(2012)]{Lancichinetti:2012}
Lancichinetti, A.; Fortunato, S.
\newblock Consensus clustering in complex networks.
\newblock {\em Sci. Rep.} {\bf 2012}, {\em 2}, 336.

\bibitem[Evans(2010)]{Evans:2010ga}
Evans, T.S.
\newblock {Clique graphs and overlapping communities}.
\newblock {\em J. Stat. Mech. Theory  Exp.} {\bf
  2010}, {\em 2010}, doi:10.1088/1742-5468/2010/12/P12037.

\bibitem[Girvan and Newman(2002)]{Girvan:2002ez}
Girvan, M.; Newman, M.E.J.
\newblock {Community structure in social and biological networks.}
\newblock {\em Proc. Natl. Acad. Sci. USA} {\bf 2002},
  {\em 99},~7821--7826.

\bibitem[De~Domenico \em{et~al.}(2013)De~Domenico, Sol{\'e}-Ribalta, Cozzo,
  Kivel{\"a}, Moreno, Porter, G{\'o}mez, and Arenas]{DeDomenico:2013}
De~Domenico, M.; Sol{\'e}-Ribalta, A.; Cozzo, E.; Kivel{\"a}, M.; Moreno, Y.; Porter,
  M.A.; G{\'o}mez, S.; Arenas, A.
\newblock Mathematical Formulation of Multilayer Networks.
\newblock {\em Phys. Rev. X} {\bf 2013}, {\em 3},~041022.

\bibitem[Bazzi \em{et~al.}(2016)Bazzi, Porter, Williams, McDonald, Fenn, and
  Howison]{Bazzi:2016}
Bazzi, M.; Porter, M.A.; Williams, S.; McDonald, M.; Fenn, D.J.; Howison, S.D.
\newblock Community Detection in Temporal Multilayer Networks, with an
  Application to Correlation Networks.
\newblock {\em Multiscale Modeling  Simul.} {\bf 2016}, {\em 14},~1--41.

\bibitem[Jeub \em{et~al.}(2011--2016)Jeub, Bazzi, Jutla, and Mucha]{GenLouvain}
Jeub, L.G.S.; Bazzi, M.; Jutla, I.S.; Mucha, P.J.
\newblock A generalized {Louvain} method for community detection implemented in
  MATLAB,  2011--2016.
\newblock Available online: \url{http://netwiki.amath.unc.edu/GenLouvain} (accessed on 16 August 2017). 

\bibitem[Blondel \em{et~al.}(2008)Blondel, Guillaume, Lambiotte, and
  Lefebvre]{Blondel:2008vn}
Blondel, V.D.; Guillaume, J.L.; Lambiotte, R.; Lefebvre, E.
\newblock {Fast unfolding of communities in large networks}.
\newblock {\em J. Stat. Mech. Theory  Exp.} {\bf
  2008}, {\em 2008}, 155--168.

\bibitem[qhu()]{qhull}
Qhull. Available online: \url{http://www.qhull.org/} (accessed on 16 August 2017). 


\bibitem[pyh()]{pyhull}
Pyhull. Available online: \url{http://pythonhosted.org/pyhull/} (accessed on 16 August 2017). 


\bibitem[Barber \em{et~al.}(1996)Barber, Dobkin, and Huhdanpaa]{Barber:1996iv}
Barber, C.B.; Dobkin, D.P.; Huhdanpaa, H.
\newblock The Quickhull Algorithm for Convex Hulls.
\newblock {\em ACM Trans. Math. Softw.} {\bf 1996}, {\em 22},~469--483.

\bibitem[Weir(2017)]{champ_software}
Weir, W.; Gibson, R.; Mucha, P.J.
\newblock CHAMP package: Convex Hull of Admissible Modularity Partitions in Python and MATLAB,  2017.
\newblock Available online:  \url{https://github.com/wweir827/CHAMP} (accessed on 16 August 2017).

\bibitem[Traag()]{traag:louvain}
Traag, V.
\newblock Louvain igraph.
\newblock Available online: \url{http://github.com/vtraag/louvain-igraph} (accessed on 16~August 2017).

\bibitem[Vinh {et~al.}(2009)Vinh, Epps, and Bailey]{Vinh:2009}
Vinh, N.X.; Epps, J.; Bailey, J.
\newblock Information Theoretic Measures for Clusterings Comparison: Is a
  Correction for Chance Necessary? In Proceedings of the 26th Annual International Conference on Machine Learning, Montreal, QC, Canada, 14--18 June  2009;
\newblock pp. 1073--1080.

\bibitem[Hubert and Arabie(1985)]{Hubert1985}
Hubert, L.; Arabie, P.
\newblock Comparing partitions.
\newblock {\em J. Classif.} {\bf 1985}, {\em 2},~193--218.

\bibitem[Meil{\u{a}}(2003)]{Meila2003}
Meil{\u{a}}, M. Comparing Clusterings by the Variation of Information.
\newblock In  {\em Proceedings of the 16th Annual Conference
  on Learning Theory and 7th Kernel Workshop, COLT/Kernel 2003, Learning Theory and Kernel Machine, Washington, DC,
  USA, 24--27 August 2003}; Sch{\"o}lkopf, B., Warmuth, M.K.,
  Eds.; Springer Berlin Heidelberg: Berlin/Heidelberg, Gernany, 2003; pp. 173--187.

\bibitem[Rijsbergen(1979)]{Rijsbergen:1979}
Rijsbergen, C.J.V.
\newblock {\em Information Retrieval}, 2nd ed.; Butterworth-Heinemann: Newton,
  MA, USA,  1979.

\bibitem[Jacomy \em{et~al.}(2014)Jacomy, Venturini, Heymann, and
  Bastian]{jacomy_et_al}
Jacomy, M.; Venturini, T.; Heymann, S.; Bastian, M.
\newblock ForceAtlas2, a Continuous Graph Layout Algorithm for Handy Network
  Visualization Designed for the Gephi Software.
\newblock {\em PLoS ONE} {\bf 2014}, {\em 9}, doi:10.1371/journal.pone.0098679.

\bibitem[Peixoto(2014)]{peixoto_graph-tool_2014}
Peixoto, T.P.
\newblock The graph-tool python library,  2014.
\newblock Available online: \url{http://figshare.com/articles/graph_tool/1164194} (accessed on 16 August 2017). 



\bibitem[Joshi-Tope \em{et~al.}(2005)Joshi-Tope, Gillespie, Vastrik,
  D'Eustachio, Schmidt, de~Bono, Jassal, Gopinath, Wu, Matthews, Lewis, Birney,
  and Stein]{JoshiTope:2005eh}
Joshi-Tope, G.; Gillespie, M.; Vastrik, I.; D'Eustachio, P.; Schmidt, E.;
  de~Bono, B.; Jassal, B.; Gopinath, G.R.; Wu, G.R.; Matthews, L.; et al.
\newblock {Reactome: A knowledgebase of biological pathways.}
\newblock {\em Nucleic Acids Res.} {\bf 2005}, {\em 33},~D428--D432.

\bibitem[Kunegis(2013)]{Kunegis:2013ih}
Kunegis, J.
\newblock {\em {KONECT: The Koblenz Network Collection}}; The Koblenz Network
  Collection; ACM: New York, NY, USA,  2013.

\bibitem[Waugh {et~al.}(2009)Waugh, Pei, Fowler, Mucha, and
  Porter]{Waugh:2009vz}
Waugh, A.S.; Pei, L.; Fowler, J.H.; Mucha, P.J.; Porter, M.A.
\newblock {Party Polarization in Congress: A Network Science Approach}. 
\newblock {\em arXiv} {\bf 2009}, arXiv:0907.3509.

\bibitem[Mucha and Porter(2010)]{MuchaPorter:2010}
Mucha, P.J.; Porter, M.A.
\newblock Communities in multislice voting networks.
\newblock {\em Chaos} {\bf 2010}, {\em 20},~041108.

\bibitem[De~Montgolfier {et~al.}(2011)De~Montgolfier, Soto, and
  Viennot]{DeMontgolfier:2011jp}
De~Montgolfier, F.; Soto, M.; Viennot, L.
\newblock {Asymptotic modularity of some graph classes}.
\newblock {\em Int. Symp. Algorithms Comput.} {\bf 2011}, {\em 7074}, {435--444}.

\bibitem[Bagrow(2012)]{Bagrow:2012hy}
Bagrow, J.P.
\newblock {Communities and bottlenecks: Trees and treelike networks have high
  modularity.}
\newblock {\em Phys. Rev. E} {\bf 2012}, {\em 85},~066118.

\end{thebibliography}
\end{document}